\documentclass{jfm}

\usepackage{graphicx}
\usepackage{subcaption}
\usepackage{amsmath,bm}
\usepackage{natbib}
\usepackage{multirow}
\usepackage{hhline}
\usepackage{booktabs}
\usepackage{makecell}
\usepackage{tikz}
\usepackage{wasysym}
\usepackage{enumerate,enumitem}

\ifCUPmtlplainloaded \else
  \checkfont{eurm10}
  \iffontfound
    \IfFileExists{upmath.sty}
      {\typeout{^^JFound AMS Euler Roman fonts on the system,
                   using the 'upmath' package.^^J}%
       \usepackage{upmath}}
      {\typeout{^^JFound AMS Euler Roman fonts on the system, but you
                   dont seem to have the}%
       \typeout{'upmath' package installed. JFM.cls can take advantage
                 of these fonts,^^Jif you use 'upmath' package.^^J}%
      }
  \else
  \fi
\fi

\ifCUPmtlplainloaded \else
  \checkfont{msam10}
  \iffontfound
    \IfFileExists{amssymb.sty}
      {\typeout{^^JFound AMS Symbol fonts on the system, using the
                'amssymb' package.^^J}%
       \usepackage{amssymb}%
       \let\le=\leqslant  
         
      }{}
  \fi
\fi

\ifCUPmtlplainloaded \else
  \IfFileExists{amsbsy.sty}
    {\typeout{^^JFound the 'amsbsy' package on the system, using it.^^J}%
     \usepackage{amsbsy}}
    {}
\fi

\newsavebox{\astrutbox}
\sbox{\astrutbox}{\rule[-5pt]{0pt}{20pt}}

\title[]{Invariant states in inclined layer convection. Part 1. Temporal transitions along dynamical connections between invariant states}

\author[F. Reetz \textit{et al.}]%
{FLORIAN REETZ$^1$ \ns \& \ns TOBIAS M.\ns SCHNEIDER$^1$ \thanks{Email address for correspondence: tobias.schneider@epfl.ch}}
\affiliation{$^1$ Emergent Complexity in Physical Systems Laboratory (ECPS), Ecole Polytechnique F{\'e}d{\'e}rale de Lausanne, CH-1015, Switzerland}

\date{?? and in revised form ??}

\begin{document}

\maketitle

\begin{abstract}
Thermal convection in an inclined layer between two parallel walls kept at different fixed temperatures is studied for fixed Prandtl number $\mathrm{Pr}=1.07$. Depending on the angle of inclination and the imposed temperature difference, the flow exhibits a large variety of self-organized spatio-temporal convection patterns. Close to onset, these patterns have been explained in terms of linear stability analysis of primary and secondary flow states. At larger temperature difference, far beyond onset, experiments and simulations show complex, dynamically evolving patterns that are not described by stability analysis and remain to be explained. Here we employ a dynamical systems approach. We construct stable and unstable exact invariant states, including equilibria and periodic orbits of the fully nonlinear three-dimensional Oberbeck-Boussinesq equations. These invariant states underlie the observed convection patterns beyond their onset. We identify state-space trajectories that, starting from the unstable laminar flow, follow a sequence of dynamical connections between unstable invariant states until the dynamics approaches a stable attractor. Together, the network of dynamically connected invariant states mediates temporal transitions between coexisting invariant states and thereby supports the observed complex time-dependent dynamics in inclined layer convection.

\end{abstract}

\section{Introduction}
Fluids in spatially extended wall-bounded domains can form regular flow patterns when driven by external forces \citep{Cross1993}. Even when the flow exhibits spatio-temporal chaos or is weakly turbulent, regular patterns may form. Prominent examples are chaotic spirals in thermal convection \citep{Morris1993}, or oblique turbulent-laminar stripes in shear flows \citep{Prigent2002}. These patterns emerge in dissipative systems that are not in thermodynamic equilibrium. Consequently, the formation of sustained patterns depends crucially on the strength and nature of the energy supplying external driving forces. 

A fluid system where not only the strength but also the nature of the driving force can be controlled and changed smoothly is inclined layer convection (ILC), the flow between two parallel walls maintained at different temperatures and inclined against gravity. Here, the angle of inclination defines the ratio between the wall-normal and the wall-parallel buoyancy force. The former drives a lift-up mechanism, by which buoyancy may directly destabilize the flow as in the non-inclined Rayleigh-B\'enard system. The latter generates shear forces between upward and downward driven flow, leading to shear instabilities. Many different convection patterns have been observed in ILC by systematically changing the angle of inclination from horizontal layer convection to vertical layer convection and beyond \citep{Daniels2000}. These observations also reveal complex spatio-temporal dynamics of convection patterns, such as intermittent bursting \citep{Busse2000,Daniels2003} or spatial competition between patterns \citep{Daniels2002a,Daniels2008}. While the onset of several convection patterns has been explained using stability analysis, the mechanisms underlying the complex dynamics far above onset are not well understood.

First experiments on ILC focused on heat transfer properties in an inclined layer of air at Prandtl number $\mathrm{Pr}\approx 0.7$ \citep{Nusselt1908,DeGraaf1953,Hollands1973,Ruth1980a}. Qualitative changes in the heat transfer were related to instabilities in the flow. Early linear stability analysis of laminar ILC at different $\mathrm{Pr}$ found two different primary instabilities \citep{Gershuni1969,Chen1989}. Depending on the angle of inclination, laminar flow becomes unstable to convection rolls with either longitudinal orientation, at small inclinations, or with transverse orientation, at large inclinations. This result was confirmed by systematic experimental surveys using water at $\mathrm{Pr}\approx 7$ \citep{Hart1971a} as well as experiments using liquid crystals at high $\mathrm{Pr}$ \citep{Shadid1990}. Observations of modulated longitudinal rolls \citep{Hart1971a,Hart1971b} were compared and related to secondary instabilities of longitudinal rolls calculated using stability analysis \citep{Clever1977a}. Similar primary and secondary instabilities have also been found in other shear flows with imposed temperature gradients \citep[see][for a review]{Kelly1994}.

Systematic experimental explorations of self-organized patterns in large aspect ratio domains of ILC under changing control parameters report on ten different convection patterns in compressed $CO_2$ at $\mathrm{Pr}=1.07$ \citep{Daniels2000,Daniels2002a,Daniels2003,Daniels2008}. While some of the observed patterns are sufficiently regular to resemble patterns linked to instabilities that had been described previously for other $\mathrm{Pr}$, most observations indicate complex dynamics including spatio-temporal chaos. Exploring the same parameter space studied by Daniels, Bodenschatz, Pesch and collaborators, \citet{Subramanian2016} identified five secondary instabilities using Floquet analysis. These instabilities were calculated at the critical values of the control parameters for the onset of the pattern and related to the dynamics observed in experiments and numerical simulations above these critical parameters using Galerkin methods \citep{Subramanian2016}. In summary, pattern formation in ILC has been studied extensively at different control parameter values using experiments, numerical simulations, and stability analysis.

Relating a pattern forming instability identified by stability analysis at a critical value of the control parameter to experimental or numerical observations above the critical value requires a particular underlying bifurcation structure: At a critical value of the control parameter, an attracting state $A$ loses stability to a forward bifurcating stable branch $B$. Above the critical control parameter value, the unstable pattern $A$ has lost dynamical relevance and the dynamics approaches the attracting state $B$ that has emerged at the critical control parameter value. Attracting state $B$ remains observable in the flow until it undergoes another bifurcation and itself loses stability. Explaining the succession of patterns observed in ILC and other flows based on stability analysis thus relies on two conditions: First, a forward bifurcating stable branch continues to values of the control parameters where the pattern is observed without undergoing another bifurcation. Second, both stable states, the one existing before and the one emerging in the bifurcation, are attracting the long-term dynamics at the respective values of the control parameter. This way, the states involved in the bifurcation control the observed dynamics. Under these conditions, a sequence of patterns can be described by a succession of single-state attractors arranged in a forward bifurcation sequence. However, such a `sequence of bifurcations'-approach \citep{Busse1996}, envisioning a forward bifurcating scenario, is not applicable \emph{a priori}. Rather, in order to describe observed patterns via sequences of forward bifurcations, the bifurcation structure needs to be confirmed by following the fully nonlinear bifurcation branches. Moreover there might not be a single attracting state as evidenced by observations of complex non-saturated temporally evolving dynamics in large domains. The time-dependent, complex dynamics was speculated to be a consequence of experimental imperfections \citep{Clever1995,Busse1996} but have also been observed in direct numerical simulations in the absence of such imperfections \citep{Subramanian2016}. Consequently, an alternative approach is required to explain those complex patterns beyond onset.

Recent studies of subcritical shear flows have demonstrated the dynamical relevance of unstable exact invariant states, also called exact coherent states \citep[][and references therein]{Kawahara2012}. Invariant states are numerically fully resolved exact solutions of the governing nonlinear Navier-Stokes equations representing non-trivial flow structures or patterns in the flow as either steady equilibrium states or exact periodic orbits. The dynamical relevance of weakly unstable invariant states follows from their ability to transiently attract and repel the dynamics along their stable and unstable manifolds \citep{Gibson2008,Halcrow2009,Chandler2013,Suri2017,Farano2019}. Whenever invariant states are transiently approached by the dynamics, they become transiently observable in the flow \citep{Hof2004}. These results support a dynamical systems description of turbulent flow where invariant states and their stable and unstable manifolds form a dynamical network embedded in the `strange' state space attractor generating the complex turbulent dynamics \citep{Lanford1982}. Likewise, within this nonlinear dynamical systems approach, we expect unstable invariant states in ILC representing pattern motifs to support the complex pattern dynamics observed in experiments and simulations.

Shortly after the discovery of the first unstable invariant state in Couette flow \citep{Nagata1990,Clever1992,Waleffe1998}, invariant states were also identified in ILC. \citet{Busse1992} revisited their analysis of the wavy instability of longitudinal rolls \citep{Clever1977a}, and constructed stable and unstable finite amplitude states corresponding to wavy rolls combining a Galerkin method with Newton-Raphson iteration. \citet{Clever1995} applied the same approach to tertiary and quarternary states for convection in a vertical layer, where shear forces dominate over buoyancy. Since then, invariant states have not been studied in ILC. In pure shear flows however, the significance of invariant states for the temporal transition between subcritical laminar and turbulent shear flows was extensively investigated \citep{Kerswell2005,Eckhardt2007,Kawahara2012}. In linearly stable shear flows, the transition to turbulence requires finite amplitude perturbations of the stable laminar flow that cross the edge of chaos between laminar and turbulent attractors in state space. This edge is spanned by the stable manifold of invariant states with a single unstable direction, a so-called edge state \citep{Skufca2006,Schneider2007b}, such that the edge separates the coexisting attractors of turbulent and laminar flow \citep{Schneider2008b}. Consequently, invariant edge states guide the transition to turbulence for linearly stable flows. In contrast to canonical subcritical shear flows, the laminar flow in ILC undergoes a linear instability so that infinitesimal perturbations are sufficient to trigger temporal transitions away from laminar flow. The role of invariant states for the dynamics leaving the unstable laminar flow and their significance for the observed complex dynamics has not been investigated in ILC. They may act as transiently visited unstable states or serve as asymptotic attractors.

In the present article we numerically study three-dimensional ILC at $\mathrm{Pr}=1.07$ in minimal doubly periodic domains and identify stable and unstable invariant states underlying different convection patterns at selected values of the control parameters where these basic convection patterns are observed in simulations and experiments. Temporal transitions from unstable laminar flow are characterized using a phase portrait analysis of the state space trajectories describing the temporal evolution. For seven different combinations of inclination angle and imposed temperature difference transient visits to unstable invariant states are observed before the dynamics approaches attracting stable invariant states. 

Depending on the inclination angle, the instability of the laminar flow is either driven by buoyancy or shear \citep{Chen1989,Daniels2000}. At small inclinations, shear forces are negligible in the laminar state so that the emerging longitudinal convection rolls are associated with a buoyancy driven instability. At large inclinations, the wall-normal lift-up mechanism due to buoyancy is negligible so that the instability giving rise to transverse convection rolls is shear driven. Disentangling the role of buoyancy and shear for higher order instabilities driving the dynamics away from non-trivial unstable states is not straightforward as even at low inclinations, the flow field of any type of convection roll will produce significant shear, and at any inclination, temperature gradients aligned with gravity will lead to buoyant forcing. We demonstrate that phase portraits based on energy transport rates provide a systematic approach for clearly characterising any instability of an equilibrium state as either shear or buoyancy driven.

The article has the following structure. Section \ref{sec:obe} introduces the governing equations for ILC, symmetries of the system and equations for energy transfer. Numerical methods for a dynamical systems description are introduced in Section \ref{sec:numerics}. Temporal transitions between invariant states are presented in seven phase portraits in Section \ref{sec:transitions} and discussed in Section \ref{sec:discussion}.

\section{Oberbeck-Boussinesq equations for inclined layers}
\label{sec:obe}

\begin{figure}
\centering
\includegraphics[width=0.7\linewidth,trim={9cm 7cm 8cm 3cm},clip]{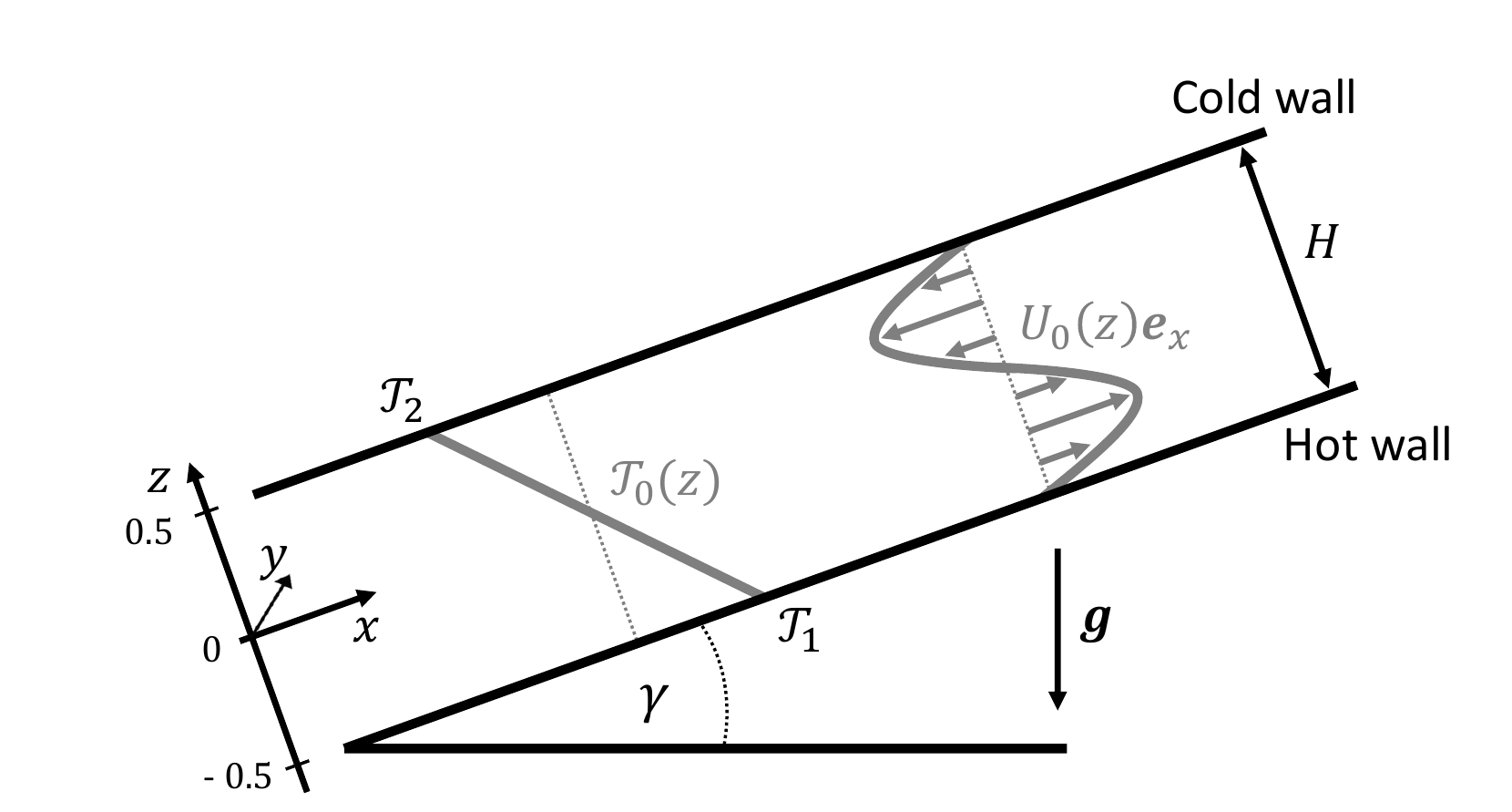}
\caption{\label{fig:ilcbase} Schematic of inclined layer convection. Streamwise, spanwise and wall-normal dimensions are indicated by $x$, $y$ and $z$, respectively. A layer of an incompressible Newtonian fluid is confined between a lower hot and an upper cold wall. The layer is inclined against gravity $\bm{g}$ at angle $\gamma$. Hot fluid flows up the hot wall while cold fluid descends along the cold wall generating a laminar base flow (\ref{eq:baseU}-\ref{eq:baseT}) with linear temperature profile $\mathcal{T}_0(z)$ and cubic velocity profile $U_0(z)$, as outlined by grey lines. The competition of buoyancy and shear gives rise to a variety of intricate convection patterns when the three control parameters, inclination $\gamma$, thermal driving $\mathrm{Ra}$ and Prandtl number $\mathrm{Pr}$ are varied. }
\end{figure}

\noindent We consider thermal convection of a Newtonian fluid in an infinite layer of thickness $H$ confined between a hot and a cold wall at prescribed temperatures $\mathcal{T}_1$ and $\mathcal{T}_2$, respectively. The fluid layer is inclined against the vector of gravitational acceleration $\bm{g}$ by angle $\gamma$ (Figure \ref{fig:ilcbase}). The dynamics of the incompressible flow with velocity vector $\bm{U}=[U,V,W](x,y,z,t)$, temperature $\mathcal{T}=\mathcal{T}(x,y,z,t)$, and pressure $p=p(x,y,z,t)$ relative to the hydrostatic pressure $P=P(x,y,z,t)$, where $\nabla P=\hat{\bm{g}}$, is given by the nondimensionalised Oberbeck-Boussinesq equations
\begin{align}
 \frac{\partial \bm{U}}{\partial t} + \left(\bm{U}\cdot \nabla \right)\bm{U} &=- \nabla p + \tilde{\nu}\nabla^2 \bm{U} -\hat{\bm{g}}\,\mathcal{T} \ ,\label{eq:obe1}\\
 \frac{\partial \mathcal{T}}{\partial t} + \left(\bm{U}\cdot\nabla \right) \mathcal{T} &= \tilde{\kappa}  \nabla^2  \mathcal{T}\ , \label{eq:obe2}\\
 \nabla \cdot \bm{U} &=0 \ ,\label{eq:obe3} 
\end{align}
with $\tilde{\nu}=(\mathrm{Pr}/\mathrm{Ra})^{1/2} $ and $\tilde{\kappa}=(\mathrm{Pr}\,\mathrm{Ra})^{-1/2} $. This set of nonlinear partial differential equations has three control parameters: the angle of inclination $\gamma$ against the gravitational unit vector $\hat{\bm{g}}=-\sin(\gamma) \hat{\bm{e}}_x - \cos(\gamma)\hat{\bm{e}}_z $, the Prandtl number $\mathrm{Pr}=\nu/\kappa$, the ratio between kinematic viscosity $\nu$ and thermal diffusivity $\kappa$, and the Rayleigh number $\mathrm{Ra}=g\,\alpha\,\Delta T\,H^3/(\nu \kappa)$ where $\Delta \mathcal{T}=\mathcal{T}_1-\mathcal{T}_2$ and $\alpha$ is the thermal expansion coefficient. 

In the nondimensionalised equations (\ref{eq:obe1}-\ref{eq:obe3}), temperature is measured in units of $\Delta \mathcal{T}$ and lengths in units of $H$. To describe convective fluid motion with an appropriate scale, we choose to measure velocity in units of the free fall velocity $U_f=(g\,\alpha\,\Delta \mathcal{T}\,H)^{1/2}$ that has also been used in previous studies of Rayleigh-B\'enard convection at values of the control parameters above convection onset \citep[e.g.][]{Gray1976,Chilla2012}. The free fall velocity scale implies a free-fall time unit $T_f=(H/g\,\alpha\,\Delta \mathcal{T})^{1/2}$. Note that an alternative nondimensionalisation using the heat diffusion time scale $T_d=H^2/\kappa$ is also common in thermal convection studies \citep[e.g.][]{Subramanian2016}. The conversion factor is $T_f=T_d/\sqrt{\mathrm{Ra}\,\mathrm{Pr}}$.

The nondimensionalised boundary conditions at the walls are 
\begin{align}
 \bm{U}(z=\pm 0.5)&=0 \label{eq:ubc} \ ,\\
 \mathcal{T}(z=\pm 0.5)&=\mp 0.5 \label{eq:tbc} \ .
\end{align}

\subsection{Laminar base flow}
\label{sec:laminar}
Equations (\ref{eq:obe1}-\ref{eq:obe3}) with boundary conditions (\ref{eq:ubc}-\ref{eq:tbc}) admit a laminar solution that only depends on the wall-normal coordinate $z$ and is spatially uniform in $x$ and $y$
\begin{align}
\bm{U}_0(z)&= \frac{\sin(\gamma)}{6\,\tilde{\nu}} \left(z^3  - \frac{1}{4}z\right) \bm{e}_x\ , \label{eq:baseU}\\
\mathcal{T}_0(z)&= -z      \ ,  \label{eq:baseT}\\
p_0(z)&=   \Pi -\cos(\gamma)z^2/2    \ ,  \label{eq:baseP}
\end{align}
with arbitrary pressure constant $\Pi$. The linear temperature profile and the cubic velocity profile of this laminar base flow are sketched in Figure \ref{fig:ilcbase} (grey lines). 
Within the laminar solution, buoyancy forces caused by the linear temperature profile as well as shear forces due to the velocity gradients in the buoyancy driven cubic velocity profile are present. The former is destabilizing for $-90^{\circ}<\gamma<90^{\circ}$ while shear can lead to instabilities at all non-zero inclination angles. At sufficiently strong driving, instabilities create overturning convective motion so that the laminar solution is no longer observed and the symmetries of ILC are broken.

\subsection{Symmetries} 
\label{sec:symmetries}

ILC at zero inclination ($\gamma=0^{\circ}$) corresponds to Rayleigh-B\'enard convection with isotropy and homogeneity in the $x$-$y$-plane. At all inclinations $0^{\circ}\neq\gamma\neq 180^{\circ}$, the isotropy of the horizontal layer is broken by the wall-parallel component of gravity, driving the laminar flow along the $x$-dimension. The laminar flow in ILC is still homogeneous and thereby invariant under continuous translations
\begin{equation}
\tau'(\Delta x, \Delta y)[U,V,W,\mathcal{T}](x,y,z)=[U,V,W,\mathcal{T}](x+\Delta x,y+\Delta y,z) \ . \label{eq:symtransinf}
\end{equation}
Moreover, ILC is invariant under discrete reflections
\begin{align}
\pi_y[U,V,W,\mathcal{T}](x,y,z)&=[U,-V,W,\mathcal{T}](x,-y,z) \label{eq:sympiy}\ ,\\
\pi_{xz}[U,V,W,\mathcal{T}](x,y,z)&=[-U,V,-W,-\mathcal{T}](-x,y,-z) \label{eq:sympixz}\ .
\end{align}
The symmetry group of ILC consists of all products of the generators $\{\pi_{y}, \pi_{xz}, \tau'(\Delta x, \Delta y)\}$. We indicate this group by $S_{ilc}=\langle \pi_{y}, \pi_{xz}, \tau'(\Delta x, \Delta y) \rangle$, where angle brackets $\langle \rangle$ imply all products of elements given in the brackets. ILC has the same symmetries as plane Couette flow where analogous notation is commonly used \citep[e.g][]{Gibson2014}.

Instead of considering an infinite fluid layer, we consider a finite periodic fluid layer by imposing periodic boundary conditions in $x$ and in $y$, $[U,V,W,\mathcal{T}](x,y,z)=[U,V,W,\mathcal{T}](x+ L_x,y,z)$ and $[U,V,W,\mathcal{T}](x,y,z)=[U,V,W,\mathcal{T}](x,y+ L_y,z)$, respectively. Due to the periodic boundary conditions, we express continuous translations as
\begin{equation}
\tau(a_x,a_y)[U,V,W,\mathcal{T}](x,y,z)=[U,V,W,\mathcal{T}](x+a_x L_x,y+a_y L_y,z) \ , \label{eq:symtrans}
\end{equation}
with shift factors $a_x,a_y\in [0,1)$ scaling the spatial periods $L_x$ and $L_y$ of the periodic domain. Continuous translations in periodic domains are cyclic and shifts by $L_x$ or $L_y$ correspond to the identity operator $\tau(0,0)$. Since the streamwise direction $x$ and the spanwise direction $y$ of ILC can be rotated and reflected, the symmetry group of ILC in $x$-$y$-periodic domains is isomorphic to the direct product of two copies of the orthogonal group in two dimensions, $O(2)$, that is $O(2) \times O(2)$. In the product, one term refers to the $x$-dimension, the other to the $y$-dimension.

The relevance of the system's symmetries for the dynamics is that once a state is invariant under a symmetry transformation of the equivariance group $S_{ilc}$, $[\bm{U},\mathcal{T}] = \sigma [\bm{U},\mathcal{T}]$ with $\sigma \in S_{ilc}$, the evolution under the full nonlinear governing equations  (\ref{eq:obe1}-\ref{eq:obe3}) will preserve the symmetry and the evolving trajectory will remain in the symmetry subspace of all possible states invariant under $\sigma$ \citep[e.g.][]{chaosbookSec10}. Consequently, trajectories and invariant states of the infinitely extended system without any symmetry constraints can be computed in symmetry subspaces, including those defined by the discrete translation symmetries imposed by periodic boundary conditions. To compute states in symmetry subspaces defined by a discrete symmetry $\sigma\in S_{ilc}$ satisfying $\sigma^2=1$, we impose $\sigma$ using a projection $([\bm{U},\mathcal{T}] + \sigma [\bm{U},\mathcal{T}])/2$ during simulations. Any exact solution in a symmetry subspace remains a valid solution of the full unconstrained infinite system. Imposing symmetries does not affect the state but may disallow instabilities breaking the imposed symmetries and thereby simplifies numerical access to invariant states with symmetries.

All invariant states discussed in the present article are invariant under transformations of subgroups of $S_{ilc}=\langle \pi_{y}, \pi_{xz}, \tau(a_x,a_y) \rangle$. We will specify the generators of the symmetry group $S$ of invariant states in terms of the combinations of $\pi_{y}$, $\pi_{xz}$ and $\tau(a_x,a_y)$. The choice of generators is not unique because translations $\tau(a_x,a_y)$ define conjugacy classes of group elements, corresponding to the free choice of the spatial phase of invariant states in $x$ and $y$. We choose the spatial phase such that three-dimensional inversion $\pi_{xyz}=\pi_{y}\pi_{xz}$, where applicable to invariant states, applies with respect to the domain origin at $(x,y,z)=(0,0,0)$.

\subsection{Energy transfer}
\label{sec:energy}
ILC is a thermally driven dissipative system. The externally imposed temperature difference results in the thermal energy flux that is required to sustain those temperature gradients that, together with gravity, generate buoyancy forces driving fluid flow. Thereby thermal energy is converted to kinetic energy, that is eventually dissipated by conversion into heat through internal viscous friction. The kinetic energy balance is obtained by multiplying (\ref{eq:obe1}) with $\bm{U}$ and space averaging Equation (\ref{eq:obe1}) over the entire domain volume $\Omega$, denoted by $\langle  \rangle_{\Omega}$,
\begin{align}
 \frac{1}{2}\frac{\partial}{\partial t} \left\langle U^2 \right\rangle_{\Omega} &=& \left\langle \hat{\bm{g}}\,\bm{U}\,\mathcal{T} \right\rangle_{\Omega} &- \tilde{\nu} \left\langle(\nabla \times \bm{U})^2\right\rangle_{\Omega} &=& I-D \ . \label{eq:kenergy}
\end{align}
The rate of change of kinetic energy in $\Omega$ is given by the difference between energy input $I$, the work due to buoyancy forces, and viscous dissipation $D$ \citep{Malkus1964}. These rates may be normalised by the laminar transfer rate 
\begin{align}\label{eq:laminar_energy}
I_0&=D_0=\sin^2(\gamma)/(720\,\tilde\nu) 
\end{align}
for non-zero inclination, $\gamma\neq0^{\circ}$.

Since the kinetic energy of all equilibrium states remains constant, energy transfer rates need to be balanced, implying $I=D$. A periodic orbit will be characterized by instantaneously unbalanced rates but the net energy transfer integrated over one period $T$ of the orbit vanishes, $\int_0^{T}(I-D)\, dt=0$. For equilibria with relative dissipation $D/I=1$, Equation (\ref{eq:kenergy}) allows to distinguish two destabilising mechanisms. When buoyancy forces drive an instability of an equilibrium state, $I$ increases over $D$ implying $D/I<1$ for the initial dynamics triggered by the instability. A shear driven instability of an equilibrium leads initially to $D/I>1$ because rising shear increases $D$ over $I$. Local oscillatory instabilities of equilibrium states discussed in the present paper cause oscillation amplitudes to grow symmetrically around $D/I=1$ with an exponential growth rate. The symmetry around $D/I=1$ suggests that buoyancy and shear forces contribute equally to the destabilising mechanism underlying an oscillatory instability. We will characterize all invariant states and their instabilities in terms of energy transfer.

On average, the thermal heat flux through any plane parallel to the walls is independent of the height $z$. At the walls, the transport is purely diffusive but in the center of the domain convective heat transport can be significant. To quantify the instantaneous, time-dependent, heat transport due to convective effects, we formulate the energy balance equation for heat not averaged over the full but over the lower half of the domain. This generates boundary flux terms at the midplane between the walls, where convective transport is expected to be largest. The volume average of (\ref{eq:obe2}) over the lower half of the domain volume $\Omega/2$, yields 
\begin{align}
 \frac{\partial}{\partial t} \left\langle \mathcal{T} \right\rangle_{\Omega/2} &=& \left\langle -\tilde{\kappa} \frac{\partial}{\partial z}  \mathcal{T} \right\rangle_{A(-0.5)} &- \left \langle W\,\mathcal{T}  - \tilde{\kappa} \frac{\partial}{\partial z}  \mathcal{T} \right\rangle_{A(0)}&=& J - \mathrm{Nu} \ . \label{eq:tenergy}
\end{align}
Here $\langle  \rangle_{A(z)}$ denote averages over planes at height $z$ parallel to the walls. The rate of change of thermal energy averaged over the lower half of the domain $\Omega/2$ is given by the diffusive boundary heat flux $J$ at the lower wall and the instantaneous Nusselt number $\mathrm{Nu}$ at the midplane. The laminar diffusive heat flux is
\begin{align}
J_0&=\mathrm{Nu}_0=\tilde\kappa \ .
\end{align}
As for the kinetic energy balance, equilibrium states imply $J=\mathrm{Nu}$. Periodic orbits will have unbalanced instantaneous fluxes that average to vanishing net thermal energy change over one period.

\section{Numerical approach}
\label{sec:numerics}
\noindent We perform direct numerical simulations of (\ref{eq:obe1}-\ref{eq:obe3}) in $x$-$y$-periodic domains and compute invariant states using matrix-free Newton methods. The evolution of simulated state trajectories is studied relative to invariant states in `phase portraits' defined by the net kinetic energy transfer rates in (\ref{eq:kenergy}). The technical details are introduced in the following sections, and the approach is demonstrated by explaining the transition dynamics from laminar flow to straight convection rolls.

\subsection{Direct numerical simulations}
\label{sec:dns}
The Oberbeck-Boussinesq equations for inclined layers (\ref{eq:obe1}-\ref{eq:obe3}) in a $x$-$y$-periodic domain are solved in direct numerical simulations (DNS) using a pseudo-spectral method \citep[p.133ff]{Canuto2006}. After substituting the base-fluctuation decomposition $[\bm{U},\mathcal{T}]=[\bm{U}_0,\mathcal{T}_0]+[\bm{u},\theta]$ into (\ref{eq:obe1}-\ref{eq:obe3}), the continuous field variables of the fluctuations $[\bm{u},\theta](x,y,z,t)$ are numerically approximated by Fourier and Chebyshev expansions of the form
\begin{align}
\left[\bm{u},\theta\right](\bm{x},t) &= \sum_{k_x=-K_x}^{K_x}  \; \sum_{k_y=-K_y}^{K_y} \; \sum_{j=0}^{N_z-1}
\left[\tilde{\bm{u}},\tilde{\theta}\right]_{k_x,k_y,j}(t) \; \mathcal{C}_{j}(z) \; e^{2 \pi i \left(k_x x/L_x + k_y y/L_y \right)}
\label{eq:numexpansion}
\end{align}
where $\mathcal{C}_j(z)$ is the $j$-th Chebyshev polynomial of the first kind, linearly rescaled to the interval $z \in[-0.5,0.5]$. Velocity and temperature are fixed at the walls of the domain at $\bm{u}(z=\pm 0.5)=0$ and $\theta(z=\pm 0.5)=0$, as the inhomogeneous boundary conditions are absorbed in $T_0(z)$. Owing to incompressibility, the pressure $p$ is a dependent variable and fully determined by $\bm{u}$. The pressure is obtained by solving a $\tau$-problem with the influence matrix method \citep{Kleiser1980,Canuto1986}. To completely specify the problem with periodic boundary conditions, an integral constraint on either pressure gradient or mean flux is required. We keep the mean-pressure gradient along the $x$- and the $y$-direction constant, specifically $\int_0^{L_y} \int_0^{L_x} \nabla p\, \mathrm{d}x\mathrm{d}y=0$. Technically, we modify pressure as $p=p'+\bm{U}^2/2$ which allows expressing the nonlinear term in (\ref{eq:obe1}) in rotational form $\bm{U}\times(\nabla\times\bm{U})=\left(\bm{U}\cdot \nabla \right)\bm{U} -\bm{U}^2/2$. After evaluation of the nonlinear terms in (\ref{eq:obe1}) and (\ref{eq:obe2}) in physical space, the products are transformed to a spectral representation using the FFTW library \citep{Frigo2005} and dealiased using the 2/3 rule \citep[p.133f]{Canuto2006}. Due to dealiasing, a grid of size $N_x \times N_y \times N_z$ in physical space implies spectral summation bounds of $K_x = N_x/3-1$ and $K_y = N_y/3 - 1$ in (\ref{eq:numexpansion}). We use e.g. $[N_x,N_y,N_z]=[32,32,25]$ to resolve a single pair of convection rolls in a domain of extent $[L_x,L_y]=[2.2211,2.0162]$. This choice is discussed in Section \ref{sec:lrtr}. For time-marching, an implicit-explicit multistep algorithm of 3rd order is implemented solving the diffusion terms and the pressure term fully implicitly, and the nonlinear terms and the buoyancy term explicitly. See Appendix \ref{sec:timestepper} for the details of the time-stepping algorithm. The code is written in $C$++ as an extension module to the MPI-parallel software \emph{Channelflow 2.0} \citep{Gibson2019}.

The numerical implementation of the extension module \emph{Channelflow-ILC} has been validated by reproducing three key results with different levels of importance of nonlinear effects. First, a highly resolved critical threshold for the linear onset of convection in \citet{Subramanian2016} is accurately reproduced (see Section \ref{sec:lrtr}). Second, numerical continuations in $\gamma$ and $\mathrm{Ra}$ of invariant states underlying longitudinal convection rolls reproduce an analytic scaling invariance of the nonlinear Oberbeck-Boussinesq equations (\ref{eq:obe1}-\ref{eq:obe3}), as discussed in Section 3 of the accompanying article \citep{Reetz2020b}. Third, the statistics of fully turbulent Rayleigh-B\'enard convection match previous results on the scaling of $\mathrm{Nu}\sim \mathrm{Ra}$ (Appendix \ref{sec:scaling}).

\subsection{Computation of invariant states}
\label{sec:ecs}
We not only simulate the time evolution of ILC but also compute invariant states. Any state of ILC can be expressed as a state vector $\bm{x}(t)=[\bm{u},\theta](x,y,z,t)$ in a state space of ILC for given boundary conditions. The unique time evolution of state vectors $\bm{x}(t)$ is computed using DNS. Invariant states are defined as particular state vectors $\bm{x}^*$ such that
\begin{equation}
\mathcal{G}(\bm{x}^*)=\sigma \mathcal{F}^T(\bm{x}^*) - \bm{x}^* = 0 \ , \label{eq:ecs_def}
\end{equation}
where $\mathcal{F}^T$ is the evolution operator for equations (\ref{eq:obe1}-\ref{eq:obe3}) over a finite time period $T$ defining a dynamical system for ILC. Operator $\sigma$ is an element of the symmetry group $S_{\mathrm{ilc}}$ and applies a discrete coordinate transformation in terms of (\ref{eq:sympiy}-\ref{eq:symtrans}). Since equations (\ref{eq:obe1}-\ref{eq:obe3}) are partial differential equations, the state space of this dynamical system is of infinite dimension. The numerical representation of ILC discussed in Section \ref{sec:dns} renders the state space dimension finite. The spatially discretised partial differential equations correspond to a set of coupled ordinary differential equations, one for each of the four fields $[u,v,w,\theta]$ at each spatial collocation point. Thus, the dynamical system has a state space with $N=4\times N_x\times N_y\times N_z\times 4/9$ dimensions. The factor $4/9$ accounts for the cut-off wavenumbers due to dealiasing. To solve (\ref{eq:ecs_def}) efficiently in an $N$-dimensional state space, \emph{Channelflow-ILC} employs a matrix-free Newton-Raphson iteration, based on GMRES to construct a Krylov subspace, together with a Hookstep trust region optimization \citep{Viswanath2007}. The trust region optimization increases the radius of convergence. To be within a radius of convergence of the Newton-Raphson method, the initial state of the iteration must be close to an invariant state. Full convergence within double-precision arithmetic requires the residual of (\ref{eq:ecs_def}) to be $||\mathcal{G}(\bm{x})||_2=\mathcal{O}(10^{-16})$. Here, we define the normalized $L_2$-norm of state vectors as 
\begin{equation}
 ||\bm{x}||_2= \left(\frac{1}{L_x L_y}\int\limits_0^{L_x}\int\limits_0^{L_y}\int\limits_{-0.5}^{0.5} u^2+v^2+w^2+\theta^2 dx\, dy\, dz\right)^{1/2} \ . \label{eq:statenorm}
\end{equation}
Once invariant states have converged in a Newton iteration, their spectrum of eigenvalues can be computed using Arnoldi iteration \citep{Viswanath2007} and bifurcation branches can be computed using continuation methods \citep[see][for a review]{Dijkstra2014}.

We distinguish two types of invariant states, namely equilibrium states (EQ) and periodic orbits (PO). If the period $T$ in (\ref{eq:ecs_def}) can be arbitrarily chosen \emph{a priori}, then invariant states are steady states or EQs. We use $T=20$ to compute an EQ. If invariant states require $T$ to match a specific time period, the state is unsteady but exactly recurrent over $T$ and the invariant state is a PO. The period $T$ of a PO is determined in the Newton iteration. There are additional classifications of EQs and POs. If $\sigma \in S_{ilc}$ in (\ref{eq:ecs_def}) with $\sigma\neq 1$, the invariant state is a \emph{relative} invariant state. Relative EQs are traveling wave states (TW) that are steady states in a co-moving frame of reference. TWs satisfy (\ref{eq:ecs_def}) with $\sigma=\tau(a_x,a_y)$, where shift factors $a_x$ and $a_y$ must be determined in the Newton iteration. A relative PO might also travel over its period $T$ requiring a specific $\sigma=\tau(a_x,a_y)$. Some periodic orbits that have $\sigma=1$ after a full period $T$ still may exploit a discrete symmetry operation $\sigma\neq 1$ after a relative period $T'=T/n$ with $n\in \mathbb{N}$. This type of relative PO is a `pre-periodic orbit' \citep[see e.g.][]{Budanur2017}.

Where possible, we name invariant states according to the existing names of observed convection patterns and instabilities in \citet{Subramanian2016}. We will show that specific nonlinear invariant states underlie specific convection patterns and that the specific states can be linked, in most cases, to bifurcation points corresponding to specific instabilities. To reflect the link between observed patterns, invariant states and instabilities but also to clearly distinguish between the three distinct objects, we use different symbols/fonts to indicate: An observed ``pattern X'' as $\mathcal{PX}$, instabilities linked to this pattern as $PX_i$, and exact invariant states underlying the pattern as $PX$.

\subsection{Straight convection rolls as equilibrium states}
\label{sec:lrtr}

%
\begin{figure}
        \begin{tikzpicture}
    	\draw (0, 0) node[inner sep=0]{\includegraphics[width=0.95\linewidth,trim={0.5cm 6.7cm 6.0cm 0.5cm},clip]{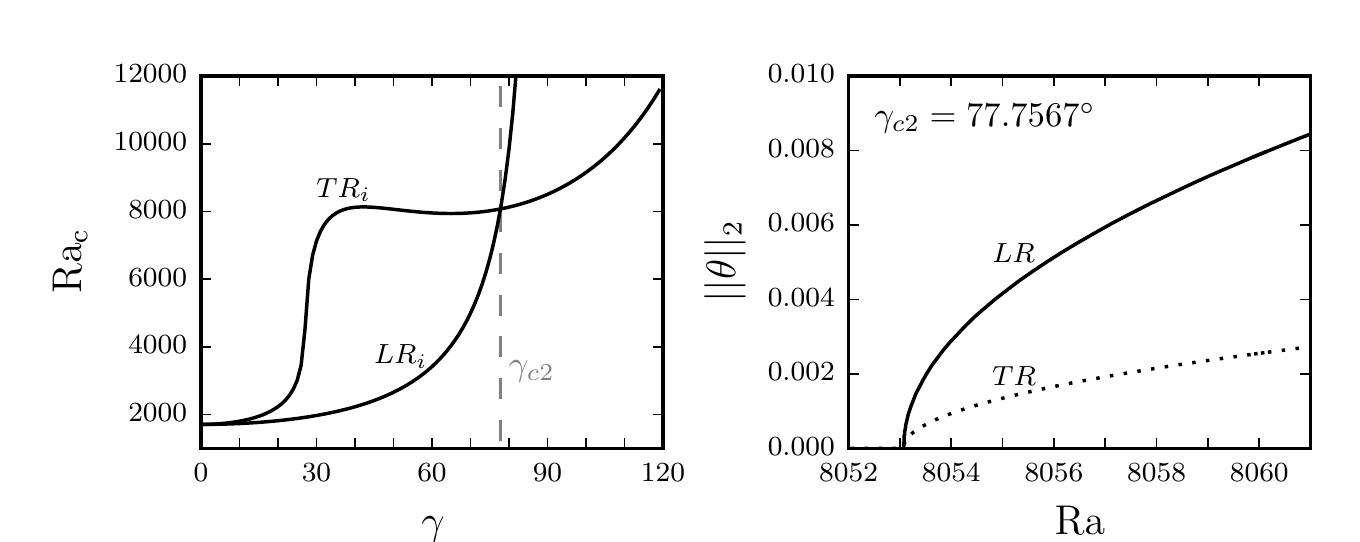}};
    	\draw (-5.5,-2.3) node {\textbf{(a)}};
    	\draw (0.5,-2.3) node {\textbf{(b)}};
		\end{tikzpicture}
\caption{\label{fig:codim2} \textbf{(a)} Critical thresholds $\mathrm{Ra}_c(\gamma)$ for the instabilities to longitudinal rolls ($LR_i$) and transverse rollse ($TR_i$) from linear stability analysis of $B$ at $\mathrm{Pr}=1.07$ \citep{Subramanian2016}. \textbf{(b)} Bifurcation branches of invariant states $LR$ and $TR$ at $\gamma_{c2}$ bifurcate together from $B$ at $\mathrm{Ra}_{c2}=8053.1$. When computing $LR$ and $TR$ in a minimal domain of size $[L_x,L_y]=[\lambda_x,\lambda_y]$, $LR$ is stable (solid line) and $TR$ is unstable (dotted line).  }
\end{figure}
The simplest invariant state in ILC is the laminar base flow (\ref{eq:baseU}-\ref{eq:baseP}), denoted as $B$ and representing a zero-state for the fluctuations $[\bm{u},\theta]=0$. When $B$ becomes dynamically unstable, straight convection rolls may form. In ILC at $\gamma=0^{\circ}$, the case corresponding to Rayleigh-B\'enard convection, the critical threshold for the onset of convection is $\mathrm{Ra}_c(\gamma=0^{\circ})=1707.76$ \citep{Busse1978a}. In ILC at $\gamma\neq 0^{\circ}$, two types of straight rolls can emerge from the primary instability, either longitudinal rolls (LR), with orientation along $x$, or transverse rolls (TR), with orientation along $y$. The type of rolls to which $B$ becomes first unstable when $\mathrm{Pr}$ is fixed and $\mathrm{Ra}$ is increased depends on $\gamma$ \citep{Gershuni1969,Hart1971a}. Figure \ref{fig:codim2}a shows the curves for critical thresholds $\mathrm{Ra}_c(\gamma)$ at $\mathrm{Pr}=1.07$. The point in the $\gamma$-$\mathrm{Ra}$-plane where $LR_i$ and $TR_i$ have the same critical threshold is a codimension-2 point. We reproduce this point at $\gamma=77.7567^{\circ}\equiv \gamma_{c2}$ and $\mathrm{Ra}_{c}(\gamma_{c2})=8053.1 \equiv \mathrm{Ra}_{c2}$ via numerical continuation of equilibrium states $LR$ and $TR$ down in $\mathrm{Ra}$ to their exact bifurcation point from $B$ (Fig. \ref{fig:codim2}b). $LR$ is invariant under the symmetry group $S_{\mathrm{lr}}=\langle\pi_{xz}\tau(0,0.5),\pi_y\tau(0,0.5),\tau(a_x,0)\rangle$, and $TR$ is invariant under $S_{\mathrm{tr}}=\langle \pi_{xz},\pi_y,\tau(0,a_y)\rangle$. Both equilibrium states, $LR$ and $TR$, are numerically fully converged to satisfy (\ref{eq:ecs_def}). Useful initial states for the Newton iteration are obtained from a `phase portrait' analysis as explained in the following section. 

Linear stability analysis suggests streamwise and spanwise wavelengths for the primary instability to longitudinal and transverse rolls at the codimension-2 point \citep{Subramanian2016}. Accordingly we choose the periodic domain to match wavelengths $\lambda_x=2.2211$ and $\lambda_y=2.0162$ of instabilities $TR_i$ and $LR_i$, respectively. For the domain size $[L_x,L_y]=[\lambda_x,\lambda_y]$, we have reproduced the codimension-2 point at $[\gamma_{c2},\mathrm{Ra}_{c2}]=[77.7567^{\circ},8053.1]$. We confirmed that all given digits are significant. The step-size of the continuation was chosen sufficiently small to indicate the bifurcation point at this accuracy. Moreover, increasing the grid resolution beyond $[n_x,n_y,n_z]=[32,32,25]$ does not change the result. We fix $\lambda_x=2.2211$ and $\lambda_y=2.0162$ as constants in this paper, and choose all periodic domains to be periodic over $[L_x,L_y]=[l\,\lambda_x,m\,\lambda_y]$ and to be discretised with $[N_x,N_y,N_z]=[l\,n_x,m\,n_y,n_z]$ with $l,m\in \mathbb{N}$. Thus, the invariant states discussed here have prescribed pattern wavelengths, unlike pattern forming instabilities calculated using a Floquet analysis.

\subsection{Phase portrait analysis}
\label{sec:ppa}
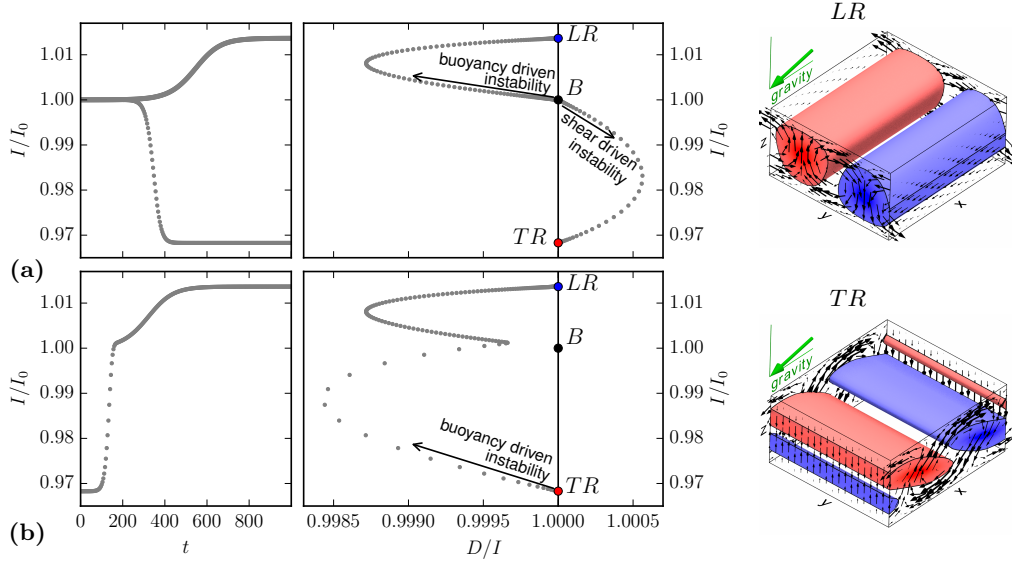
\begin{figure}
		\begin{subfigure}[b]{0.74\textwidth}
        \begin{tikzpicture}
    	\draw (0, 0) node[inner sep=0]{\includegraphics[width=0.99\linewidth,trim={1.cm 2.cm 3.3cm 0.5cm},clip]{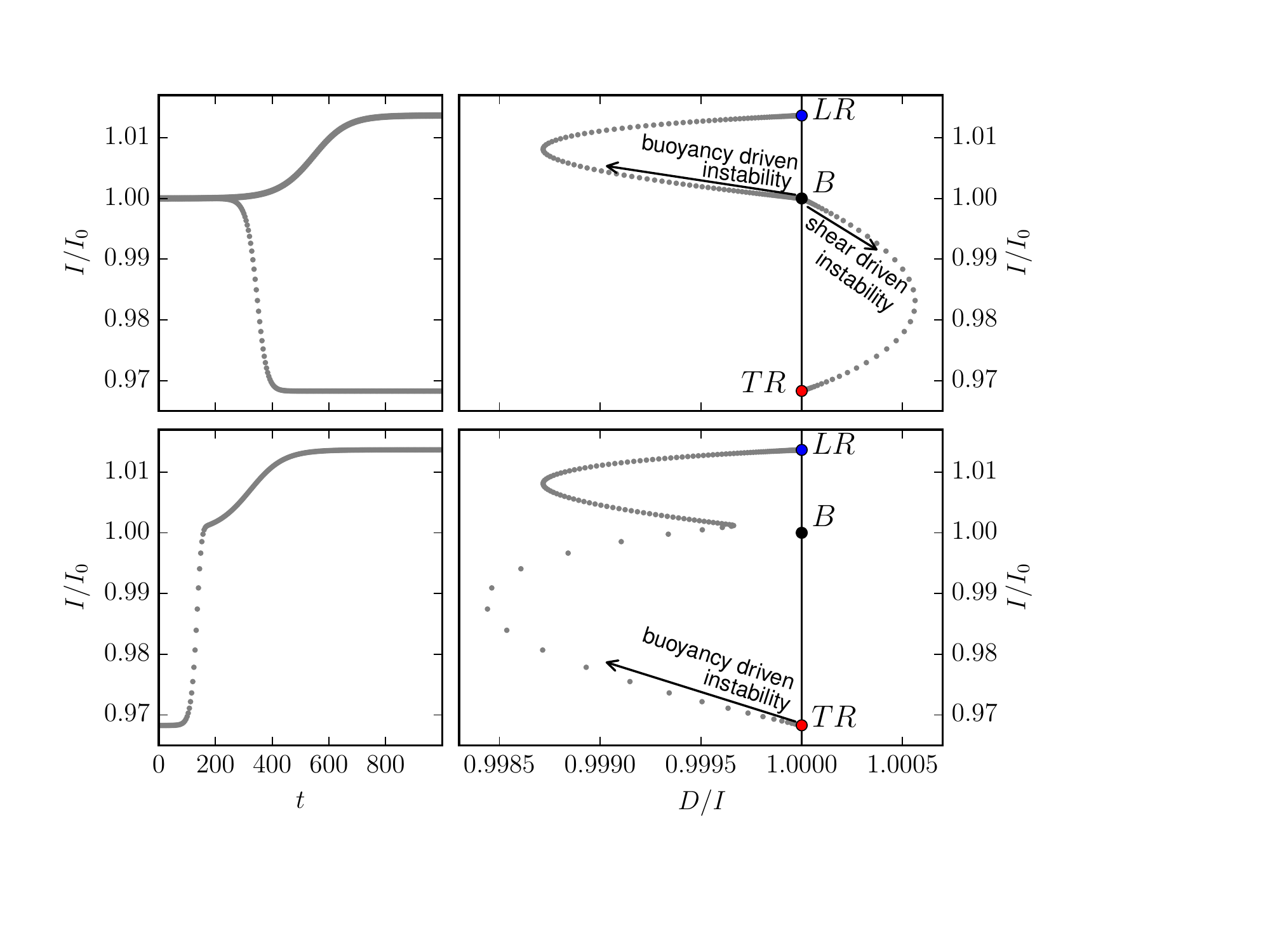}};
    	\draw (-4.7,0) node {\textbf{(a)}};
    	\draw (-4.7,-3.5) node {\textbf{(b)}};
		\end{tikzpicture}
		\end{subfigure}
		\begin{subfigure}[b]{0.25\textwidth}
		\begin{subfigure}[b]{0.99\textwidth}
        \begin{tikzpicture}
    	\draw (0, 0) node[inner sep=0]{\includegraphics[width=\linewidth,trim={0.1cm -6.6cm 1.1cm 0.6cm},clip]{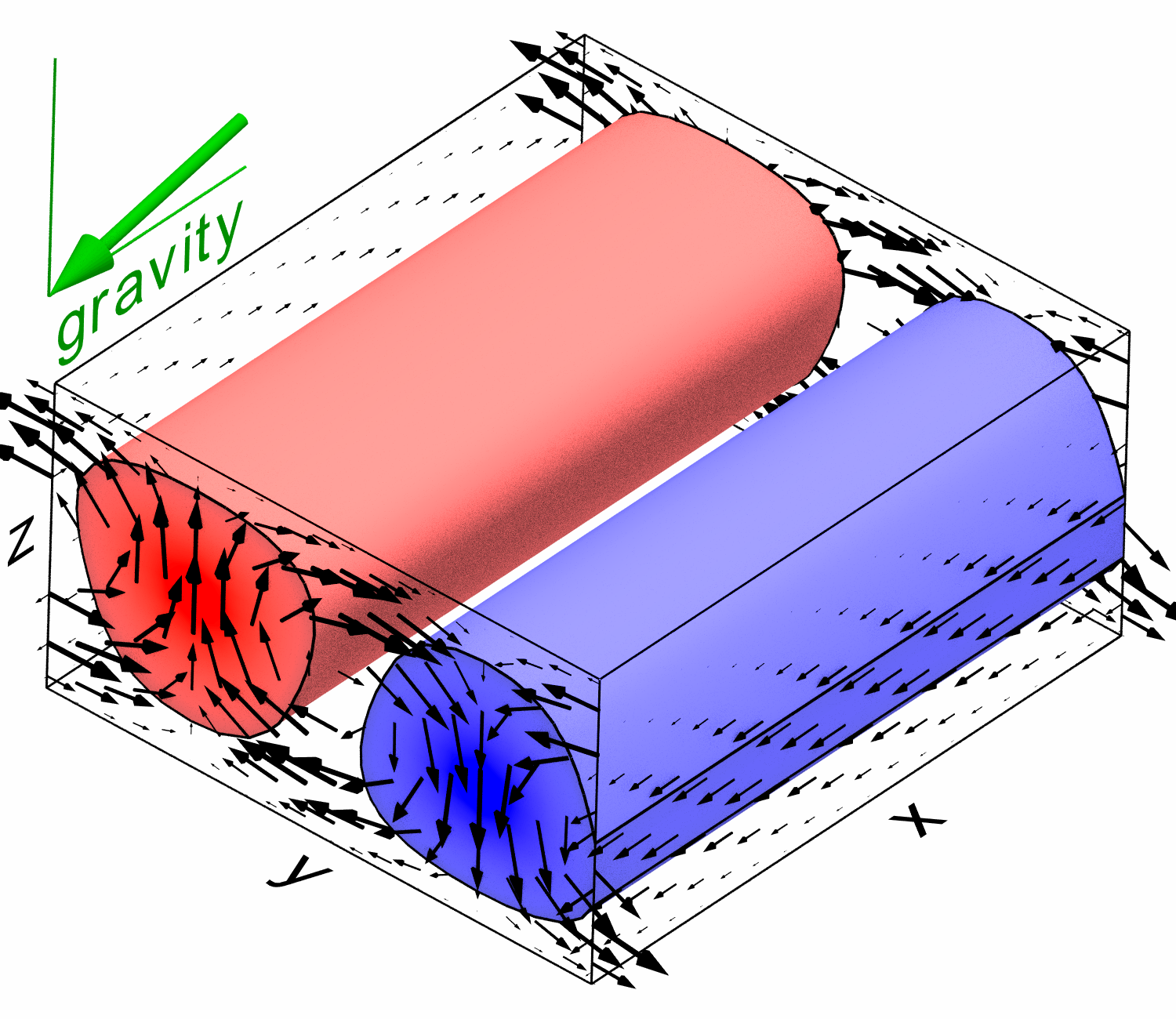}};
    	\draw (-0.5,1.9) node {$LR$};
		\end{tikzpicture}
		\end{subfigure}
		\begin{subfigure}[b]{0.99\textwidth}
        \begin{tikzpicture}
    	\draw (0, 0) node[inner sep=0]{\includegraphics[width=\linewidth,trim={0.1cm -6.6cm 1.1cm 0.6cm},clip]{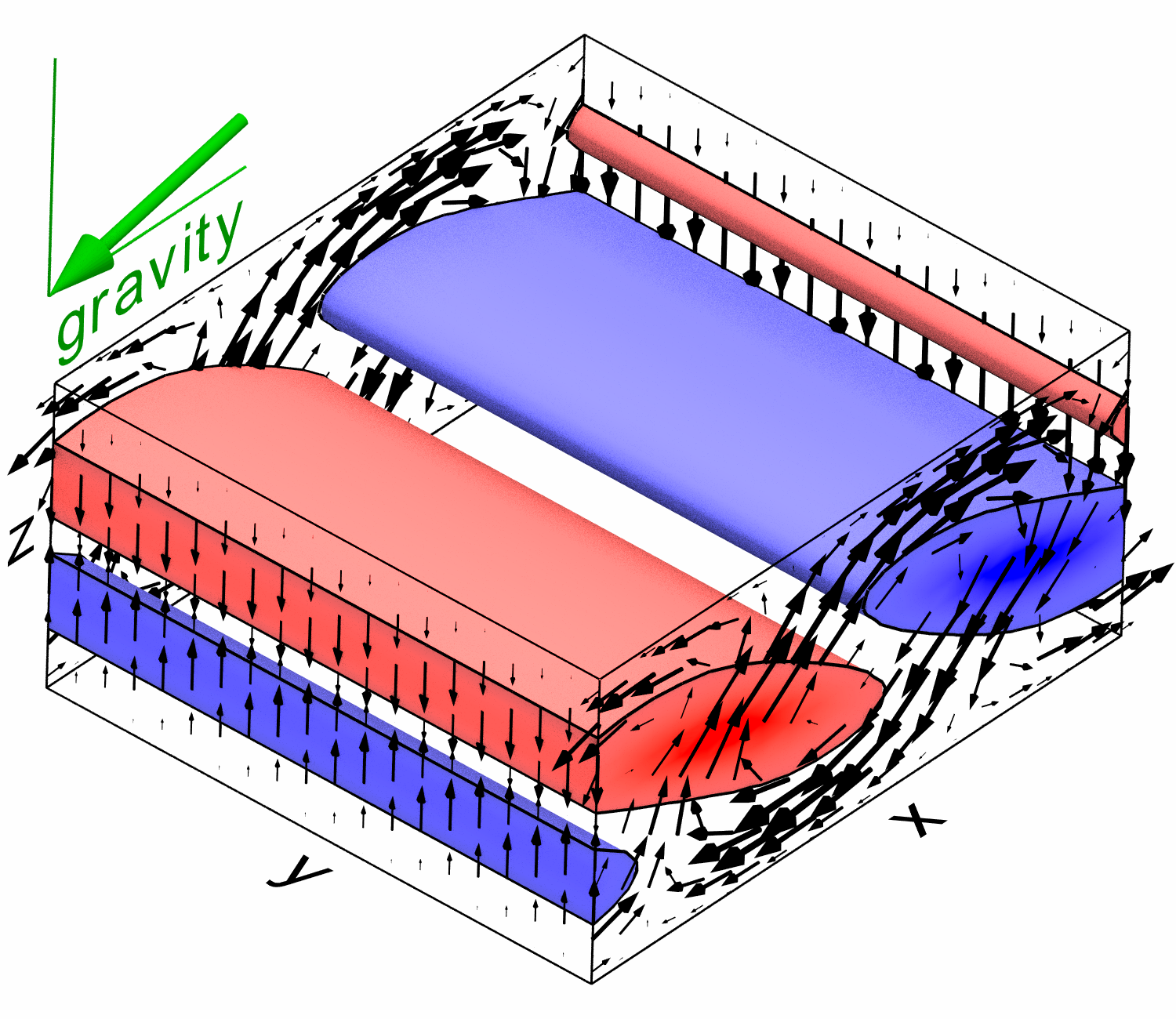}};
    	\draw (-0.5,1.9) node {$TR$};
		\end{tikzpicture}
		\end{subfigure}
		\end{subfigure}
\caption{\label{fig:codim2pplane} \textbf{(a)} Simulated state trajectories (grey dots) evolving from noise around the unstable laminar base flow $B$ at $\gamma_{c2}$ and $\mathrm{Ra}=8500$ over time $t$ (left), and plotted as phase portraits in a plane of normalised kinetic energy input $I/I_0$ and relative dissipation $D/I$ (right). The DNS is confined to either $S_{\mathrm{lr}}$ and $S_{\mathrm{tr}}$, allowing either a buoyancy driven instability to initiate a temporal transition to $LR$ or a shear driven instability to initiate a temporal transition to $TR$. Arrows indicate the direction of the evolution. Exact equilibrium states $LR$ and $TR$ are visualized by 3D contours at $1/3[\min(\theta),\max(\theta)]$ and the inplane components of $\bm{u}$ at the domain sides. \textbf{(b)} Without imposing discrete symmetries, $TR$ is dynamically unstable. Perturbing $TR$ initiates a dynamical connection to $LR$ with fast dynamics near the unstable manifold of $TR$ and slow dynamics near the stable manifold of $LR$.}
\end{figure}

Temporal transitions from laminar flow to longitudinal or transverse rolls are studied by initialising a simulation with small perturbations around the dynamically unstable base state $B$ and visualizing the time evolution in a state space projection representing a `phase portrait'. Two state vector trajectories $\bm{x}(t)$ are simulated just above the codimension-2 point, at $\gamma_{c2}$ and $\mathrm{Ra}=8500>\mathrm{Ra}_{c2}$. Each trajectory starts from $B$ perturbed by small amplitude noise of $\mathcal{O}(10^{-5})$. The evolution of $\bm{x}(t)$ is simulated in the symmetry subspace of $(\lambda_x,\lambda_y)$-periodicity, corresponding to the domain size, and either $S_{\mathrm{lr}}$ or $S_{\mathrm{tr}}$. Imposing either $S_{\mathrm{lr}}$ or $S_{\mathrm{tr}}$ causes $\bm{x}(t)$ to remain in the symmetry subspace since (\ref{eq:obe1}-\ref{eq:obe3}) are equivariant under $S_{\mathrm{lr}}$ and $S_{\mathrm{tr}}$. Each symmetry subspace contains only one type of straight convection rolls. Thus, the choice of either $S_{\mathrm{lr}}$ or $S_{\mathrm{tr}}$ selects whether longitudinal or transverse rolls emerge.  \\
The longitudinal and the transverse state trajectories are analysed in a `phase plane' spanned by kinetic energy input $I$ and relative viscous dissipation $D/I$ defined in $(\ref{eq:kenergy})$. The $D/I$-axis allows to distinguish two types of instabilities of equilibrium states in ILC satisfying $D=I$. The transition towards $LR$ is triggered by a buoyancy driven instability of $B$ that initially increases $I$ over $D$. The transition towards $TR$ is triggered by a shear driven instability of $B$ that initially increases $D$ over $I$. The phase portrait illustrates that the state $LR$ is reached via a temporal transition from a buoyancy driven instability of $B$, and $TR$ is reached via a temporal transition from a shear driven instability of $B$ (Figure \ref{fig:codim2pplane}a).\\
%
The phase portrait analysis not only characterises the forces driving an instability but also helps to identify good initial guesses for Newton iterations that may converge to invariant states. After a stage of exponential growth in the transition from $B$, the two state trajectories saturate and the dynamics slows down exponentially (Figure \ref{fig:codim2pplane}a). Exponential slow-down near $D/I=1$ suggests the presence of an equilibrium state, and indeed, the two final state vectors $\bm{x}(t=1000)$ are close to invariant states and good initial guesses for a Newton iteration. They converge to $LR$ and $TR$, respectively, and provide the starting point for the numerical continuation shown in Figure \ref{fig:codim2}b. Consequently, the phase portrait analysis is useful for finding invariant states during temporal transitions. Moreover, the phase portrait clearly illustrates how the dynamics follow dynamical connections between invariant states, in this case $B\rightarrow LR$ and $B\rightarrow TR$. We use the term `dynamical connection' for state trajectories connecting the state space neighborhood of two invariant states in a finite time. Dynamical connections indicate the existence of a nearby heteroclinic connection requiring infinite time to be traversed \citep[e.g.][]{Farano2019}.\\
The dynamical stability of $LR$ and $TR$ at $\gamma_{c2}$ and $\mathrm{Ra}=8500$ is characterised using Arnoldi iteration in the symmetry subspace of the $(\lambda_x,\lambda_y)$-periodic domain. $LR$ is stable and $TR$ is weakly unstable with respect to two shift-symmetry related three-dimensional, longitudinally oriented, eigenmodes with linear growth rate of $\omega_r=0.044$. These unstable eigenmodes of $TR$ do not exist in the symmetry subspace defined by $S_{\mathrm{tr}}$ where the temporal transition to stable $TR$ was simulated. However, the simulated dynamical connection $B\rightarrow TR$ also exists in the larger subspace where $S_{\mathrm{tr}}$ is not imposed and $B\rightarrow TR$ connects two unstable invariant states. When perturbing unstable $TR$, a buoyancy driven instability triggers a dynamical connection $TR\rightarrow LR$. Along this connection the dynamics undergo a rapid slow-down suggesting a transition from the fast unstable manifold of $TR$ to the slow stable manifold of $LR$ whose leading eigenvalue is $\omega_r=-0.016$ (Figure \ref{fig:codim2pplane}b).\\
In summary, the phase portrait analysis serves three purposes. First, high-dimensional state space trajectories can be visualised in a two-dimensional projection. Second, close approaches to invariant states and a slow-down of the dynamics provide useful initial guesses for Newton iterations to converge. Thus, the phase portrait analysis gives access to invariant states. Third, the type of instability triggering a transition from an equilibrium state can be characterized as either buoyancy driven or shear driven via the departure from the $D/I=1$ line in the phase portrait. 

\section{Transitions to tertiary states}
\label{sec:transitions}

On increasing $\mathrm{Ra}$, secondary patterns of regular straight convection rolls give way to five different tertiary convection patterns \citep{Daniels2000}. These patterns have been associated with five different secondary instabilities \citep{Subramanian2016}. The type of convection pattern emerging when increasing $\mathrm{Ra}$ depends on the inclination angle $\gamma$. Following the cited work, we study the five tertiary convection patterns in ILC at $\mathrm{Pr}=1.07$.  In the parameter space $\gamma \in [0^{\circ},120^{\circ}]$ and $\epsilon \in [0,2]$, where $\epsilon=(\mathrm{Ra}-\mathrm{Ra}_c)/\mathrm{Ra}_c$ is a normalized Rayleigh number relative to the critical threshold $\mathrm{Ra}_c(\gamma)$ for convection onset (Figure \ref{fig:codim2}a). We select specific values of the control parameters where the patterns have been observed.\\
The following sections apply the phase portrait analysis outlined in Section \ref{sec:ppa} to each convection pattern individually. Instead of discussing the five patterns in order of increasing angle of inclination $\gamma$, we choose to order the patterns in terms of the complexity of the transition dynamics towards the attractive invariant state underlying the pattern: first, transitions to equilibrium states (Section \ref{sec:equilibria}), second, transitions to periodic orbits (Section \ref{sec:orbits}), and third, transition dynamics in the absence of an attractive tertiary state (Section \ref{sec:svtransient}).

\subsection{Transitions with equilibrium state attractor}
\label{sec:equilibria}

\subsubsection{Wavy rolls}
\label{sec:wr}
The convection pattern of wavy rolls ($\mathcal{WR}$) has been observed in early experiments \citep{Hart1971a} and associated to the wavy instability $WR_i$ of $LR$ \citep{Clever1977a}, also found for longitudinal rolls in B\'enard-Couette flow \citep{Clever1977b}. \citet{Hart1971b} already hypothesised a relation between $\mathcal{WR}$ and wavy vortex flow in Taylor-Couette experiments \citep{Coles1965}. Such a relation was later found to exist, and exploited in the first constructions of invariant states underlying wavy velocity streaks in subcritical shear flows \citep{Nagata1990,Clever1992}. $\mathcal{WR}$ are observed in ILC at control parameter values $[\gamma,\epsilon,\mathrm{Pr}]=[40^{\circ},0.07,1.07]$ \citep{Daniels2000}. We simulate a temporal transition starting from small-amplitude noise around the unstable base flow at these values of the control parameters. The size of the periodic domain is $[2\lambda_x,\lambda_y]$ and no additional discrete symmetries are imposed.

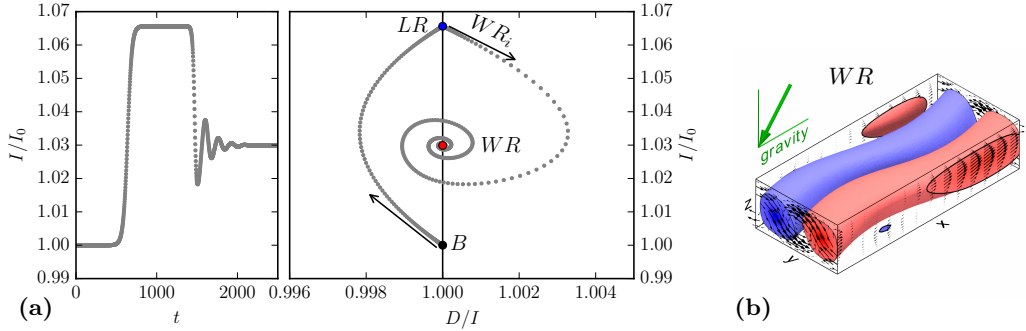
\begin{figure}
\center
		\begin{subfigure}[b]{0.7\textwidth}
        \begin{tikzpicture}
    	\draw (0, 0) node[inner sep=0]{\includegraphics[width=0.99\linewidth,trim={1.0cm 2.0cm 3.5cm 0.5cm},clip]{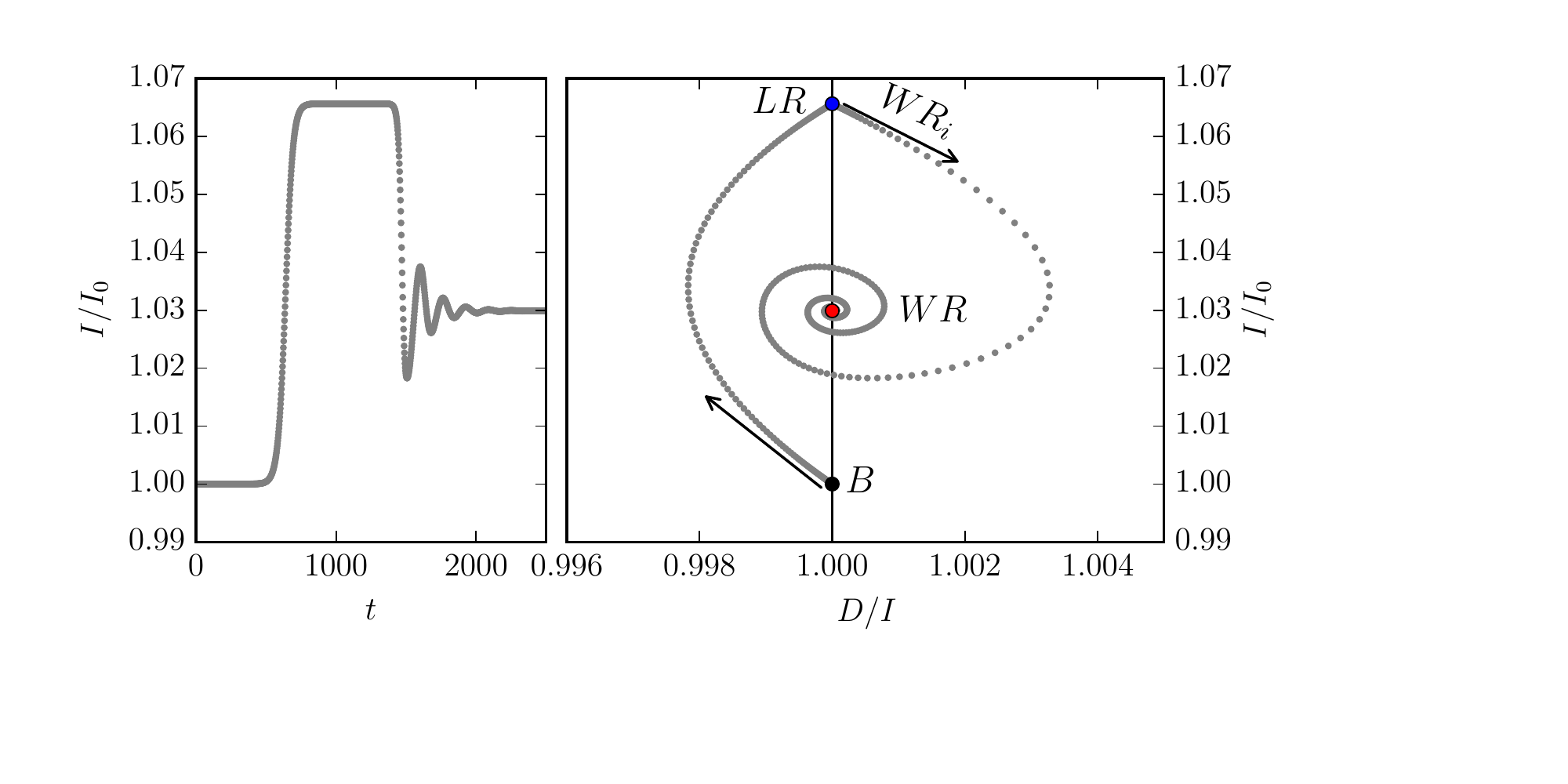}};
    	\draw (-4.3,-2.0) node {\textbf{(a)}};
		\end{tikzpicture}
		\end{subfigure}
		\begin{subfigure}[b]{0.29\textwidth}
        \begin{tikzpicture}
    	\draw (0, 0) node[inner sep=0]{\includegraphics[width=\linewidth,trim={5.1cm 5.5cm 6.8cm 5.5cm},clip]{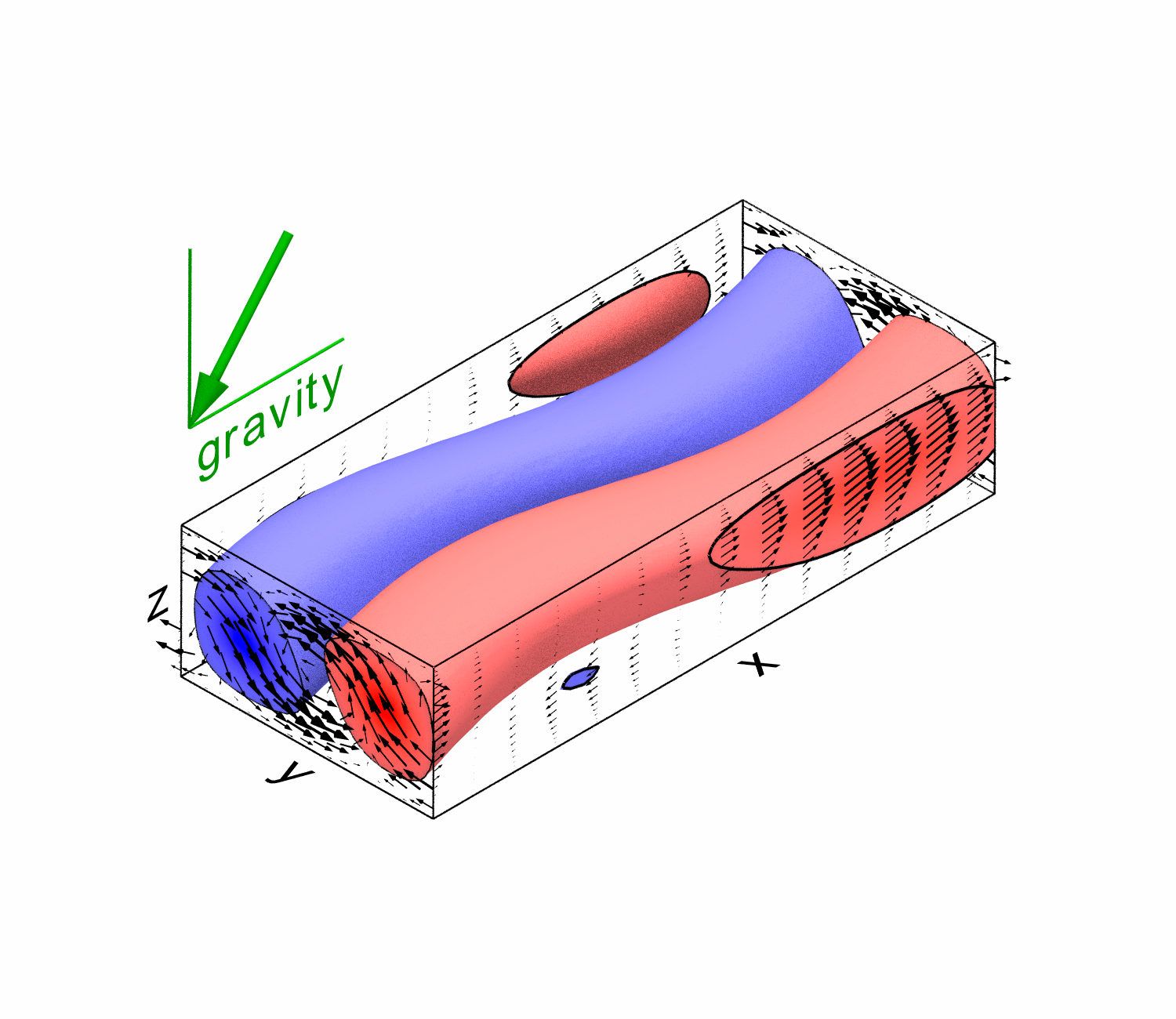}};
    	\draw (-1.7,-1.8) node {\textbf{(b)}};
    	\draw (-0.4,1.3) node {$WR$};
		\end{tikzpicture}
		\end{subfigure}
\caption{\label{fig:wr} \textbf{(a)} State trajectory evolution from the unstable base state $B$ at $\gamma=40^{\circ}$ and $\mathrm{Ra}=2385$ ($\epsilon=0.07$). After a transient in the vicinity of $LR$, the shear driven instability $WR_i$ of $LR$ makes the trajectory follow a stable spiral towards equilibrium state $WR$. \textbf{(b)} Flow structure of $WR$ in a periodic domain of size $[2\lambda_x,\lambda_y]$.}
\end{figure}

The phase portrait reveals a two-stage transition form the base flow $B$ to wavy rolls. First, a buoyancy driven instability of $B$ leads to a slow transient over $700< t < 1\,400$ in the vicinity of $LR$. Second, a shear driven instability of $LR$ leads to a spiraling trajectory on which the dynamics approach the equilibrium state $WR$ (Figure \ref{fig:wr}a). The spectrum of eigenvalues of $WR$ has the complex pair $(\omega_r,\pm\omega_i)=(-0.007,\pm0.039)$ closest to the imaginary axis. The imaginary part suggests an oscillation period on the spiraling trajectory of $T=2\pi/\omega_i=161$. The decaying oscillations of the trajectory for $ t > 1\,500$ match this period. The flow structure of equilibrium state $WR$ shows the characteristic wavy modulations observed in experiments and simulations (Figure \ref{fig:wr}b). $WR$ are invariant under shift-reflect and shift-rotate symmetries $S_{\mathrm{wr}}=\langle \pi_y\tau(0.5,0.5),\pi_{xz}\tau(0.5,0.5)\rangle$. These symmetries are analogous to the symmetries of wavy velocity streaks in plane Couette flow \citep[e.g.][]{Gibson2008}.

\begin{figure}
\center
        \begin{tikzpicture}
    	\draw (0, 0) node[inner sep=0]{\includegraphics[width=0.99\linewidth,trim={1.9cm 1cm 1.5cm 0.5cm},clip]{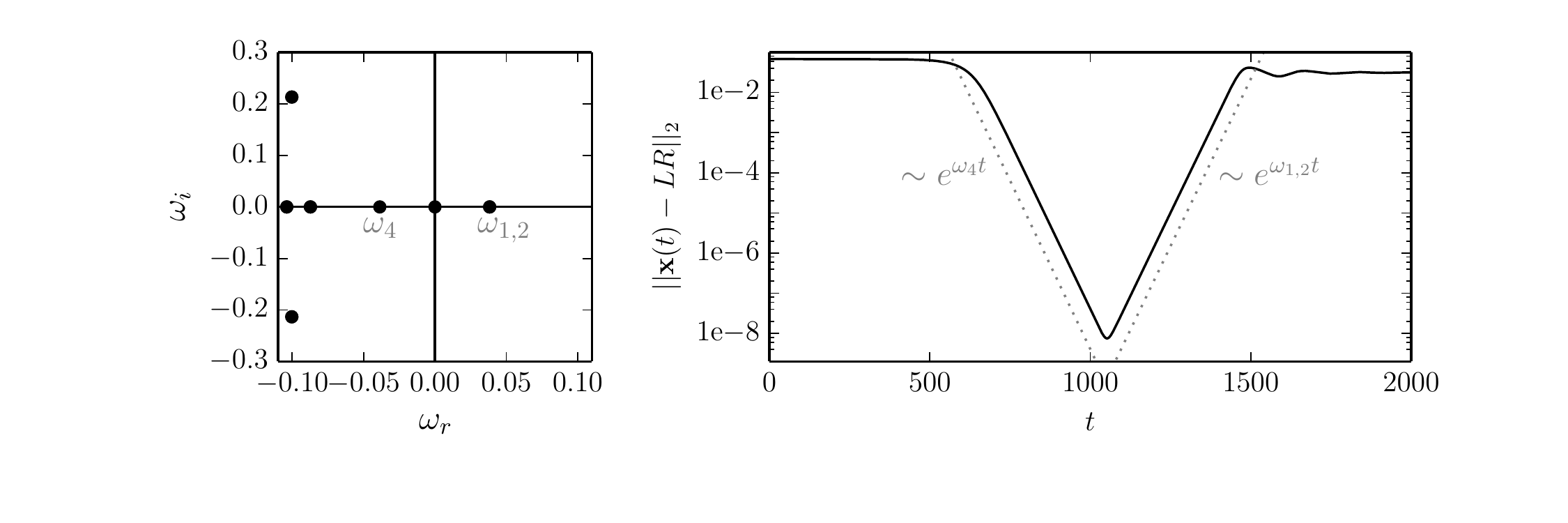}};
    	\draw (-6.1,-1.8) node {\textbf{(a)}};
    	\draw (-1.3,-1.8) node {\textbf{(b)}};
		\end{tikzpicture}
\caption{\label{fig:visitingLR} \textbf{(a)} Spectrum of leading eigenvalues of $LR$ explains exponential approach and escape rates relative to $LR$. \textbf{(b)} $L_2$-distance between $LR$ and the state trajectory shown in Figure \ref{fig:wr}a illustrates exponential approach and escape dynamics in the state space neighborhood of $LR$. The dotted line marks the exponential rates given by the leading stable and unstable eigenvalues of $LR$.}
\end{figure}

The state trajectory follows a sequence of dynamical connections $B\rightarrow LR\rightarrow WR$. The transient close to $LR$, a saddle point with stable and unstable eigendirections, follows exponential dynamics. The leading eigenvalues of $LR$ are real, $[\omega_{1,2},\omega_3,\omega_4]=[0.038,10^{-9},-0.039]$, and define the exponential rate of approach, $\sim e^{\omega_4 t}$, and escape, $\sim e^{\omega_{1,2} t}$ (Figure \ref{fig:visitingLR}). The double multiplicity of the positive eigenvalue is a result of the continuous translation symmetry $\tau(a_x,0)$ of $LR$ allowing two orthonormal eigenmodes of arbitrary $x$-phase. Continuous translations $\tau(0,a_y)$ are not an invariance of $LR$ but change the $y$-phase of $LR$ and generate a continuum of equivalent states. The Goldstone mode corresponding to the translation invariance of $LR$ is the marginally stable eigenmode with eigenvalue $\omega_3$. Therefore, the pitchfork bifurcation creating $LR$ is a circle pitchfork bifurcation\citep{Crawford1991}. The non-zero finite value of $\omega_3$ is a measure for the accuracy of the Arnoldi iteration. The minimal distance of the state trajectory to $LR$ is $||\bm{x}(t=1050)-LR||_2/||LR||_2\approx 10^{-8}$. Consequently, the transition dynamics from $B$ generate a trajectory transiently visiting the state space neighborhood of the unstable equilibrium state $LR$, as already suggested by the phase portrait. 

When increasing thermal driving, the $\mathcal{WR}$ pattern is succeeded by weakly turbulent wavy rolls, also called `crawling rolls' \citep{Daniels2000}. A DNS at $\epsilon=0.5$ leads to a much more complicated phase portrait than at $\epsilon=0.07$. At these control parameter values, the state trajectory initially still undergoes the transition sequence $B\rightarrow LR \rightarrow WR$. However, $WR$ is now unstable. After a transient visit close to $WR$ at $t=500$, a buoyancy driven instability of $WR$ leads to a sequence of large-amplitude, fast oscillations before the state trajectory is attracted to a small-amplitude, slow bursting cycle (Figure \ref{fig:obwr_portrait}). The fast transient oscillations have clockwise revolving trajectories in the $D/I$-$I$-plane and dominate the phase portrait. We suspect the existence of an unstable periodic orbit with similar shaped phase portrait underlying the transient oscillations. Finding and analysing this periodic orbit is beyond the scope of this work and discussed elsewhere \citep{Reetz2020d}. Here, we discuss the slow dynamical attractor at these values of the control parameters. 

\begin{figure}
\center
		\begin{subfigure}[b]{0.99\textwidth}
        \begin{tikzpicture}
    	\draw (0, 0) node[inner sep=0]{\includegraphics[width=0.99\linewidth,trim={1.0cm 1.5cm 0.5cm 0.5cm},clip]{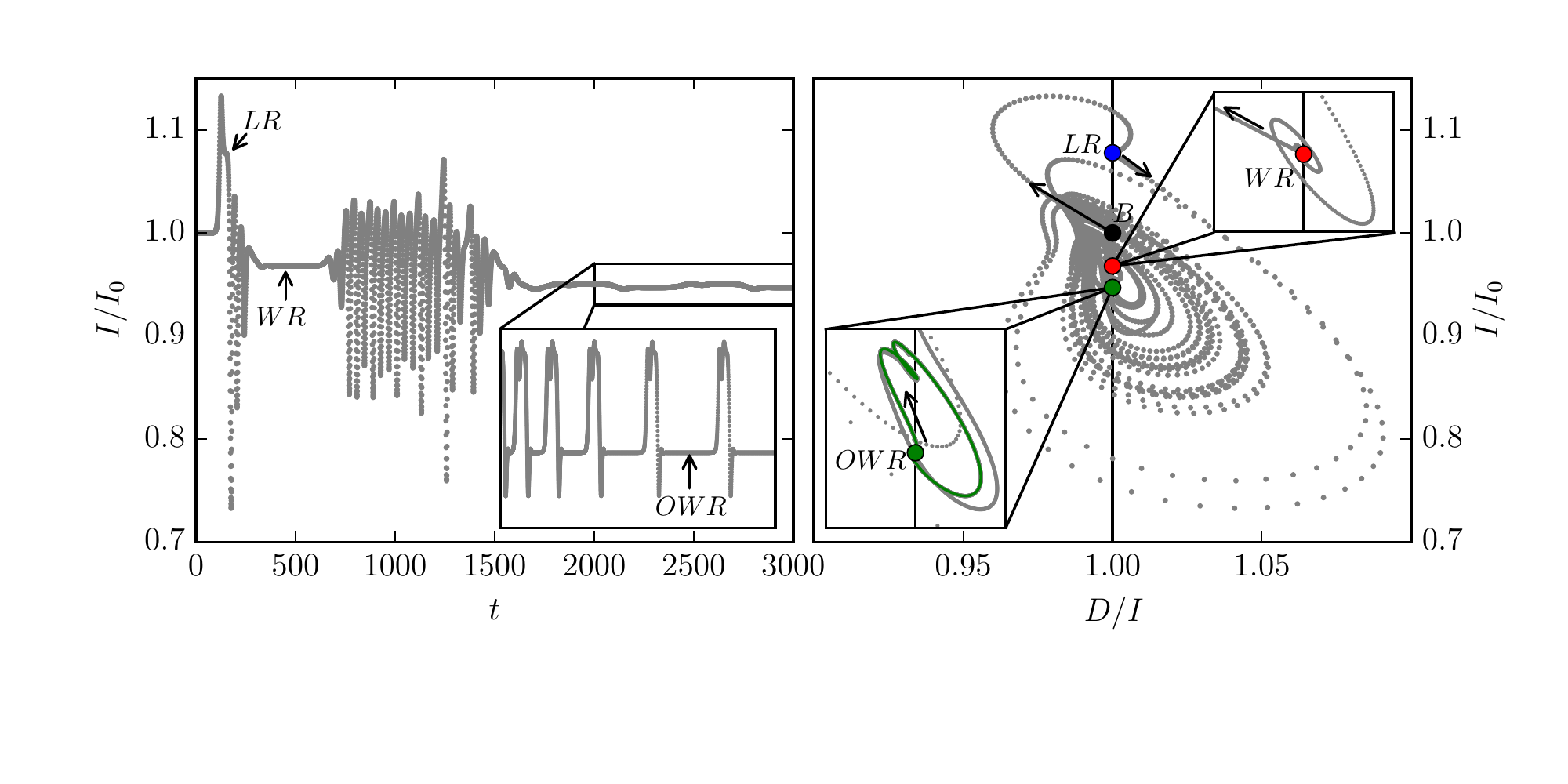}};
    	\draw (-2.5,-1.4) node {\textbf{(a)}};
    	\draw (0.5,-1.4) node {\textbf{(b)}};
    	\draw (4.0,1.3) node {\textbf{(c)}};
		\end{tikzpicture}
		\end{subfigure}
\caption{\label{fig:obwr_portrait}  State trajectory evolution from the unstable base state $B$ at $\gamma=40^{\circ}$ and $\mathrm{Ra}=3344$ ($\epsilon=0.5$). The trajectory visits unstable $LR$, followed by a transient visit of $WR$ (inset \textbf{(c)}). Subsequently, the trajectory undergoes a sequence of rapid oscillations and is finally attracted to a heteroclinic cycle between equilibrium state $OWR$ and a symmetry related equilibrium. The time series for $2000<t<10\,000$ in inset \textbf{(a)} indicates the increasing time spent near the equilibrium states. The phase portrait of the heteroclinic cycle is magnified in inset \textbf{(b)}. Since the energy transfer rates do not differ for symmetry related states, the heteroclinic cycle appears as a homoclinic cycle. See Figure \ref{fig:obwr_cycle}a for a projection that distinguishes the two equilibrium states.}
\end{figure}
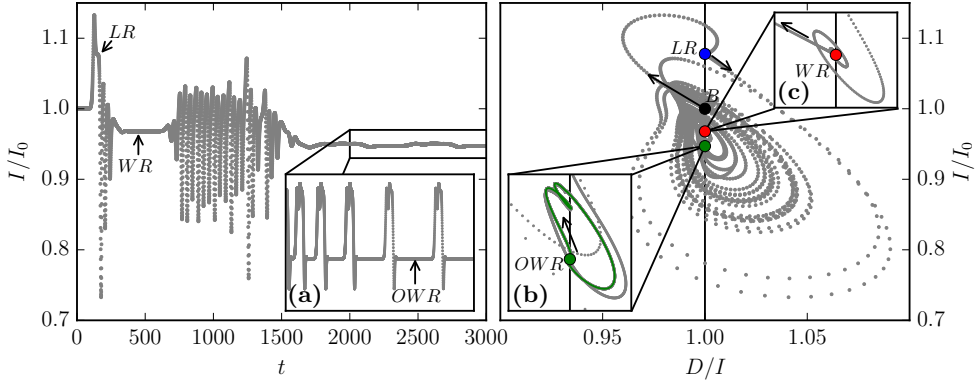
\begin{figure}
\center
		\begin{subfigure}[b]{0.49\textwidth}
        \begin{tikzpicture}
    	\draw (0, 0) node[inner sep=0]
    	{\includegraphics[width=\linewidth,trim={1.1cm 8.9cm 11.8cm 1.5cm},clip]{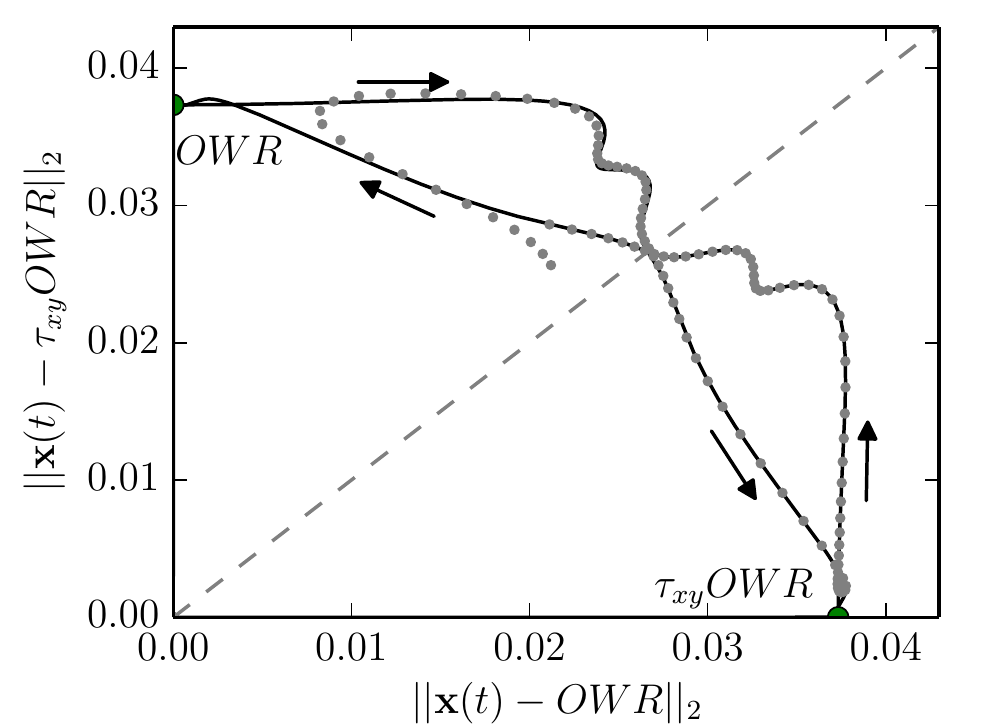}};
    	\draw (-2.9,-2.3) node {\textbf{(a)}};
		\end{tikzpicture}
		\end{subfigure}
		\begin{subfigure}[b]{0.49\textwidth}
        \begin{tikzpicture}
    	\draw (0, 0) node[inner sep=0]
    	{\includegraphics[width=\linewidth,trim={0.1cm 0.1cm 0.1cm 0.1cm},clip]{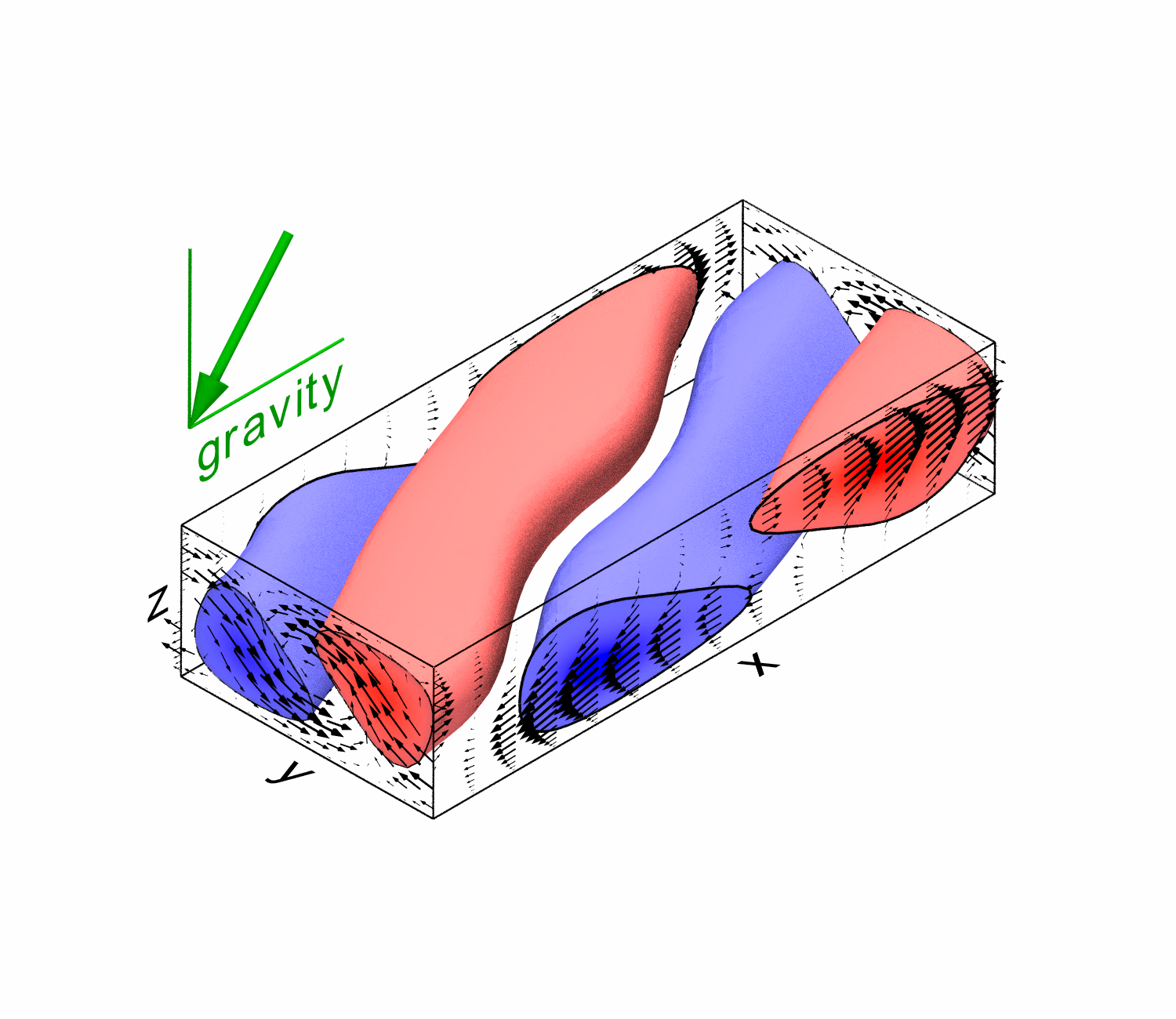}};
    	\draw (-2.5,-2.5) node {\textbf{(b)}};
    	\draw (-0.5,1.6) node {$OWR$};
		\end{tikzpicture}
		\end{subfigure}
		\begin{subfigure}[b]{0.99\textwidth}
        \begin{tikzpicture}
    	\draw (0, 0) node[inner sep=0]
    	{\includegraphics[width=\linewidth,trim={0.0cm 0.1cm 0.0cm -0.9cm},clip]{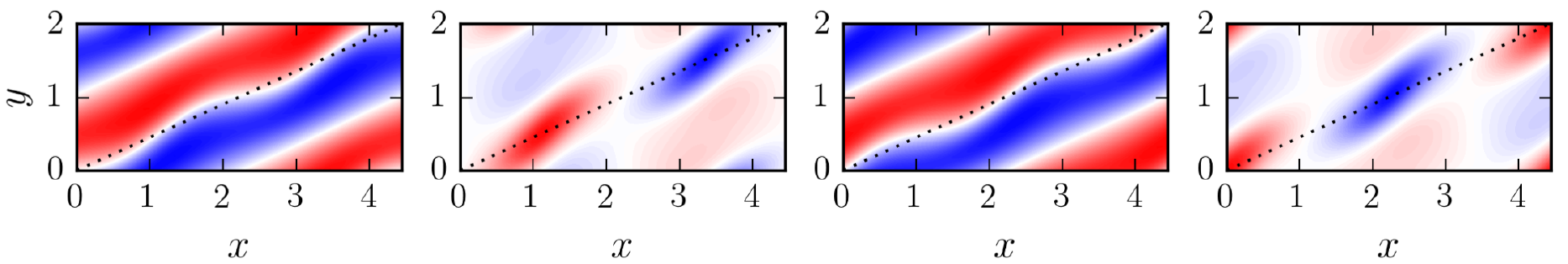}};
    	\draw (-6.3,-1.2) node {\textbf{(c)}};
    	\draw (-4.8,0.9) node {$OWR$};
    	\draw (-1.5,0.9) node {$\bm{e}^u$};
    	\draw (1.8,0.9) node {$\tau_{xy}OWR$};
    	\draw (5.0,0.9) node {$\tau_{xy}\bm{e}^u$};
		\end{tikzpicture}
		\end{subfigure}
\caption{\label{fig:obwr_cycle} Robust heteroclinic cycle between two symmetry related equilibrium states at $\gamma=40^{\circ}$ and $\mathrm{Ra}=3344$ ($\epsilon=0.5$). \textbf{(a)} $L_2$-distance relative to the two symmetry related equilibrium states, $OWR$ and $\tau_{xy}OWR$, visualises how the simulated state trajectory (grey dots) approaches the heteroclinic cycle (black line). The direction of the dynamics is indicated by black arrows. \textbf{(b)} Flow structure of $OWR$ in a periodic domain of size $[2\lambda_x,\lambda_y]$. \textbf{(c)} Temperature contours at midplane illustrate the spatial phase of $OWR$, $\tau_{xy}OWR$, and their unstable eigenmodes $\bm{e}^u$ and $\tau_{xy}\bm{e}^u$, respectively, along the diagonal of the domain (dotted line). The pattern wavenumbers of equilibria and unstable eigenmodes along the domain diagonals suggest a nearby 1:2 resonance.}
\end{figure}
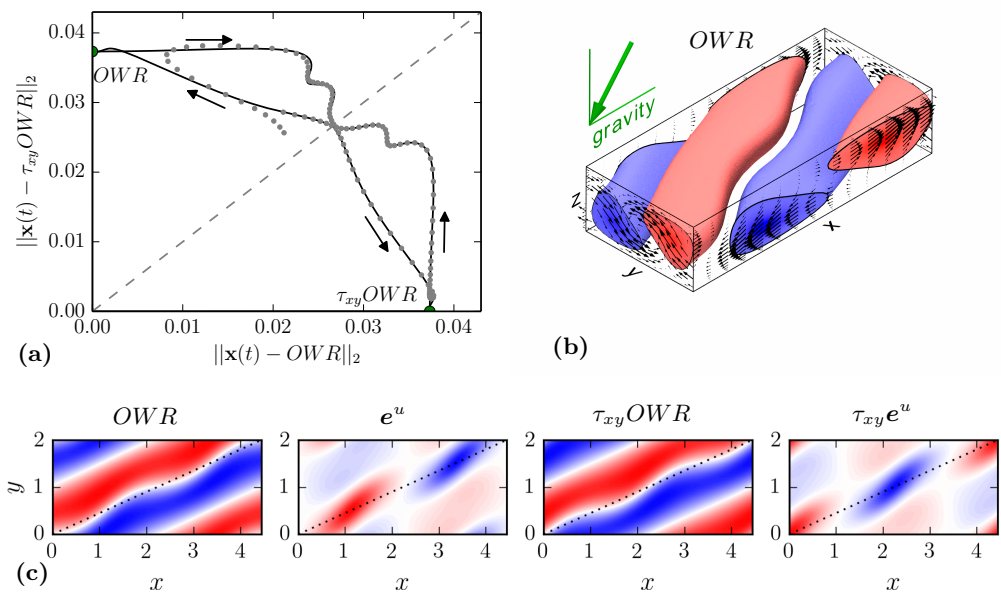

The temporal dynamics of ILC at $[\gamma,\epsilon,\mathrm{Pr}]=[40^{\circ},0.5,1.07]$ in a periodic domain of size $[2\lambda_x,\lambda_y]$ is attracted to a heteroclinic cycle. The cycle dynamically connects an equilibrium state, which we name oblique wavy rolls ($OWR$), with a symmetry related equilibrium state $\tau_{xy}OWR$ (Figure \ref{fig:obwr_cycle}). Here, $\tau_{xy}=\tau(0.25,0.25)$ is a shift operator translating the wavy flow structure by half a pattern wavelength in the direction of the domain diagonal along which the wavy convection rolls of $OWR$ are aligned. $OWR$ and $\tau_{xy}OWR$ show two spatial periods of the wavy pattern along the domain diagonal and both are invariant under transformations of $S_{\mathrm{owr}}=\langle \pi_{xyz},\tau(0.5,0.5)\rangle$. Without imposing the symmetries in $S_{\mathrm{owr}}$, $OWR$ and $\tau_{xy}OWR$ have each a single purely real unstable eigenmode, denoted as state vector $\bm{e}^u$ and $\tau_{xy}\bm{e}^u$, respectively, that breaks the $\tau(0.5,0.5)$-symmetry by having only one spatial period along the domain diagonal (Figure \ref{fig:obwr_cycle}c). Perturbing $OWR$ with $\bm{e}^u$ initiates the temporal transition $OWR\rightarrow\tau_{xy}OWR$. Perturbing $\tau_{xy}OWR$ with $\tau_{xy}\bm{e}^u$ initiates the temporal transition $\tau_{xy}OWR\rightarrow OWR$. Together, the two dynamical connections form a heteroclinic cycle. The two involved unstable eigenmodes $\bm{e}^u$ and $\tau_{xy}\bm{e}^u$ preserve the symmetries $\pi_{xyz}$ and $\pi_{xyz}\tau(0.5,0.5)$, respectively. Thus, they are analogous to \emph{sine}- and \emph{cosine}-eigenmodes in that they first, are orthogonal such that the $L_2$ inner product is $\langle \bm{e}^u,\tau_{xy}\bm{e}^u \rangle=0$ and second, have a reflection symmetry with respect to different reflection points, namely $\pi_{xyz}$ and $\pi_{xyz}\tau(0.5,0.5)$ (Figure \ref{fig:obwr_cycle}c). When imposing $\pi_{xyz}$-symmetry, the unstable eigenmode $\tau_{xy}\bm{e}^u$ is disallowed, $\tau_{xy}OWR$ becomes dynamically stable, and the associated symmetry subspace $\Sigma_1$ contains only the dynamical connection $OWR\rightarrow \tau_{xy}OWR$. When imposing $\pi_{xyz}\tau(0.5,0.5)$-symmetry, the unstable eigenmode $\bm{e}^u$ is disallowed, $OWR$ becomes dynamically stable, and the associated symmetry subspace $\Sigma_{\tau}$ contains only the dynamical connection $\tau_{xy} OWR\rightarrow OWR$. Hence, the heteroclinic cycle satisfies all three conditions for the existence of a robust heteroclinic cycle between two symmetry related equilibrium states \citep{Krupa1997}:
\begin{itemize}[leftmargin=0.9cm,itemindent=0.0cm]
\setlength\itemsep{0.2cm}
\item[(i) ] $OWR$ is a saddle and $\tau_{xy}OWR$ is an attractor (or sink) in a symmetry subspace $\Sigma_1$ of the entire state space of the $[2\lambda_x,\lambda_y]$-periodic domain.
\item[(ii) ] There is a saddle-attractor connection $OWR \rightarrow \tau_{xy}OWR$ in $\Sigma_1$.
\item[(iii) ] There is a symmetry relation between the two equilibrium states, mediated by $\tau_{xy} \in S_{\mathrm{ilc}}$.
\end{itemize}
Robust heteroclinic cycles of this type have been previously described in systems with $O(2)$-symmetry that are near a codimension-2 point where bifurcating eigenmodes show a spatial 1:2 resonance \citep{Armbruster1988,Proctor1988,Mercader2002,Nore2003}. In the present case, the existence of $OWR$ with wavenumber $m=2$ along the domain diagonal and with an instability of  wavenumber $m=1$ suggests a nearby codimension-2 point where oblique straight rolls become simultaneously unstable to $m=1$ and $m=2$ wavy modulations. Oblique straight rolls are not discussed here but are a known instability of laminar ILC \citep{Gershuni1969}. \citet{Reetz2020b} demonstrate that $OWR$ of both wavenumbers, $m=1$ and $m=2$, indeed bifurcate off oblique straight rolls in two pitchfork bifurcations at only slightly different values of the control parameters, suggesting a 1:2 resonance.

The robust heteroclinic cycle is numerically identified as an attractor of the dynamics suggesting its dynamical stability. The stability of robust heteroclinic cycles depends on the leading eigenvalues of the involved equilibrium states \citep{Krupa1995}. The leading eigenvalues of $OWR$ without imposing additional discrete symmetries are $[\omega_1,\omega_2,\omega_{3,4}]=[0.016,-0.023,-0.037\pm 0.050]$. When imposing $\pi_{xyz}$, the contracting eigenvalue $\omega_2$ vanishes. When imposing $\pi_{xyz}\tau(0.5,0.5)$, the expanding eigenvalue $\omega_1$ vanishes. Thus, the leading expanding and contracting eigenvalues belong to two different symmetry subspaces. The complex eigenvalue $\omega_{3,4}$ is radial as it belongs to both subspaces and does not influence the stability of the cycle. Since $|\omega_1|/|\omega_2|<1$, the heteroclinic cycle is dynamically stable \citep{Krupa1995}. 

We do not expect this heteroclinic cycle to be stable in larger domains. However, oblique wavy rolls are observed to evolve slowly in experiments and simulations at lower thermal driving \citep{Daniels2008}. At the control parameter values selected here, observations in larger domains indicate chaotic dynamics on a faster time scale than the time scale of the approach to the heteroclinic cycle \citep{Reetz2020d}. The time period $\Delta t$ over which the state trajectory remains close to an equilibrium increases with time (Figure \ref{fig:obwr_portrait}a). It should eventually diverge but here saturates at $\Delta t\approx \mathcal{O}(10^3)$. This saturation effect is due to the numerical double-precision of the DNS. The unstable eigenvalue $\omega_1=0.016$ of $OWR$ amplifies the numerical noise on a time scale of $ \log(10^{16})/\omega_1= \mathcal{O}(10^3)$.

\subsubsection{Knots}
\label{sec:kn}
%
The convection pattern of knots ($\mathcal{KN}$) is experimentally observed as `knotted' superposition of $\mathcal{TR}$ and $\mathcal{LR}$ just above inclination $\gamma_{c2}$ \citep{Daniels2000}. Stability analysis confirms the existence of a $KN_i$ instability of $TR$ \citep{Fujimura1993}. We refer to experimental and numerical observations of $\mathcal{KN}$ at $[\gamma,\epsilon,\mathrm{Pr}]=[80^{\circ},0.05,1.07]$ \citep{Subramanian2016}. At these control parameter values, a temporal transition from the noise-perturbed unstable base flow is simulated in a periodic domain of size $[\lambda_x,\lambda_y]$. No additional discrete symmetries are imposed.

After the initial shear driven transition $B\rightarrow TR$, the buoyancy driven instability $KN_i$ of $TR$ leads to a stable spiral approaching $KN$, an equilibrium state underlying the observed $\mathcal{KN}$ pattern (Figure \ref{fig:kn}). The spectrum of eigenvalues of stable $KN$ has a complex pair $(\omega_r,\omega_i)=(-0.0085,\pm 0.0304)$ closest to the imaginary axis. The linear period of $T=2\pi/\omega_i=207$ matches the simulated oscillations on the stable spiral trajectory. The flow structure of $KN$ shows the characteristic bimodal mix of longitudinal and transverse modes, here, with a stronger transverse contribution (Figure \ref{fig:kn}b). $KN$ is invariant under symmetries $S_{\mathrm{kn}}=\langle \pi_y\tau(0,0.5), \pi_{xyz}\rangle$.

Close to the values of the control parameters where stationary $\mathcal{KN}$ are observed, \citet{Daniels2000} report on bursting dynamics. When simulating a transition at increased $\epsilon=0.15$ ($\mathrm{Ra}=9338$), the state trajectory, after transiently visiting $TR$, does not approach $KN$ that has become unstable. Instead, the trajectory visits again the laminar base flow ($TR\rightarrow B$) from where it approaches a stable periodic orbit with period $T=251$ and $S_{\mathrm{kn}}$ symmetries. We call this orbit bursting knots ($BKN$). The $BKN$ orbit describes a bursting cycle with slow dynamics near $B$ and fast dynamics along a clockwise revolving trajectory in the $D/I$-plane (Figure \ref{fig:knorb}). The fast stage shows growth of a transient $\mathcal{KN}$ pattern that ultimately forms decaying longitudinal plumes (Figure \ref{fig:knorb}b). Longitudinal modes decay because $LR_i$ exists only at higher $\epsilon$ at $\gamma=80^{\circ}$. During the slow stage, the phase portrait of the orbit shows sharp turns near $B$ suggesting an influence of the stable and unstable manifold of $B$ on the orbit (inset in Figure \ref{fig:knorb}). The bursting dynamics of this specific periodic orbit appears similar to a nonlinear limit cycle found in natural doubly diffusive convection \citep{Bergeon2002} but does not match the traveling dynamics of the longitudinal bursts observed in ILC at these control parameter values \citet{Daniels2000}. The next section discusses two periodic orbits clearly underlying experimental observations.

\begin{figure}
\center
		\begin{subfigure}[b]{0.7\textwidth}
        \begin{tikzpicture}
    	\draw (0, 0) node[inner sep=0]{\includegraphics[width=0.99\linewidth,trim={0.5cm 2.0cm 3.0cm 0.5cm},clip]{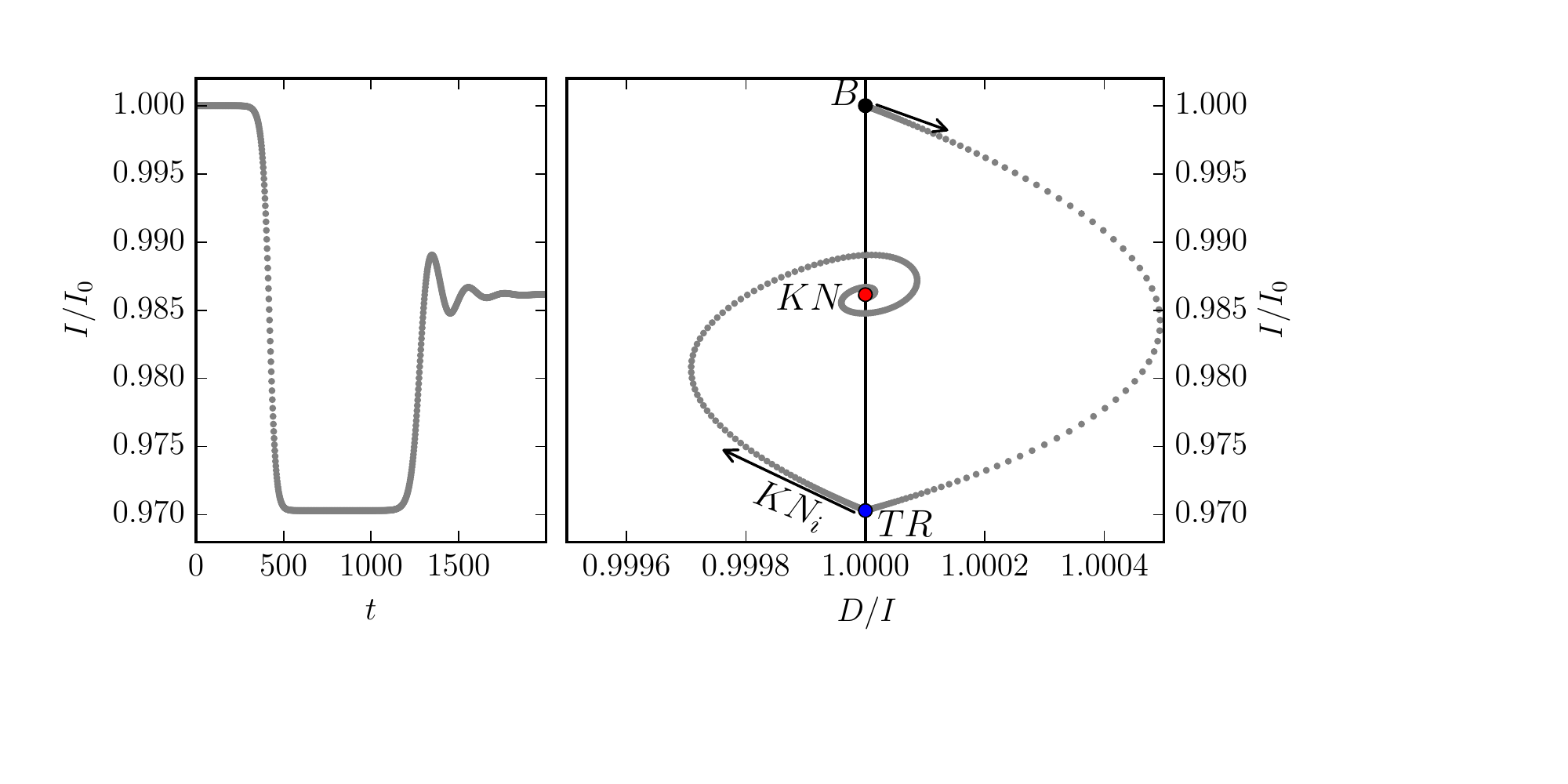}};
    	\draw (-4.3,-1.8) node {\textbf{(a)}};
		\end{tikzpicture}
		\end{subfigure}
		\begin{subfigure}[b]{0.29\textwidth}
        \begin{tikzpicture}
    	\draw (0, 0) node[inner sep=0]{\includegraphics[width=\linewidth,trim={0.0cm 0.0cm 0.5cm 0.1cm},clip]{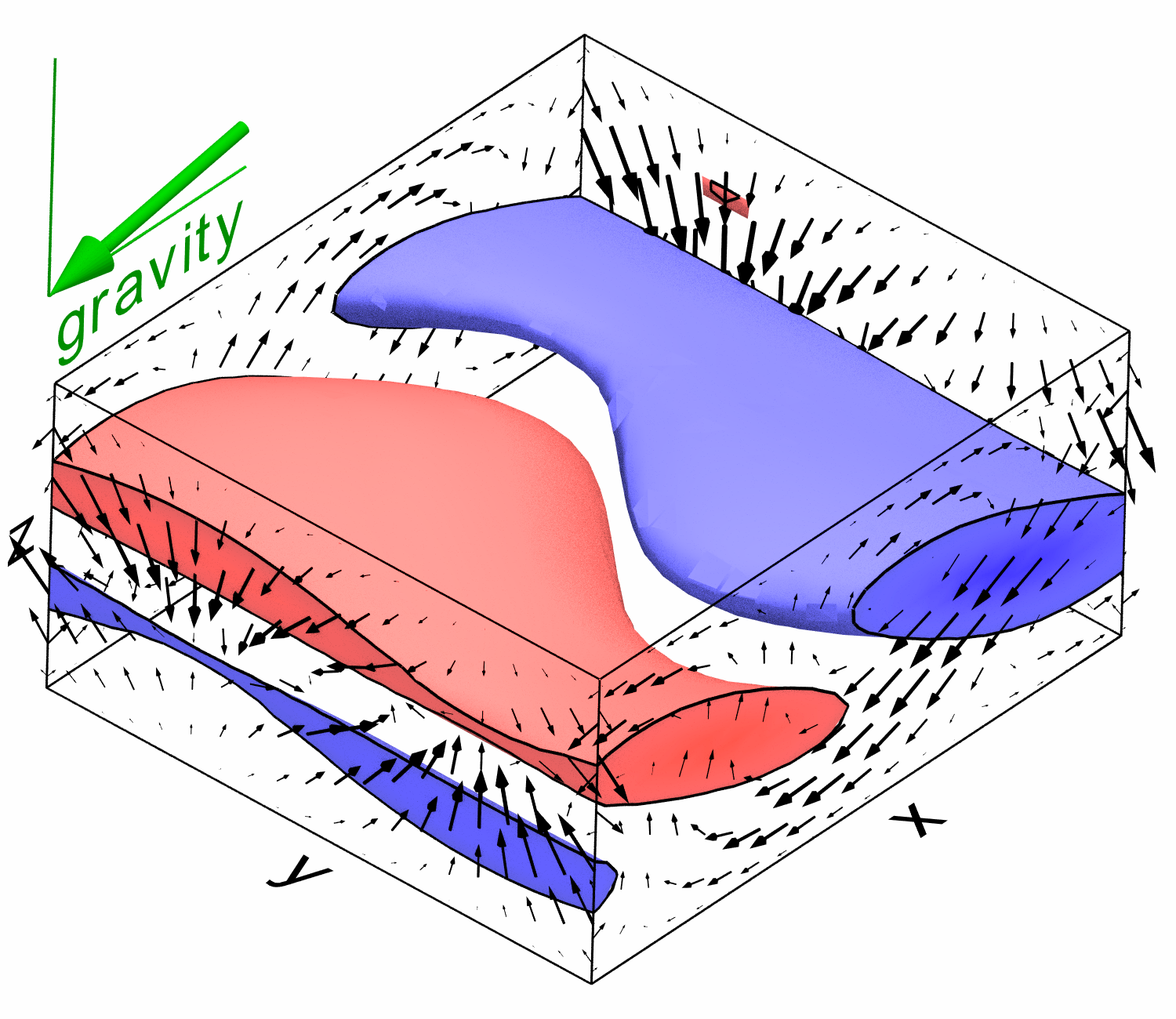}};
    	\draw (-1.8,-1.8) node {\textbf{(b)}};
    	\draw (-0.5,1.7) node {$KN$};
		\end{tikzpicture}
		\end{subfigure}
\caption{\label{fig:kn} \textbf{(a)} State trajectory evolution from the unstable base state $B$ at $\gamma=80^{\circ}$ and $\epsilon=0.05$ ($\mathrm{Ra}=8525$). After a transient in the vicinity of $TR$, the buoyancy driven instability $KN_i$ causes the trajectory to follow a stable spiral towards equilibrium state $KN$. \textbf{(b)} Flow structure of $KN$ in a periodic domain of size $[\lambda_x,\lambda_y]$.}
\end{figure}
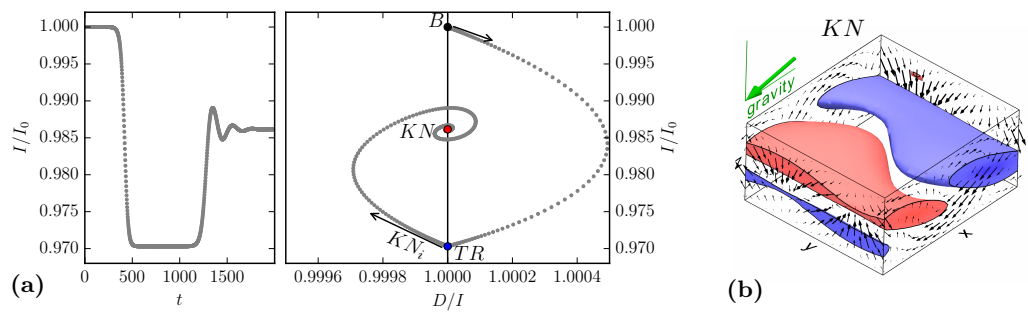
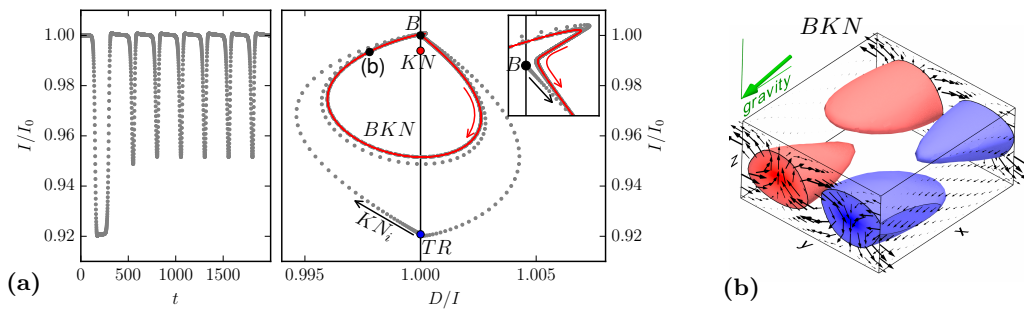
\begin{figure}
		\begin{subfigure}[b]{0.7\textwidth}
        \begin{tikzpicture}
    	\draw (0, 0) node[inner sep=0]{\includegraphics[width=0.99\linewidth,trim={0.5cm 2.0cm 3.0cm 0.5cm},clip]{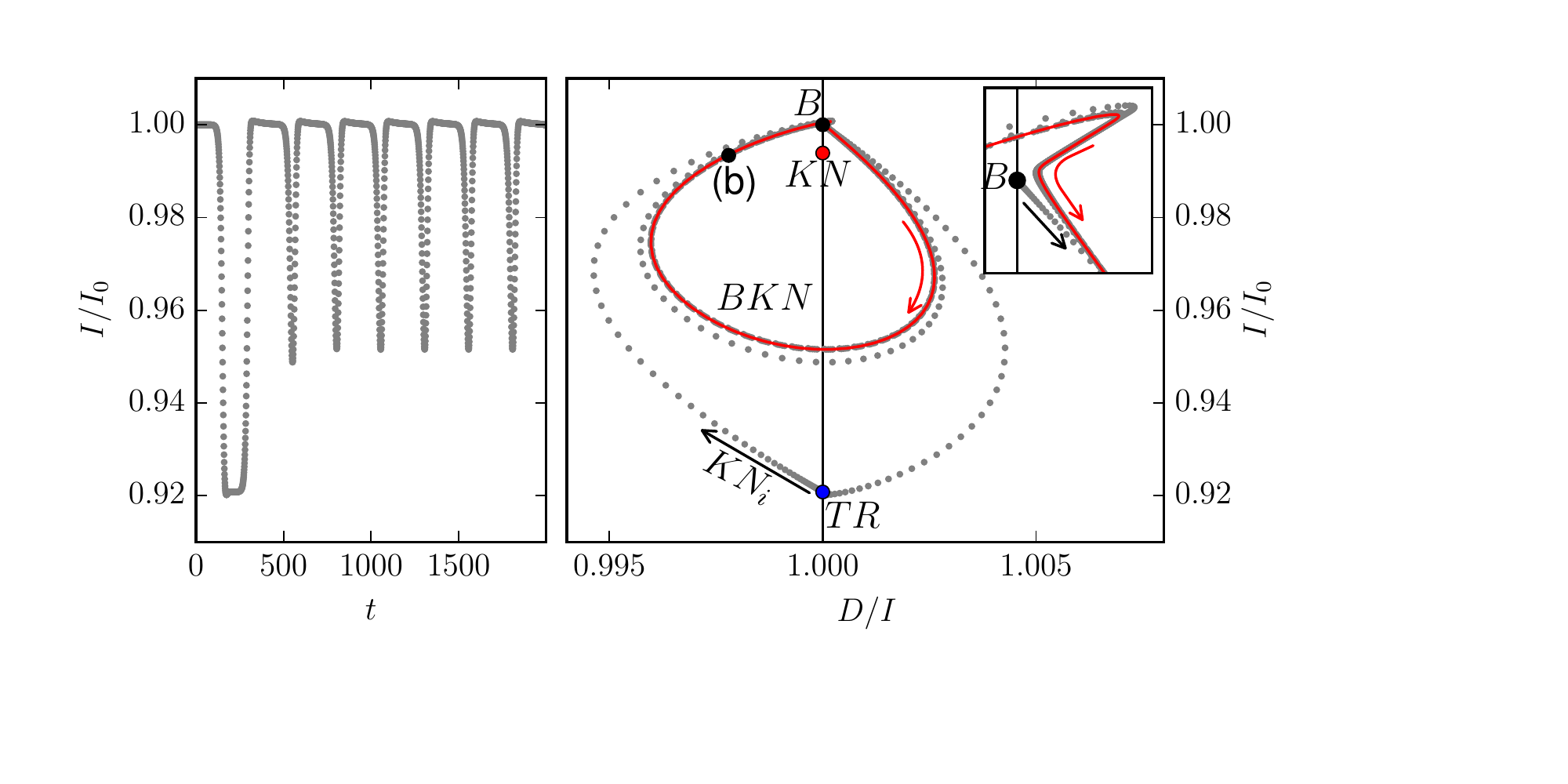}};
    	\draw (-4.3,-1.8) node {\textbf{(a)}};
		\end{tikzpicture}
		\end{subfigure}
		\begin{subfigure}[b]{0.29\textwidth}
        \begin{tikzpicture}
    	\draw (0, 0) node[inner sep=0]{\includegraphics[width=\linewidth,trim={0.0cm 0.0cm 0.5cm 0.1cm},clip]{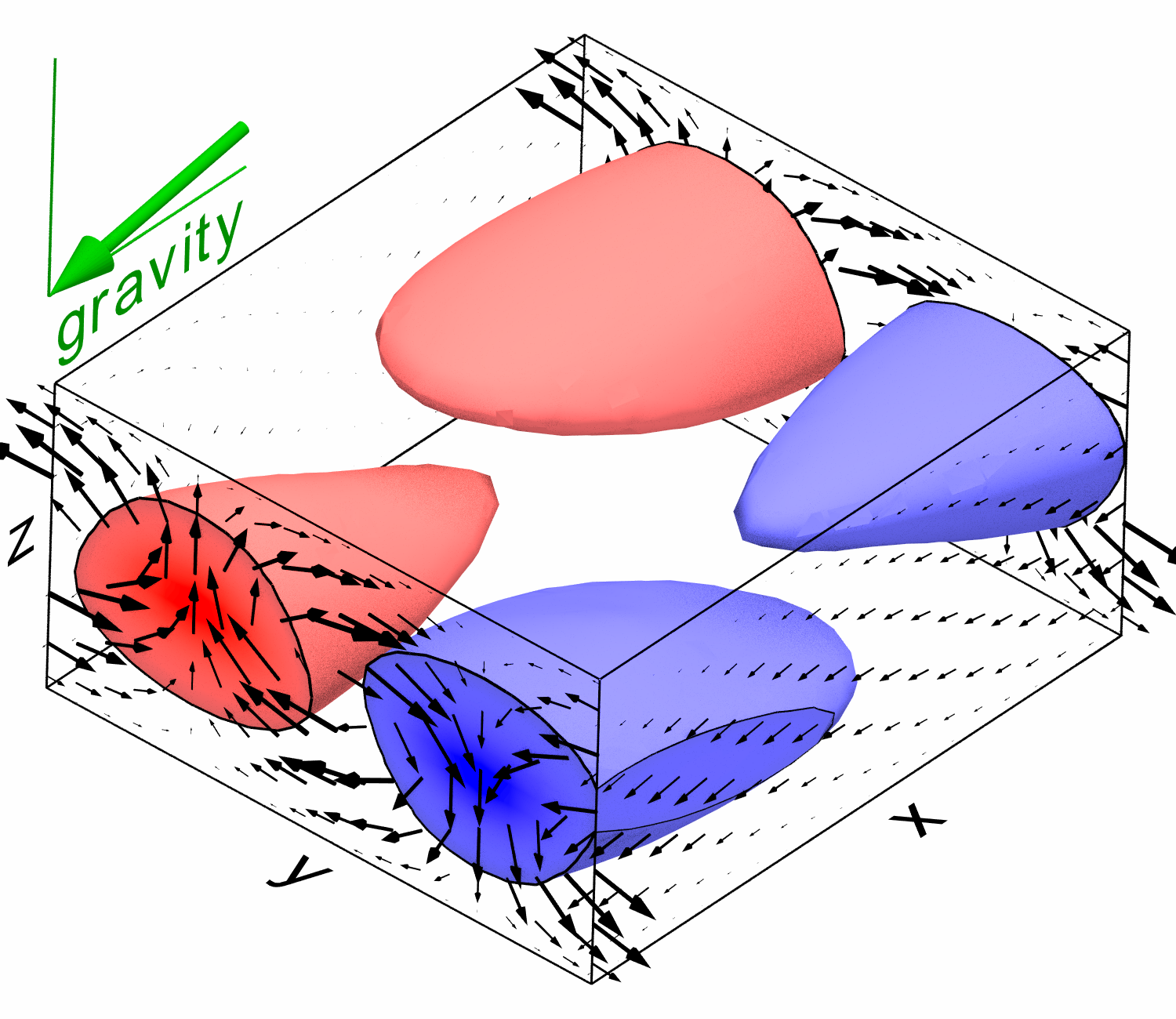}};
    	\draw (-1.8,-1.8) node {\textbf{(b)}};
    	\draw (-0.6,1.7) node {$BKN$};
		\end{tikzpicture}
		\end{subfigure}
\caption{\label{fig:knorb} \textbf{(a)} State trajectory evolution from the unstable base state $B$ at $\gamma=80^{\circ}$, like in Figure \ref{fig:kn}, but at increased $\epsilon=0.15$ ($\mathrm{Ra}=9338$). Instead of terminating in a stable spiral, the trajectory returns to laminar flow from where it approaches the stable periodic orbit $BKN$ along which knots emerge as bursts. The inset magnifies the phase portrait close to the laminar base state $B$. \textbf{(b)} Flow structure of an instance along the periodic orbit $BKN$ illustrates decaying longitudinal plumes that bring the state trajectory close to laminar flow.}
\end{figure}

\subsection{Transitions with periodic attractors}
\label{sec:orbits}

\subsubsection{Subharmonic oscillations}
\label{sec:lso}
An oscillatory instability of $LR$ at small inclinations gives rise to a convection pattern of spatially subharmonic oscillations observed in experiments at $\mathrm{Pr}=1.07$ \citep{Daniels2000} and studied using Floquet analysis at $\mathrm{Pr}=0.71$ \citep{Busse2000}. Here, we depart from our convention to follow the naming of \citet{Subramanian2016} and name this pattern subharmonic standing wave ($\mathcal{SSW}$) instead of longitudinal subharmonic oscillations to stress the standing wave nature of the pattern. We refer to observations of $\mathcal{SSW}$ at $[\gamma,\epsilon,\mathrm{Pr}]=[17^{\circ},1.5,1.07]$ \citep{Daniels2000,Subramanian2016}. A temporal transition is simulated in a periodic domain of size $[2\lambda_x,2\lambda_y]$, well matching the results of the Floquet analysis in \citet{Subramanian2016}. Unexpectedly, the dynamics transiently exhibits spatially subharmonic oscillations but does not saturate at a stable oscillatory pattern at these values of the control parameters. Therefore, we reduce the angle of inclinations to $\gamma=15^{\circ}$. No additional discrete symmetries are imposed.

\begin{figure}
\center
\includegraphics[width=0.99\linewidth,trim={1.2cm 2.0cm 0.5cm 0.5cm},clip]{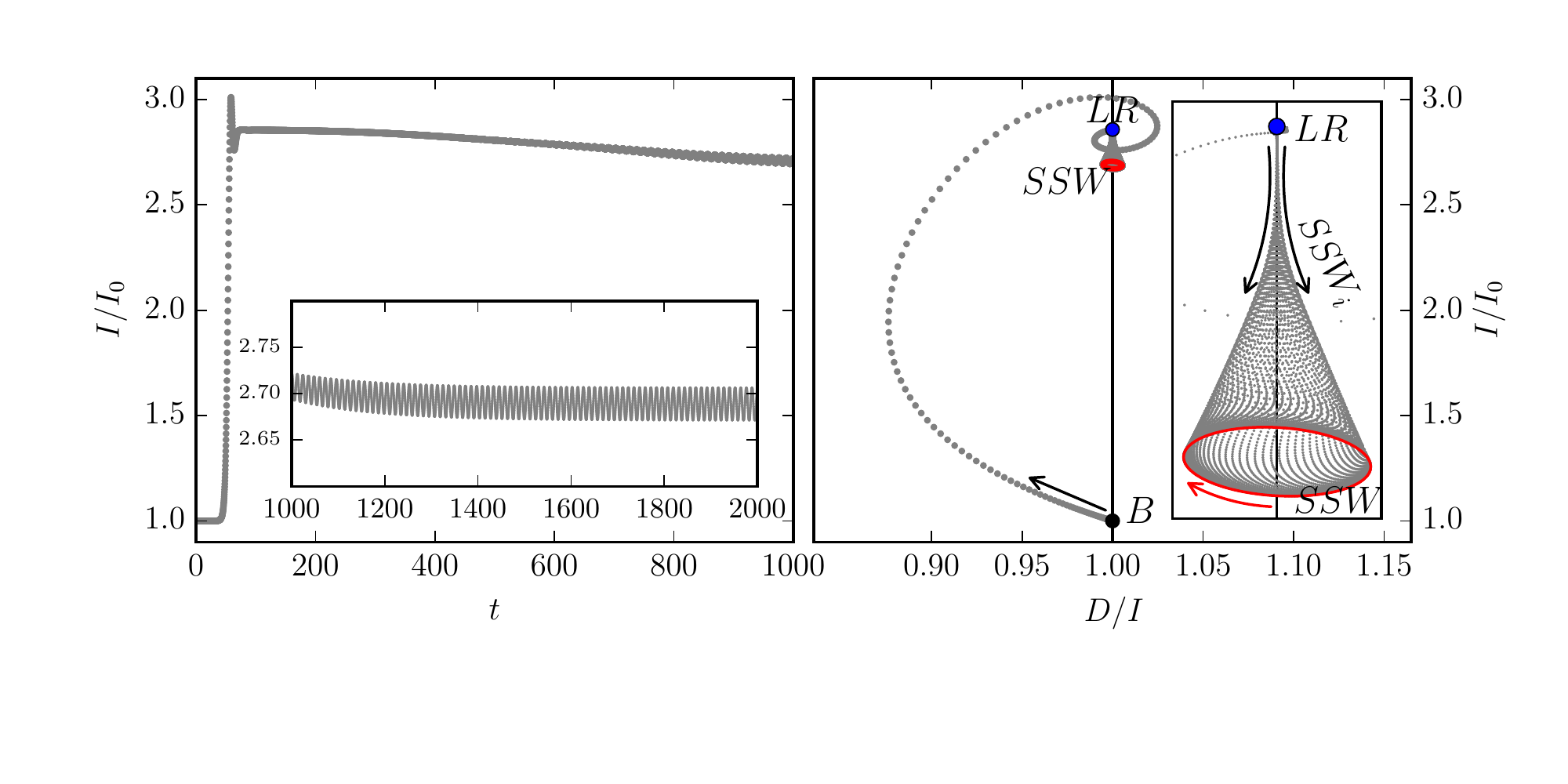}
\caption{\label{fig:pport_lso} State trajectory evolution from the unstable base state $B$ at $\gamma=15^{\circ}$ and $\epsilon=1.5$ ($\mathrm{Ra}=4420$). The phase portrait illustrates how the instability $SSW_i$ leads to an oscillatory transition $LR\rightarrow SSW$ along a symmetric unstable spiral approaching periodic orbit $SSW$ (red solid line). The inset magnifies the phase portrait of the transition $LR\rightarrow SSW$. }
\end{figure}
%

After a rapid transient visit near $LR$, the dynamics along a drifting spiral trajectory is attracted to a stable limit cycle, a periodic orbit named $SSW$ (Figure \ref{fig:pport_lso}). The initial part of the transition $LR\rightarrow SSW$ is symmetric around $D/I=1$ indicating that buoyancy and shear forces equally drive this instability. The periodic orbit $SSW$ revolves clockwise in the $D/I$-plane. $SSW$ is also a pre-periodic orbit satisfying condition ($\ref{eq:ecs_def}$) with $\sigma_{\mathrm{ssw}}=\pi_y \tau(0.25,0.25)$ and a pre-period of $T'_{\mathrm{ssw}}=12.03$. The local oscillatory instability of $LR$ suggests $T=2\pi/\omega_i=45.11$, close to the observed full orbit period of $T_{\mathrm{ssw}}=4T'_{\mathrm{ssw}}=48.12$. After $2T'_{\mathrm{ssw}}$, condition (\ref{eq:ecs_def}) requires $\sigma=\tau(0.5,0)$. The orbit is invariant under inversion and half-box shifts $S_{\mathrm{ssw}}=\langle\pi_{xyz},\tau(0.5,0.5)\rangle$. 

\begin{figure}
		\center
        \begin{subfigure}[b]{0.99\textwidth}
                \includegraphics[width=\linewidth,trim={0cm 0.0cm 0.0cm 0.2cm},clip]{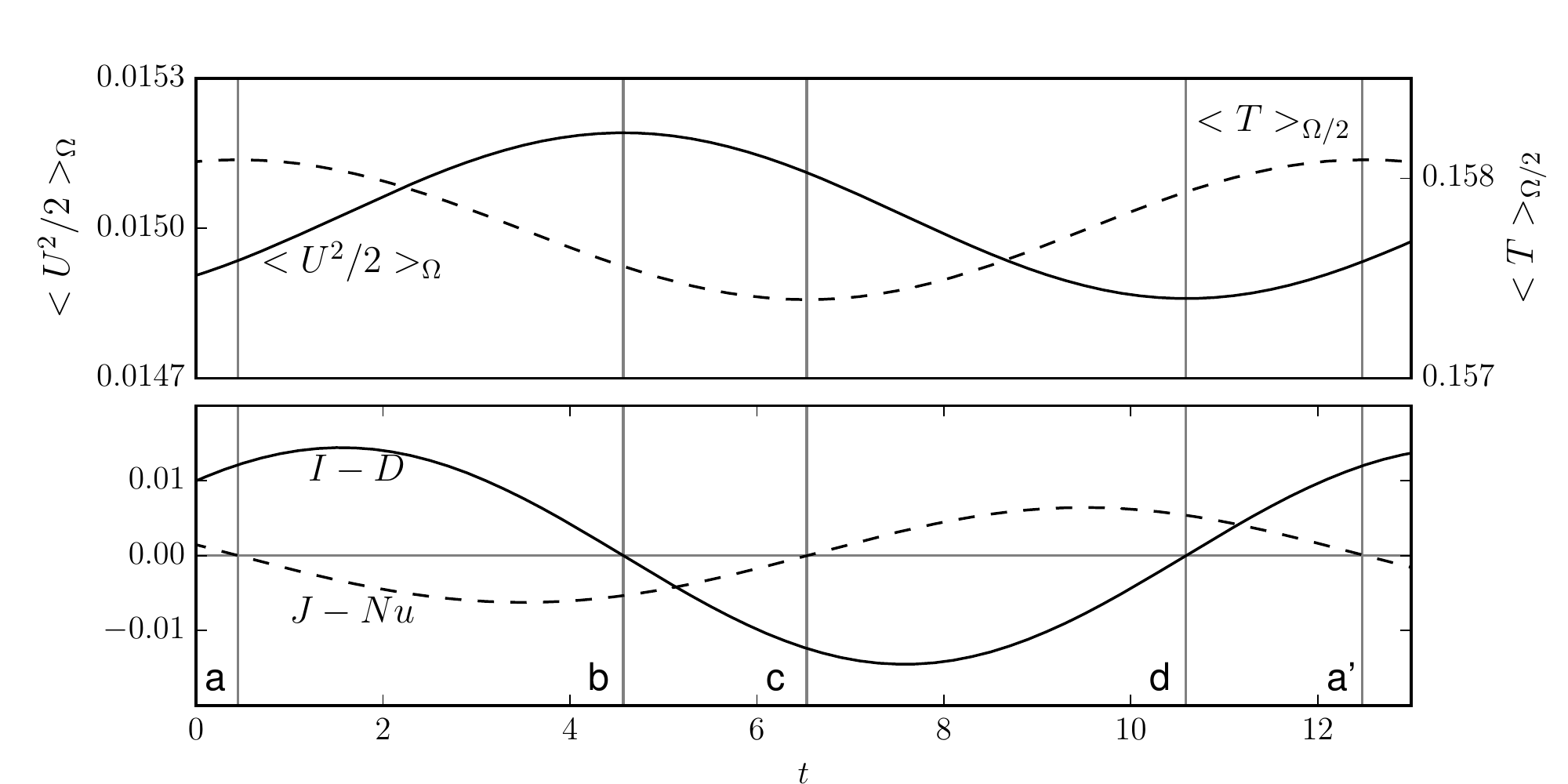}
                \label{fig:lso:phys}
        \end{subfigure}
        \begin{subfigure}[b]{0.221\textwidth}
                \includegraphics[width=\linewidth,trim={0.1cm 0.1cm 0cm 0.2cm},clip]{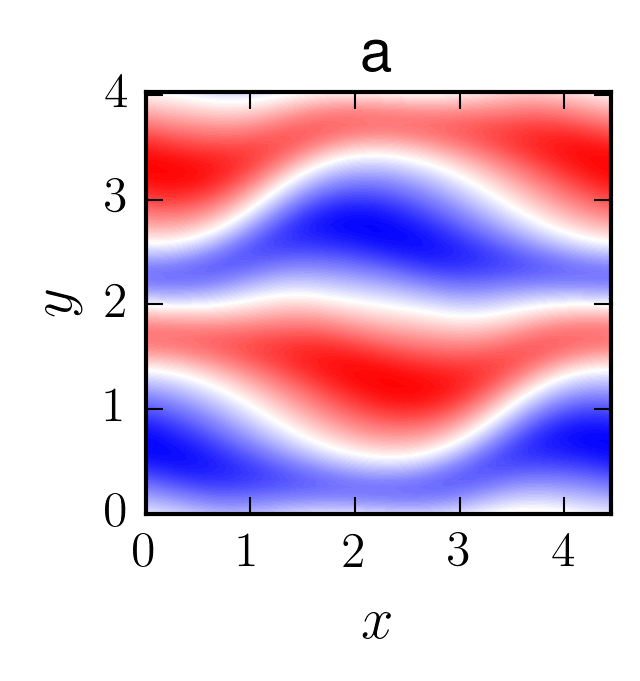}
        \end{subfigure}
        \begin{subfigure}[b]{0.185\textwidth}
                \includegraphics[width=\linewidth,trim={0.1cm 0.1cm 0cm 0.2cm},clip]{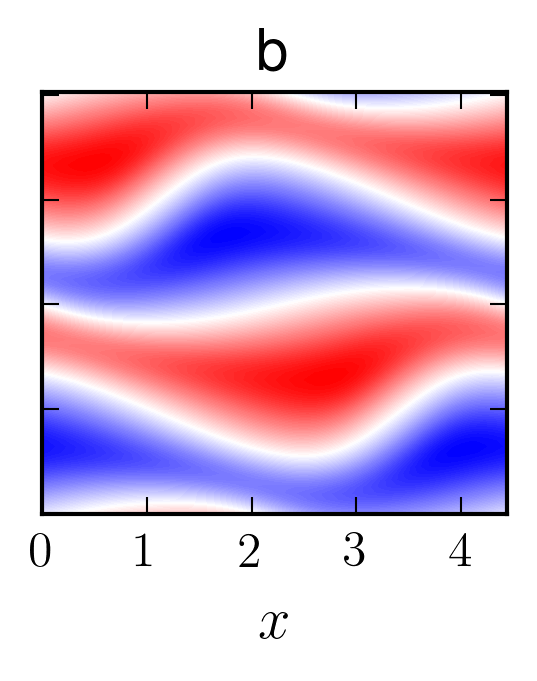}
        \end{subfigure}
        \begin{subfigure}[b]{0.185\textwidth}
                \includegraphics[width=\linewidth,trim={0.1cm 0.1cm 0cm 0.2cm},clip]{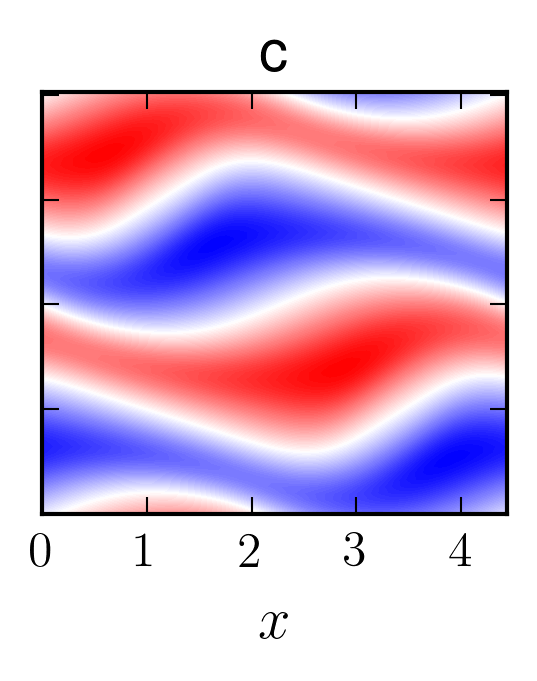}
        \end{subfigure}
        \begin{subfigure}[b]{0.185\textwidth}
                \includegraphics[width=\linewidth,trim={0.1cm 0.1cm 0cm 0.2cm},clip]{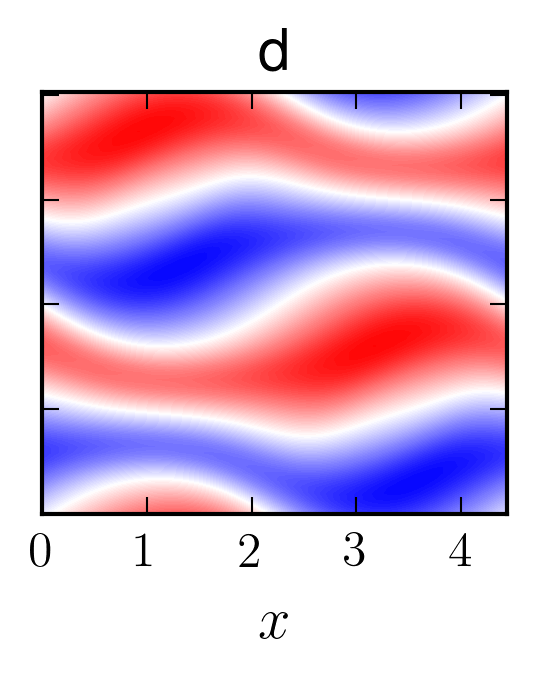}
        \end{subfigure}
        \begin{subfigure}[b]{0.185\textwidth}
                \includegraphics[width=\linewidth,trim={0.1cm 0.1cm 0cm 0.2cm},clip]{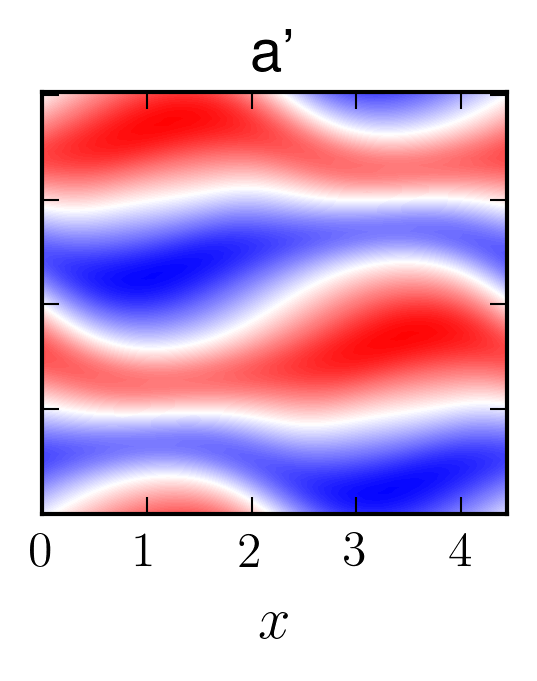}
        \end{subfigure}
		\caption{\label{fig:lso_orb} Transfer rates of kinetic energy (solid) and thermal energy (dashed) over one pre-period $T'_{\mathrm{ssw}}=12.03$ of the $SSW$ orbit at $\gamma=15^{\circ}$ and $\epsilon=1.5$ ($\mathrm{Ra}=4420$). Temperature contours at midplane show instances of kinetic energy balance $I-D=0$ (b,d) or thermal energy balance $J-\mathrm{Nu}=0$ (a,c,a') along the orbit. States $a$ and $a'$ are related by symmetry transformation $\sigma=\pi_y \tau(0.25,0.25)$.}
\end{figure}
%
Periodic orbits in ILC must exactly balance net transfer of kinetic energy and heat over one period $\int_0^{T'_{\mathrm{ssw}}}(I-D)\, dt=0$. The terms in the energy equations (\ref{eq:kenergy}-\ref{eq:tenergy}) oscillate approximately harmonically with one relative period of $SSW$. Instances of $I-D=0$ or $J-\mathrm{Nu}=0$ correspond to local extrema of $<U^2/2>_{\Omega}$ and $<T>_{\Omega/2}$, respectively. The phase lag between kinetic energy and heat phase is $0.37\,T'_{\mathrm{ssw}}$ (Figure \ref{fig:lso_orb}). The Nusselt number varies between $1.88\le \mathrm{Nu} \le 1.90$, close but below the convective heat transfer of $LR$ with $\mathrm{Nu}=1.98$. The pattern of $SSW$ over one period can be described as standing wave modulation (panels in Figure \ref{fig:lso_orb}), a consequence of counter-propagating traveling waves along the hot and cold plumes of $LR$ \citep{Busse2000}. \\

\subsubsection{Transverse oscillations}
\label{sec:to}
%
Transverse oscillations ($\mathcal{TO}$) are observed experimentally as chaotic bending modulations of $\mathcal{TR}$, a pattern also named `switching diamond panes' \citep{Daniels2000}. An oscillatory $TO_i$ instability is found as a secondary instability of $TR$ in the interval $83.2^{\circ}< \gamma\le 120^{\circ}$ for $\mathrm{Pr}=1.07$ \citep{Subramanian2016}. We refer to observations of $\mathcal{TO}$ at $[\gamma,\epsilon,\mathrm{Pr}]=[100^{\circ},0.1,1.07]$ \citep{Daniels2000,Subramanian2016}, and simulate a temporal transition at these control parameter values in a periodic domain of size $[12\lambda_x,6\lambda_y]$, close to the pattern wavelengths used to simulate $TO$ in \citet{Subramanian2016}. Without imposing additional discrete symmetries, no stable periodic orbit is found. Therefore, a transition is simulated in a symmetry subspace defined by $S_{\mathrm{to}}=\langle \pi_y,\pi_{xz},\tau(0.5,0.5)\rangle$.

\begin{figure}
\center
\includegraphics[width=0.99\linewidth,trim={0.8cm 2.0cm 0.5cm 0.5cm},clip]{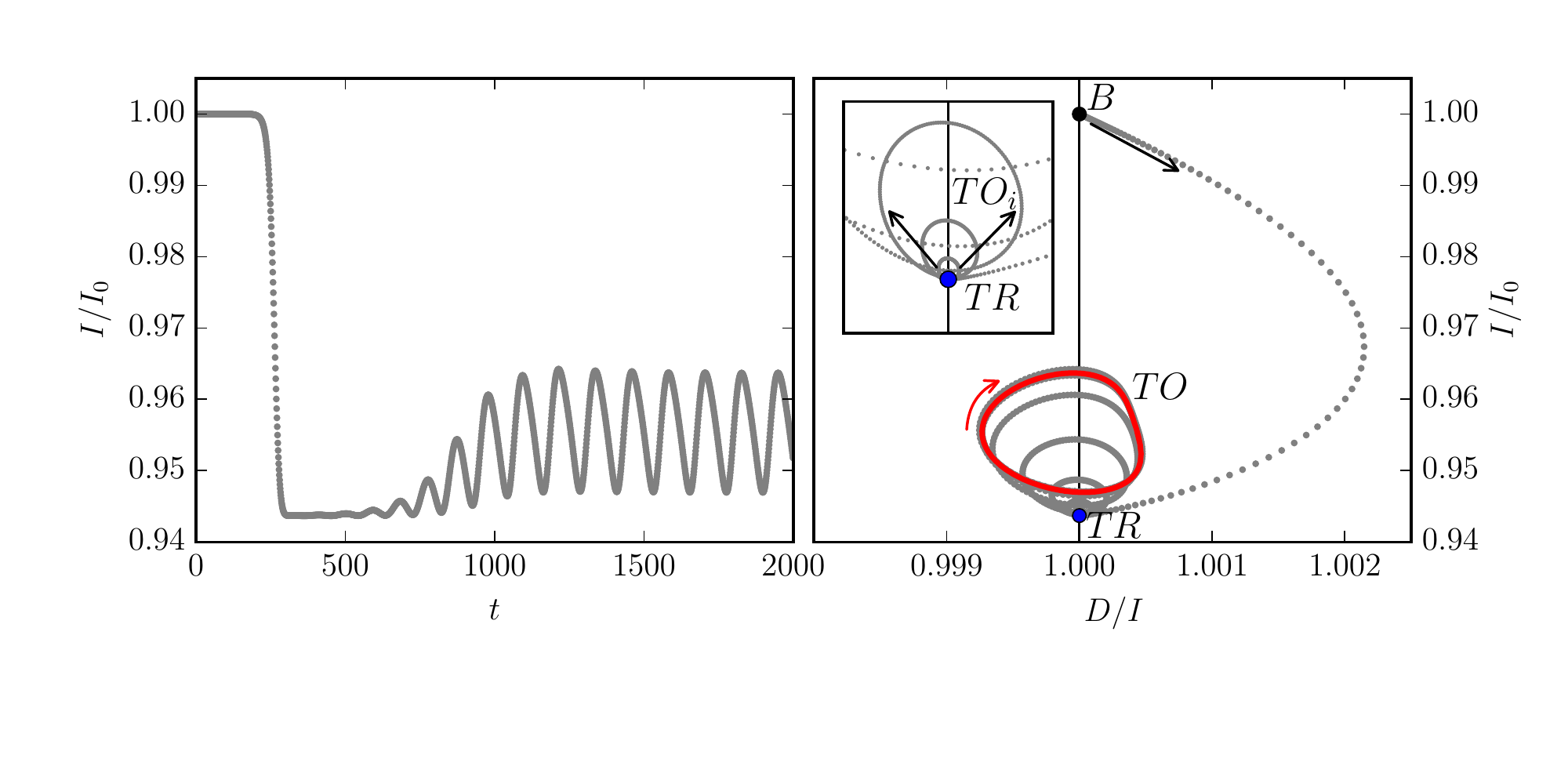}
\caption{\label{fig:pport_to} State trajectory evolution from the unstable base state $B$ at $\gamma=100^{\circ}$ and $\epsilon=0.1$ ($\mathrm{Ra}=10050$). The phase portrait illustrates how the TO instability leads to an oscillatory transition $LR\rightarrow TO$ along an unstable spiral approaching periodic orbit $TO$ (red solid line). Initial oscillations triggered by $TO_i$ are symmetric with respect to $D/I=1$. The inset magnifies the initial symmetric trajectory from $TR$.}
\end{figure}
The transition $B\rightarrow TR$ gives rise to 12 pairs of straight transverse rolls before slow and weak bending modulations set in. Like $SSW_i$, the instability $TO_i$ of $TR$ generates a state trajectory that is initially symmetric around $D/I=1$ suggesting that buoyancy and shear forces drive this instability equally. The state trajectory from $TR$ approaches the stable periodic orbit $TO$ within a few oscillation periods (Figure \ref{fig:pport_to}). $TO$ is a pre-periodic orbit solving condition (\ref{eq:ecs_def}) with  $\sigma_{\mathrm{to}}=\tau(0.5,0)$ and a pre-period of $T'_{\mathrm{to}}=122.1$, i.e. half of the full period. As observed experimentally \citep{Daniels2000}, the oscillation period is on the order of one diffusion time scale $\mathcal{O}(T_d)= \mathcal{O}(\sqrt{\mathrm{Ra}\,\mathrm{Pr}}\,T_f)$. Without the imposed discrete symmetries $S_{\mathrm{to}}$, $TO$ has six eigenvalues with positive real part at the given parameters. The associated eigenmodes break all symmetries in $S_{\mathrm{to}}$.

\begin{figure}
\center
        \begin{subfigure}[b]{0.99\textwidth}
                \includegraphics[width=\linewidth,trim={0cm 0.0cm 0.0cm 0.2cm},clip]{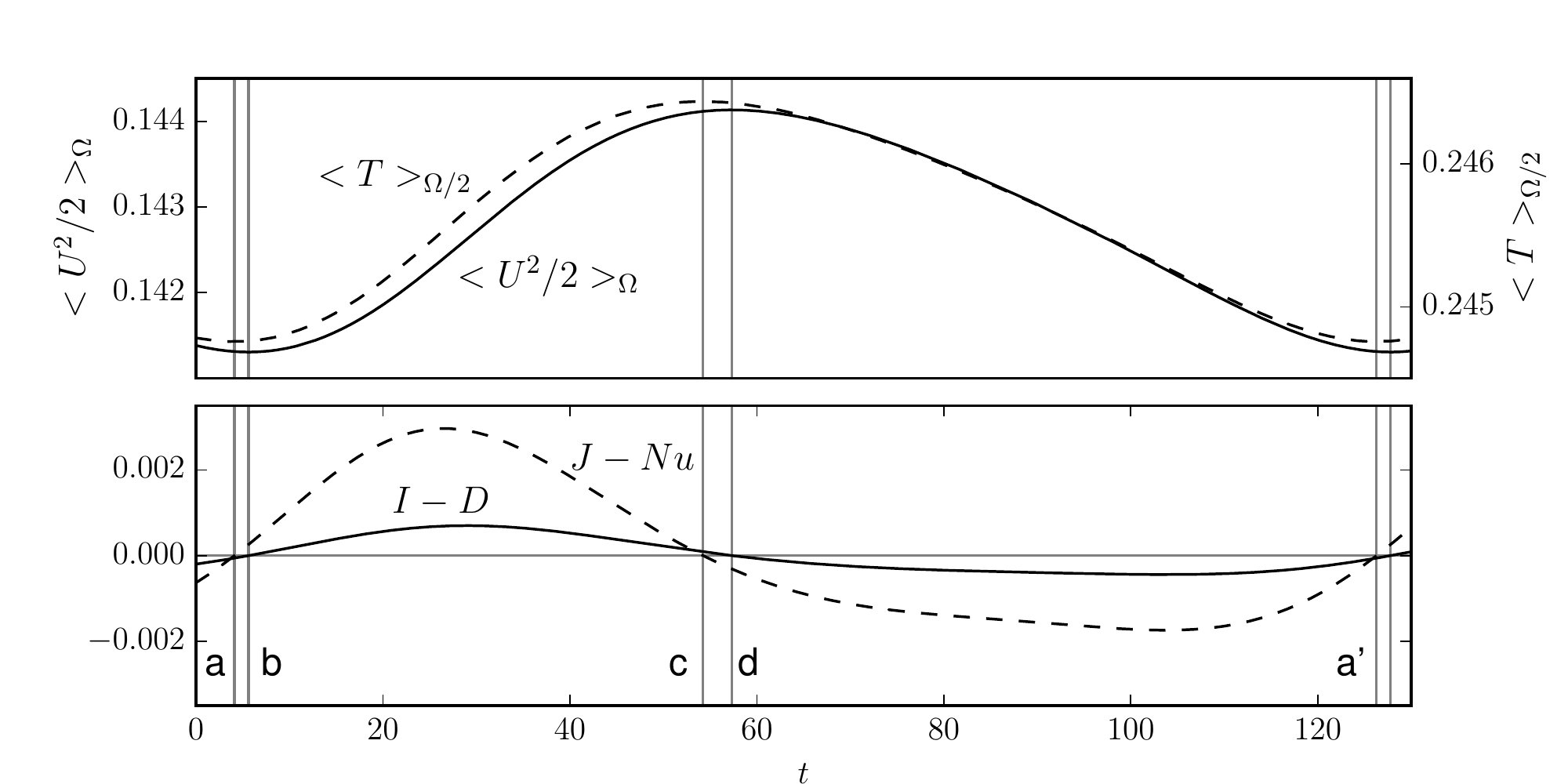}
        \end{subfigure}
        \begin{subfigure}[b]{0.362\textwidth}
                \includegraphics[width=\linewidth,trim={0.1cm 0.1cm 0cm 0cm},clip]{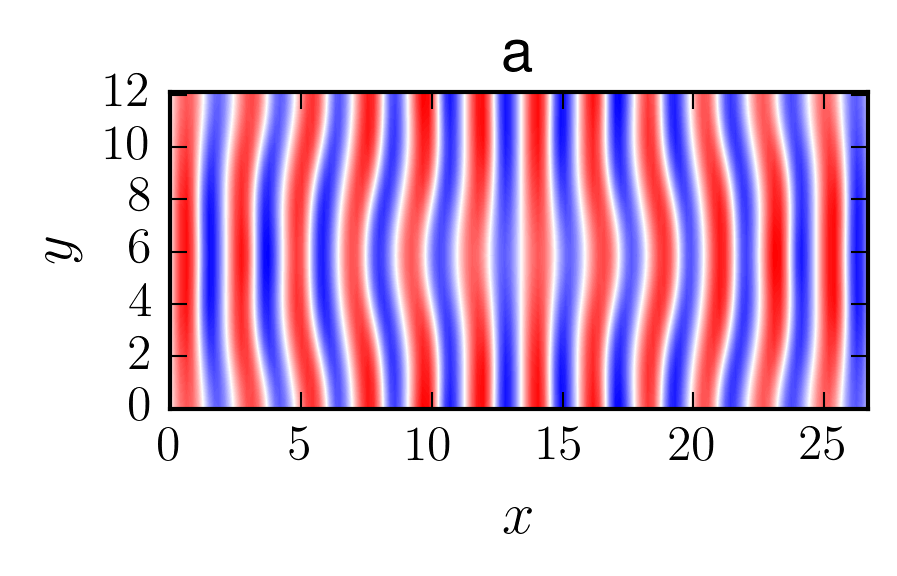}
        \end{subfigure}
        \begin{subfigure}[b]{0.31\textwidth}
                \includegraphics[width=\linewidth,trim={0.1cm 0.1cm 0cm 0cm},clip]{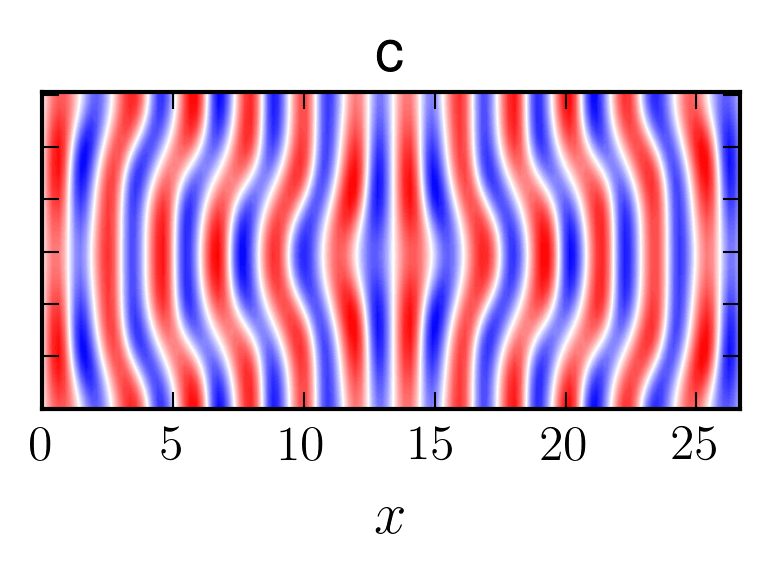}
        \end{subfigure}
        \begin{subfigure}[b]{0.31\textwidth}
                \includegraphics[width=\linewidth,trim={0.1cm 0.1cm 0cm 0cm},clip]{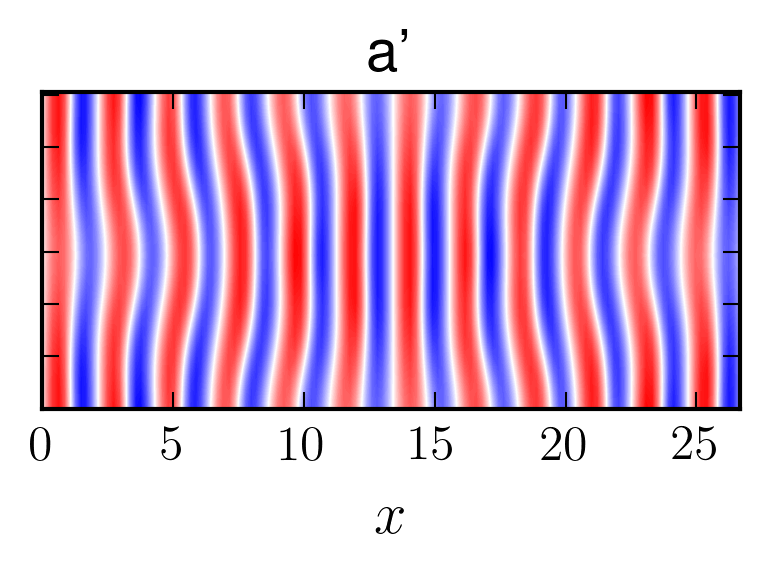}
        \end{subfigure}
		\caption{\label{fig:to_orb} Transfer rates of kinetic (solid) and thermal energy (dashed) over one pre-period $T'_{\mathrm{to}}=122.1$ of the $TO$ orbit at $\gamma=100^{\circ}$ and $\epsilon=0.1$ ($\mathrm{Ra}=10050$). Temperature contours at midplane show instances of thermal energy balance $J-\mathrm{Nu}=0$. States $a$ and $a'$ are related by symmetry transformation $\sigma=\tau(0.5,0)$.}
\end{figure}

Heat and kinetic energy oscillate non-harmonically and almost in phase over a relative period of $TO$ at these control parameter values. The pattern of $TO$ resembles $TR$ at the kinetic energy minimum. Near the energy maximum, the transverse rolls are maximally bent (Figure \ref{fig:to_orb}). The weak subharmonic varicose oscillations have a much larger pattern wavelength than all other invariant states discussed in the present work. In very large domains, observations show spatial-temporal chaos at these control parameter values \citep{Daniels2000,Subramanian2016}, suggesting that the periodic orbit $TO$ in larger domains is embedded in a chaotic attractor.

\subsection{Transient dynamics of the skewed varicose pattern in Rayleigh-B\'enard convection}
\label{sec:svtransient}
%
Various secondary instabilities are known in Rayleigh-B\'enard convection, namely Eckhaus, zigzag, knot, skewed varicose, cross rolls, and oscillatory instability \citep{Busse1978a}. At $\mathrm{Pr}=1.07$, the skewed varicose instability $SV_i$ is the first to destabilise convection rolls as demonstrated by experiments \citep{Bodenschatz2000} and stability analysis \citep{Subramanian2016}. We refer to observations of the skewed varicose pattern ($\mathcal{SV}$) at $[\gamma,\epsilon,\mathrm{Pr}]=[0^{\circ},2.26,1.07]$ \citep[][Fig.7]{Bodenschatz2000}. A normalised Rayleigh number of $\epsilon=2.26$ is far above the critical threshold for $SV_i$. Here, we simulate a temporal transition at $\epsilon=1.05$, much closer to threshold. The periodic domain is of size $[4\lambda_x,4\lambda_y]$, and no additional discrete symmetries are imposed.

\begin{figure}
\begin{subfigure}[]{1\textwidth}
\includegraphics[width=0.99\linewidth,trim={0.8cm 1.5cm 0.4cm 0.5cm},clip]{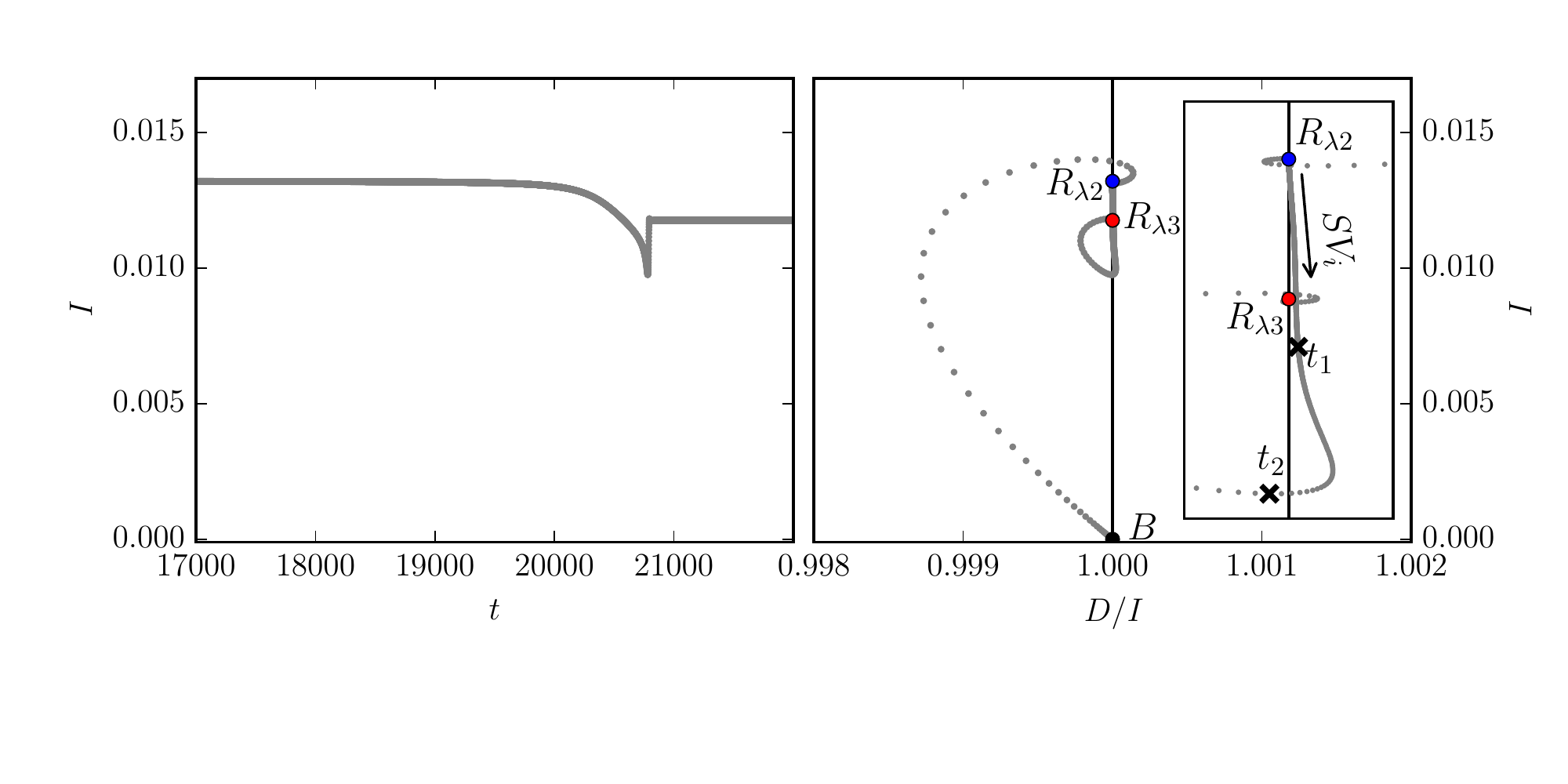}
\end{subfigure}
        \begin{subfigure}[b]{0.276\textwidth}
                \includegraphics[width=\linewidth,trim={0.1cm 0.1cm 0.1cm 0.1cm},clip]{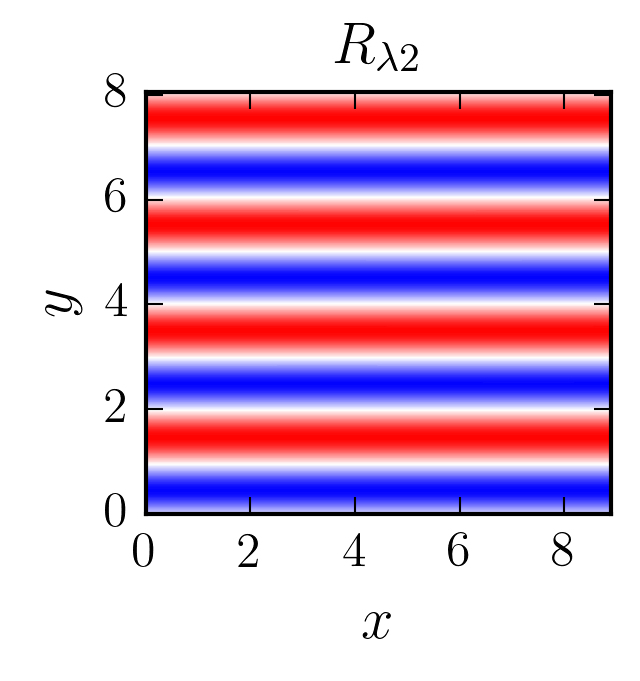}
        \end{subfigure}
        \begin{subfigure}[b]{0.23\textwidth}
                \includegraphics[width=\linewidth,trim={0.1cm 0.1cm 0.1cm 0.1cm},clip]{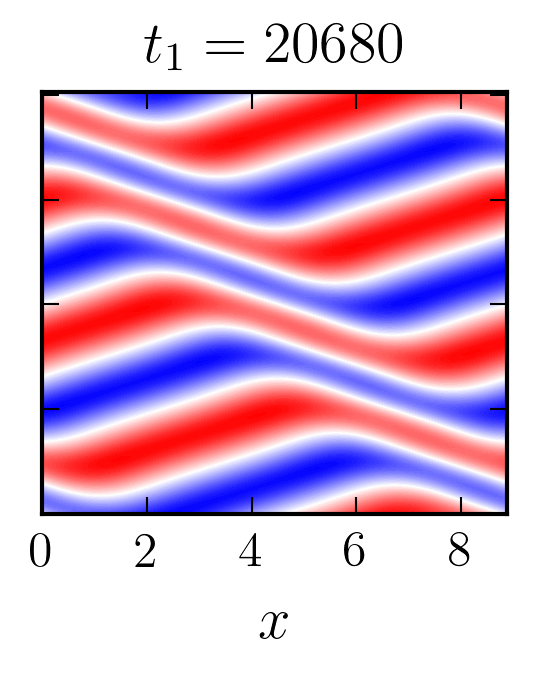}
        \end{subfigure}
        \begin{subfigure}[b]{0.23\textwidth}
                \includegraphics[width=\linewidth,trim={0.1cm 0.1cm 0.1cm 0.1cm},clip]{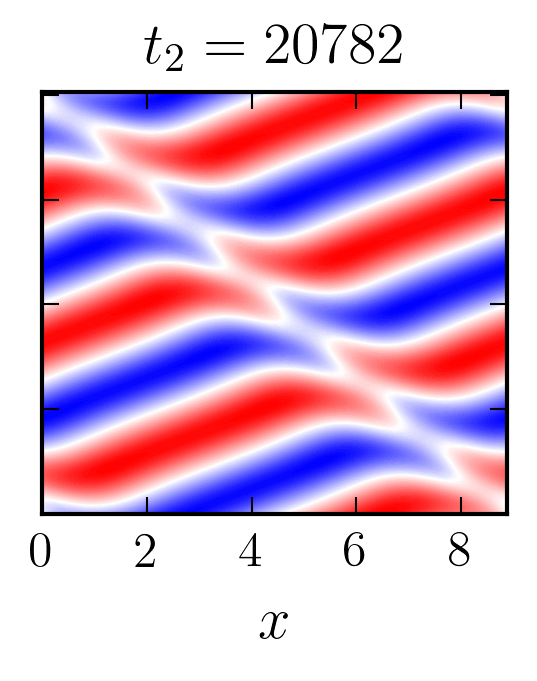}
        \end{subfigure}
        \begin{subfigure}[b]{0.23\textwidth}
                \includegraphics[width=\linewidth,trim={0.1cm 0.1cm 0.1cm 0.1cm},clip]{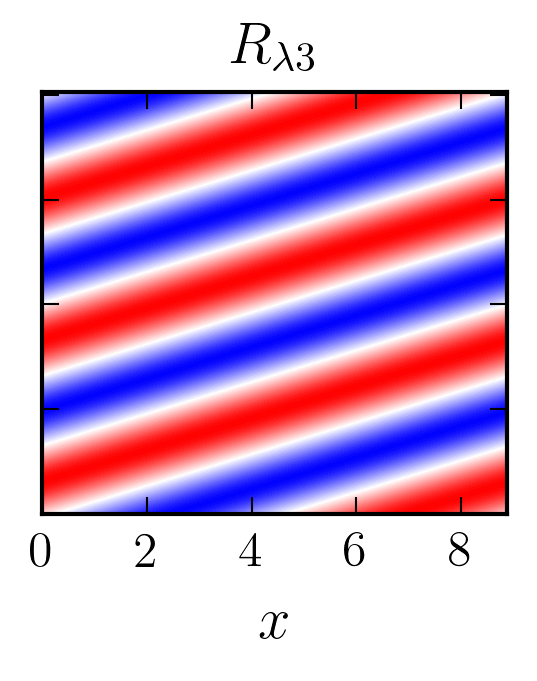}
        \end{subfigure}
\caption{\label{fig:pport_sv} State trajectory evolution from the unstable base state $B$ at $\gamma=0^{\circ}$ and $\epsilon=1.05$ ($\mathrm{Ra}=3500$). Initially for $0<t<40$, a fast transition along $B\rightarrow R_{\lambda 2}$ gives rise to straight convection rolls at wavelength $\lambda_y$ (not shown). After a very long transient close to $R_{\lambda 2}$, the shear driven instability $SV_i$ of straight convection rolls $R_{\lambda 2}$ leads to rolls $R_{\lambda 3}$ at increased wavelength. The inset magnifies the phase portrait of the transition $R_{\lambda 2}\rightarrow R_{\lambda 3}$.  During the transition $R_{\lambda 2}\rightarrow R_{\lambda 3}$ around $t_1=20\,680$, a skewed varicose pattern emerges transiently. The midplane temperature contours (bottom) illustrate different instances along the state trajectory (cross markers). Unlike in previous figures, the kinetic energy input $I$ is not normalized by the laminar input $I_0$ since $I_0=0$ at $\gamma=0^{\circ}$ (see Equation \ref{eq:laminar_energy}).}
\end{figure}

The conducting state of the Rayleigh-B\'enard system before convection onset is isotropic in the $x$-$y$-plane and rolls have no preferred orientation in the infinite system. Therefore, we denote straight convection rolls in the isotropic system as $R_{\lambda}$ with a subscript indicating the approximate pattern wavelength $\lambda$ if $\gamma=0^{\circ}$. The attributes `longitudinal' and `transverse' relate to the direction of rolls relative to the base flow and are only used in the inclined case. At control parameters $[\gamma,\epsilon,\mathrm{Pr}]=[0^{\circ},1.05,1.07]$, rolls of various wavelengths are unstable. To promote the growth of rolls at $\lambda_y$, we perturb the base state with small-amplitude rolls at $\lambda_y$ and aligned with the $x$-dimension. In addition, we add small-amplitude noise to break the translational symmetries. This perturbation of $B$ triggers the growth of four pairs of convection rolls $R_{\lambda 2}$, comparable to the pattern of $LR$ at $\gamma\neq 0^{\circ}$ in the present study. After a rapid transition $B \rightarrow R_{\lambda 2}$ over $\Delta t =40$, the shear driven instability $SV_i$ generates a slow departure from $R_{\lambda 2}$ over $\Delta t=20\,000$. The shear forces that drive $SV_i$ are generated solely by the convective motion of the secondary state, $R_{\lambda 2}$, as the primary base state, $B$, does not generate shear at $\gamma=0^{\circ}$. The exponential escape rate from $R_{\lambda 2}$ is given by the only positive real eigenvalue of $R_{\lambda 2}$, $\omega_r=3.8\cdot 10^{-4}$, with quadruple multiplicity. The associated eigenmodes show the characteristic three-dimensional oblique pattern of the skewed varicose instability \citep{Busse1979}. While escaping from $R_{\lambda 2}$, the convection rolls $R_{\lambda 2}$ start tilting and form a thin skewed region along the domain diagonal where rolls become strongly sheared ($t_1=20\,680$), pinch off ($t_2=20\,782$), reconnect and form rolls $R_{\lambda 3}$ at increased wavelength $\lambda=2.5731$ and rotated by $16.8^{\circ}$ against the $x$-direction (Figure \ref{fig:pport_sv}). $R_{\lambda 3}$ are linearly stable at these control parameter values.

The simulated sequence of dynamical connections $B\rightarrow R_{\lambda 2} \rightarrow R_{\lambda 3}$ is invariant under $S_{\mathrm{sv}}=\langle \pi_{xyz},\tau(0.25,0.25)\rangle$ and at $t_2$, resembles the experimentally observed SV pattern \citep[Figure 7 in][]{Bodenschatz2000}. However, the phase portrait does not indicate a transiently visited invariant state. The state trajectory crosses $D/I=1$ without slow-down (Figure \ref{fig:pport_sv}). Simulating the instability $SV_i$ at other values of the control parameters does not change this observation. Thus, the transient dynamics of the $\mathcal{SV}$ pattern seems to occur in the absence of an underlying invariant state.

\section{Discussion and conclusions}
\label{sec:discussion}
In this study, we identify exact invariant states of the fully nonlinear three-dimensional Oberbeck-Boussinesq equations that underlie the various convection patterns observed in ILC. At control parameter values where tertiary convection patterns have been observed in experiments and simulations, we numerically study the temporal dynamics from a perturbed unstable base flow. Table \ref{tab:transition} summarises all cases studied. Except for the transient skewed varicose pattern at $\gamma=0^{\circ}$, the dynamics asymptotically approaches stable invariant states underlying observed convection patterns. Temporal dynamics approaching attractive invariant states has been suggested by earlier works \cite{Daniels2008,Subramanian2016}. In these previous studies, numerical simulations close to known secondary instabilities, and in parts constrained to specific modal interactions, are found to approach nonlinear states emerging from the instabilities. Here, we find both stable and unstable invariant states as fixed points of a Newton iteration, numerically fully resolved and converged to machine precision. Details of the transition from the laminar state to transiently visited unstable invariant states are discussed. 

\begin{table}
  \begin{center}
    \caption{Summary of temporal transitions sequences identified at selected values of the control parameters where complex convection patterns are observed. For each transition along a dynamical connection, denoted by $\rightarrow$, we list the initial driving force of the instability (B: buoyancy, S: shear, E: equally).}
    \label{tab:transition}
    \begin{tabular}{| r | r | l | l | l | r |} 
      \toprule 
      $\gamma$ & $\mathrm{Ra}$ & $\epsilon$ & temporal transition sequence & driving forces & figure \\ \midrule
      $0^{\circ}$ & $3500$ & $1.05$ & $B\rightarrow R_{\lambda2}\rightarrow R_{\lambda3}$ & B, S & \ref{fig:pport_sv} \\
      $15^{\circ}$ & $4420$ & $1.5$ & $B\rightarrow LR\rightarrow SSW$ & B, E & \ref{fig:pport_lso} \\
      $40^{\circ}$ & $2385$ & $0.07$ & $B\rightarrow LR\rightarrow WR$ & B, S & \ref{fig:wr} \\
      $40^{\circ}$ & $3344$ & $0.5$ & $B\rightarrow LR\rightarrow WR \rightarrow OWR \rightarrow \tau_{xy}OWR \rightarrow$... & B, S, B, S, S, ... & \ref{fig:obwr_portrait} \\
      $80^{\circ}$ & $8525$ & $0.05$ & $B\rightarrow TR\rightarrow KN $ & S, B & \ref{fig:kn} \\
      $80^{\circ}$ & $9338$ & $0.15$ & $B\rightarrow TR\rightarrow B \rightarrow BKN $ & S, B, S & \ref{fig:knorb} \\
      $100^{\circ}$ & $10050$ & $0.1$ & $B\rightarrow TR\rightarrow TO$ & S, E & \ref{fig:pport_to} \\
      \bottomrule
    \end{tabular}
  \end{center}
\end{table}

State trajectories are never found to directly approach a stable tertiary or higher order state, but the dynamics first transiently visits unstable invariant states underlying the secondary convection pattern of straight convection rolls, as shown in Table \ref{tab:transition}. Approach to and escape from unstable invariant states follow the exponential dynamics along their stable and unstable manifolds, well described by the eigenvalues of the invariant states. This observation strongly suggests the existence of nearby heteroclinic connections between invariant states, located within the intersection of the unstable manifold of the initial state and the stable manifold of the final state. In one case at $\gamma=40^{\circ}$, a robust heteroclinic cycle between two symmetry-related unstable equilibrium states underlying oblique wavy rolls has been found. Thus, the present results demonstrate the dynamical relevance not only of stable but also of coexisting unstable invariant states and their dynamical connections. The network of dynamically connected invariant states clearly supports complex temporal dynamics in ILC.

Invariant states and their dynamical connections have been computed in minimal periodic domains matching single pattern wavelengths but they also exist in larger domains. The size of the domain does however change the stability properties of the invariant states. States that are stable in small domains can be unstable in larger domains \citep{Ahlers1978,Melnikov2014}. Consequently, the sequences of temporal transitions between invariant states observed here may also be observed in larger domains, but not necessarily with the same stable terminal state as in the present study. In the small domains, most transitions are unidirectional, from the base state $B$, via a secondary roll ($LR$, $TR$) to a tertiary state (Table \ref{tab:transition}). In larger domains, unstable tertiary states are expected to allow dynamical cycles that visit the same states multiple times. Examples of such cycles observed in the small domains include the robust heteroclinic cycle ($OWR\rightarrow \tau_{xy}OWR\rightarrow OWR\rightarrow$...) and the dynamics leading to the periodic orbit of bursting knots ($BKN$): After escaping from unstable laminar flow and transiently visiting $TR$, the state trajectory returns to the state space neighborhood of laminar flow $B$ (Figure \ref{fig:knorb}). Together, the connections may form a dynamical network supporting the spatio-temporally chaotic dynamics observed in experiments and large-domain simulations.

We characterise the instabilities of equilibrium states that trigger temporal transitions as buoyancy driven, shear driven or equally buoyancy-shear driven, by analysing the temporal transitions in phase portraits defined by kinetic energy input and dissipation. We thereby confirmed that $\mathcal{LR}$ emerge from a buoyancy driven instability and $\mathcal{TR}$ emerge from a shear driven instability of the base state. Secondary instabilities are never driven by the same force as the associated primary instability. If the primary instability is buoyancy dominated, the secondary one will involve shear and \emph{vice versa} (Table \ref{tab:transition}). Consequently, the temporal dynamics in ILC at all angles of inclinations may involve instabilities driven by buoyancy and shear.

We find seven invariant states that participate in sequences of temporal transitions that may be described as \emph{primary state} $\rightarrow$ \emph{secondary state} $\rightarrow$ \emph{tertiary state}. Here, the terminology refers to the order of states visited in transition sequences such that \emph{primary} transitions to \emph{secondary}, \emph{secondary} transitions to \emph{tertiary}. We expect that this order reflects the order of bifurcations that create these invariant states. However, generically the sequence of bifurcations will not prescribe the order in which states, coexisting at the same values of the control parameters, are visited during temporal evolution. An example for this is the temporal evolution triggered by the skewed varicose instability of straight convection rolls. The transition $R_{\lambda 2}\rightarrow R_{\lambda 3}$ cannot result from bifurcations of $R_{\lambda 3}$ from $R_{\lambda 2}$ because both types of straight convection rolls must bifurcate from $B$. To understand the relation between the complex temporal dynamics reported here and the corresponding bifurcation structure, the bifurcations of the invariant states visited by the dynamics must be computed. We list all invariant states found to underlie observed convection patterns in the present study in Table \ref{tab:states}. This collection represents the starting point for subsequent work where invariant states are numerically continued under changing control parameters to compute bifurcation diagrams in ILC \citep{Reetz2020b}.

\begin{table}
  \begin{center}
    \caption{Summary of invariant states underlying observed convection patterns  ($EQ$: equilibrium, $PPO$: pre-periodic orbit). The symmetries of the invariant states are given by the size of the periodic domain ($\lambda_x=2.2211$, $\lambda_y=2.0162$) and the generators of the symmetry group.}
    \label{tab:states}
    \begin{tabular}{| l | c | c | c | l |} 
      \toprule
      convection pattern & invariant state & type & domain & symmetry group generators \\ \midrule
      laminar flow & $B$ & EQ & -  & $\pi_y$, $\pi_{xz}$, $\tau(a_x,a_y)$    \\ \midrule
      isotropic rolls & $R_{\lambda}$ & EQ  & $[4\lambda_x,4\lambda_y]$ & $\pi_y$, $\pi_{xz}$, $\tau(a_x,a_y(a_x))$   \\
      long. rolls & $LR$ & EQ  & $[1\lambda_x,1\lambda_y]$ & $\pi_y$, $\pi_{xz}$, $\tau(a_x,0)$  \\
      trans. rolls & $TR$ & EQ  & $[1\lambda_x,1\lambda_y]$ & $\pi_y$, $\pi_{xz}$, $\tau(0,a_y)$  \\ \midrule
      skewed varicose & - &  -  & $[4\lambda_x,4\lambda_y]$ & $\pi_{xyz}$, $\tau(0.25,0.25)$  \\
      \makecell{subharmonic \\ standing wave} & $SSW$ & PPO   & $[2\lambda_x,2\lambda_y]$ & $\pi_{xyz}$, $\tau(0.5,0.5)$  \\
      wavy rolls & $WR$ & EQ  & $[2\lambda_x,1\lambda_y]$ &  $\pi_y\tau(0.5,0.5)$, $\pi_{xz}\tau(0.5,0.5)$ \\
      oblique wavy rolls & $OWR$ & EQ  & $[2\lambda_x,1\lambda_y]$ &  $\pi_{xyz}$, $\tau(0.5,0.5)$  \\
      knots & $KN$ & EQ  & $[1\lambda_x,1\lambda_y]$ & $\pi_y\tau(0,0.5)$, $\pi_{xz}\tau(0,0.5)$   \\
      trans. oscillations & $TO$ & PPO  & $[12\lambda_x,6\lambda_y]$ & $\pi_y$, $\pi_{xz}$, $\tau(0.5,0.5)$   \\ \bottomrule
      
    \end{tabular}
  \end{center}
\end{table}

In conclusion, temporal transitions from unstable laminar flow in ILC are found to follow sequences of dynamical connections between unstable invariant states until the dynamics approaches a stable invariant state. The stable invariant state underlies the basic pattern observed in experiments and simulations. Existence and dynamical influence of the dynamical connections between unstable invariant states support the complex dynamics observed in large domains.

\section*{Declaration of Interests}
The authors report no conflict of interest.

\section*{Acknowledgements}
\begin{acknowledgments}
We thank Laurette Tuckerman and Edgar Knobloch for many helpful discussions on the formalisation of symmetries in doubly-periodic domains, the structural stability of heteroclinic cycles and various other aspects of the work, as well as for detailed comments on the manuscript. Furthermore, we thank Priya Subramanian for regular discussions, in particluar on primary and secondary instabilities in ILC. This work was supported by the Swiss National Science Foundation (SNF) under grant no. 200021-160088.
\end{acknowledgments}


\begin{thebibliography}{72}
\expandafter\ifx\csname natexlab\endcsname\relax\def\natexlab#1{#1}\fi

\bibitem[Ahlers \& Behringer(1978)]{Ahlers1978}
{\sc Ahlers, G. \& Behringer, R.~P.} 1978 {Evolution of Turbulence from the
  Rayleigh-B{\'{e}}nard Instability}. {\em Physical Review Letters\/} {\bf
  40}~(11), 712--716.

\bibitem[Armbruster {\em et~al.\/}(1988)Armbruster, Guckenheimer \&
  Holmes]{Armbruster1988}
{\sc Armbruster, D., Guckenheimer, J. \& Holmes, P.} 1988 {Heteroclinic cycles
  and modulated travelling waves in systems with O(2) symmetry}. {\em Physica
  D: Nonlinear Phenomena\/} {\bf 29}~(3), 257--282.

\bibitem[Bergeon \& Knobloch(2002)]{Bergeon2002}
{\sc Bergeon, A. \& Knobloch, E.} 2002 {Natural doubly diffusive convection in
  three-dimensional enclosures}. {\em Physics of Fluids\/} {\bf 14}~(9),
  3233--3250.

\bibitem[Bodenschatz {\em et~al.\/}(2000)Bodenschatz, Pesch \&
  Ahlers]{Bodenschatz2000}
{\sc Bodenschatz, E., Pesch, W. \& Ahlers, G.} 2000 {Recent Developments in
  Rayleigh-B{\'{e}}nard Convection}. {\em Annual Review of Fluid Mechanics\/}
  {\bf 32}~(1), 709--778.

\bibitem[Budanur \& Cvitanovi{\'{c}}(2017)]{Budanur2017}
{\sc Budanur, N.~B. \& Cvitanovi{\'{c}}, P.} 2017 {Unstable Manifolds of
  Relative Periodic Orbits in the Symmetry-Reduced State Space of the
  Kuramoto–Sivashinsky System}. {\em Journal of Statistical Physics\/} {\bf
  167}~(3-4), 636--655.

\bibitem[Busse(1978)]{Busse1978a}
{\sc Busse, F.} 1978 {Non-linear properties of thermal convection}. {\em Rep.
  Prog. Phys.\/} {\bf 41}~(41).

\bibitem[Busse \& Clever(1979)]{Busse1979}
{\sc Busse, F.~H. \& Clever, R.~M.} 1979 {Instabilities of convection rolls in
  a fluid of moderate Prandtl number}. {\em Journal of Fluid Mechanics\/} {\bf
  91}~(2), 319--335.

\bibitem[Busse \& Clever(1992)]{Busse1992}
{\sc Busse, F.~H. \& Clever, R.~M.} 1992 {Three-dimensional convection in an
  inclined layer heated from below}. {\em Journal of Engineering Mathematics\/}
  {\bf 26}~(1), 1--19.

\bibitem[Busse \& Clever(1996)]{Busse1996}
{\sc Busse, F.~H. \& Clever, R.~M.} 1996 {The sequence-of-bifurcations approach
  towards an understanding of complex flows}. In {\em Mathematical Modeling and
  Simulation in Hydrodynamic Stability\/} (ed. D.~N. Riahi), pp. 15--34. World
  Scientific.

\bibitem[Busse \& Clever(2000)]{Busse2000}
{\sc Busse, F.~H. \& Clever, R.~M.} 2000 {Bursts in inclined layer convection}.
  {\em Physics of Fluids\/} {\bf 12}~(8), 2137--2140.

\bibitem[Canuto {\em et~al.\/}(2006)Canuto, Hussaini, Quarteroni \&
  Zang]{Canuto2006}
{\sc Canuto, C., Hussaini, M.~Y., Quarteroni, A. \& Zang, T.~A.} 2006 {\em
  {Spectral Methods: Fundamentals in Single Domains}\/}. Springer.

\bibitem[Canuto \& Landriani(1986)]{Canuto1986}
{\sc Canuto, C. \& Landriani, G.~S.} 1986 {Analysis of the Kleiser-Schumann
  method}. {\em Numerische Mathematik\/} {\bf 50}~(2), 217--243.

\bibitem[Chandler \& Kerswell(2013)]{Chandler2013}
{\sc Chandler, G.~J. \& Kerswell, R.~R.} 2013 {Invariant recurrent solutions
  embedded in a turbulent two-dimensional Kolmogorov flow}. {\em Journal of
  Fluid Mechanics\/} {\bf 722}, 554--595.

\bibitem[Chen \& Pearlstein(1989)]{Chen1989}
{\sc Chen, Y.-M. \& Pearlstein, A.~J.} 1989 {Stability of free-convection flows
  of variable-viscosity fluids in vertical and inclined slots}. {\em Journal of
  Fluid Mechanics\/} {\bf 198}~(-1), 513.

\bibitem[Chill{\`{a}} \& Schumacher(2012)]{Chilla2012}
{\sc Chill{\`{a}}, F. \& Schumacher, J.} 2012 {New perspectives in turbulent
  Rayleigh-B{\'{e}}nard convection}. {\em The European Physical Journal E\/}
  {\bf 35}~(7), 58.

\bibitem[Clever \& Busse(1995)]{Clever1995}
{\sc Clever, R. \& Busse, F.} 1995 {Tertiary and quarternary solutions for
  convection in a vertical fluid layer heated from the side}. {\em Chaos,
  Solitons {\&} Fractals\/} {\bf 5}~(10), 1795--1803.

\bibitem[Clever \& Busse(1977)]{Clever1977a}
{\sc Clever, R.~M. \& Busse, F.~H.} 1977 {Instabilities of longitudinal
  convection rolls in an inclined layer}. {\em Journal of Fluid Mechanics\/}
  {\bf 81}~(01), 107.

\bibitem[Clever \& Busse(1992)]{Clever1992}
{\sc Clever, R.~M. \& Busse, F.~H.} 1992 {Three-dimensional convection in a
  horizontal fluid layer subjected to a constant shear}. {\em Journal of Fluid
  Mechanics\/} {\bf 234}~(-1), 511.

\bibitem[Clever {\em et~al.\/}(1977)Clever, Busse \& Kelly]{Clever1977b}
{\sc Clever, R.~M., Busse, F.~H. \& Kelly, R.~E.} 1977 {Instabilities of
  longitudinal convection rolls in couette flow}. {\em Zeitschrift f{\"{u}}r
  angewandte Mathematik und Physik ZAMP\/} {\bf 28}~(5), 771--783.

\bibitem[Coles(1965)]{Coles1965}
{\sc Coles, D.} 1965 {Transition in circular Couette flow}. {\em Journal of
  Fluid Mechanics\/} {\bf 21}~(3), 385--425.

\bibitem[Crawford \& Knobloch(1991)]{Crawford1991}
{\sc Crawford, J. \& Knobloch, E.} 1991 {Symmetry And Symmetry-Breaking
  Bifurcations In Fluid Dynamics}. {\em Annual Review of Fluid Mechanics\/}
  {\bf 23}~(1), 341--387.

\bibitem[Cross \& Hohenberg(1993)]{Cross1993}
{\sc Cross, M.~C. \& Hohenberg, P.~C.} 1993 {Pattern formation outside of
  equilibrium}. {\em Reviews of Modern Physics\/} {\bf 65}~(3), 851--1112.

\bibitem[Cvitanovi{\'{c}} {\em et~al.\/}(2017)Cvitanovi{\'{c}}, Artuso,
  Mainieri, Tanner \& Vattay]{chaosbookSec10}
{\sc Cvitanovi{\'{c}}, P., Artuso, R., Mainieri, R., Tanner, G. \& Vattay, G.}
  2017 {Section 10 "Flips, slides and turns"}. In {\em Chaos: Classical and
  Quantum\/}, version 15 edn. Niels Bohr Institute, Copenhagen.

\bibitem[Daniels {\em et~al.\/}(2000)Daniels, Plapp \&
  Bodenschatz]{Daniels2000}
{\sc Daniels, K., Plapp, B. \& Bodenschatz, E.} 2000 {Pattern Formation in
  Inclined Layer Convection}. {\em Physical Review Letters\/} {\bf 84}~(23),
  5320--5323.

\bibitem[Daniels \& Bodenschatz(2002)]{Daniels2002a}
{\sc Daniels, K.~E. \& Bodenschatz, E.} 2002 {Defect Turbulence in Inclined
  Layer Convection}. {\em Physical Review Letters\/} {\bf 88}~(3), 034501.

\bibitem[Daniels {\em et~al.\/}(2008)Daniels, Brausch, Pesch \&
  Bodenschatz]{Daniels2008}
{\sc Daniels, K.~E., Brausch, O., Pesch, W. \& Bodenschatz, E.} 2008
  {Competition and bistability of ordered undulations and undulation chaos in
  inclined layer convection}. {\em Journal of Fluid Mechanics\/} {\bf 597},
  261--282.

\bibitem[Daniels {\em et~al.\/}(2003)Daniels, Wiener \&
  Bodenschatz]{Daniels2003}
{\sc Daniels, K.~E., Wiener, R.~J. \& Bodenschatz, E.} 2003 {Localized
  Transverse Bursts in Inclined Layer Convection}. {\em Physical Review
  Letters\/} {\bf 91}~(11), 114501.

\bibitem[Dijkstra {\em et~al.\/}(2014)Dijkstra, Wubs, Cliffe, Doedel, Hazel,
  Lucarini, Salinger, Phipps, Sanchez-Umbria, Schuttelaars, Tuckerman \&
  Thiele]{Dijkstra2014}
{\sc Dijkstra, H.~A., Wubs, F.~W., Cliffe, A.~K., Doedel, E., Hazel, A.~L.,
  Lucarini, V., Salinger, A.~G., Phipps, E.~T., Sanchez-Umbria, J.,
  Schuttelaars, H., Tuckerman, L.~S. \& Thiele, U.} 2014 {Numerical Bifurcation
  Methods and their Application to Fluid Dynamics: Analysis beyond Simulation}.
  {\em Communications in Computational Physics\/} {\bf 15}~(1), 1--45.

\bibitem[Eckhardt {\em et~al.\/}(2007)Eckhardt, Schneider, Hof \&
  Westerweel]{Eckhardt2007}
{\sc Eckhardt, B., Schneider, T.~M., Hof, B. \& Westerweel, J.} 2007
  {Turbulence transition in pipe flow}. {\em Annual Review of Fluid
  Mechanics\/} {\bf 39}~(1), 447--468.

\bibitem[Farano {\em et~al.\/}(2019)Farano, Cherubini, Robinet, {De Palma} \&
  Schneider]{Farano2019}
{\sc Farano, M., Cherubini, S., Robinet, J.-C., {De Palma}, P. \& Schneider,
  T.~M.} 2019 {Computing heteroclinic orbits using adjoint-based methods}. {\em
  Journal of Fluid Mechanics\/} {\bf 858}, R3.

\bibitem[Frigo \& Johnson(2005)]{Frigo2005}
{\sc Frigo, M. \& Johnson, S.} 2005 {The Design and Implementation of FFTW3}.
  {\em Proceedings of the IEEE\/} {\bf 93}~(2), 216--231.

\bibitem[Fujimura \& Kelly(1993)]{Fujimura1993}
{\sc Fujimura, K. \& Kelly, R.~E.} 1993 {Mixed mode convection in an inclined
  slot}. {\em Journal of Fluid Mechanics\/} {\bf 246}, 545--568.

\bibitem[Gershuni \& Zhukhovitskii(1969)]{Gershuni1969}
{\sc Gershuni, G.~Z. \& Zhukhovitskii, E.~M.} 1969 {Stability of plane-parallel
  convective motion with respect to spatial perturbations}. {\em Prikl. Mat. i
  Mekh.\/} {\bf 33}~(5), 855--860.

\bibitem[Gibson \& Brand(2014)]{Gibson2014}
{\sc Gibson, J.~F. \& Brand, E.} 2014 {Spanwise-localized solutions of planar
  shear flows}. {\em Journal of Fluid Mechanics\/} {\bf 745}, 25--61.

\bibitem[Gibson {\em et~al.\/}(2008)Gibson, Halcrow \&
  Cvitanovi{\'{c}}]{Gibson2008}
{\sc Gibson, J.~F., Halcrow, J. \& Cvitanovi{\'{c}}, P.} 2008 {Visualizing the
  geometry of state space in plane Couette flow}. {\em Journal of Fluid
  Mechanics\/} {\bf 611}, 107--130.

\bibitem[Gibson {\em et~al.\/}(2019)Gibson, Reetz, Azimi, Ferraro, Kreilos,
  Schrobsdorff, Farano, Yesil, Sch{\"{u}}tz, Culpo \& Schneider]{Gibson2019}
{\sc Gibson, J.~F., Reetz, F., Azimi, S., Ferraro, A., Kreilos, T.,
  Schrobsdorff, H., Farano, M., Yesil, A.~F., Sch{\"{u}}tz, S.~S., Culpo, M. \&
  Schneider, T.~M.} 2019 {Channelflow 2.0}. {\em in preparation\/} .

\bibitem[de~Graaf \& van~der Held(1953)]{DeGraaf1953}
{\sc de~Graaf, J. G.~A. \& van~der Held, E. F.~M.} 1953 {The relation between
  the heat transfer and the convection phenomena in enclosed plane air layers}.
  {\em Applied Scientific Research, Section A\/} {\bf 3}, 393--409.

\bibitem[Gray \& Giorgini(1976)]{Gray1976}
{\sc Gray, D.~D. \& Giorgini, A.} 1976 {The validity of the boussinesq
  approximation for liquids and gases}. {\em International Journal of Heat and
  Mass Transfer\/} {\bf 19}~(5), 545--551.

\bibitem[Halcrow {\em et~al.\/}(2009)Halcrow, Gibson, Cvitanovi{\'{c}} \&
  Viswanath]{Halcrow2009}
{\sc Halcrow, J., Gibson, J.~F., Cvitanovi{\'{c}}, P. \& Viswanath, D.} 2009
  {Heteroclinic connections in plane Couette flow}. {\em Journal of Fluid
  Mechanics\/} {\bf 621}, 365--376.

\bibitem[Hart(1971{\natexlab{{\em a\/}}})]{Hart1971a}
{\sc Hart, J.~E.} 1971{\natexlab{{\em a\/}}} {Stability of the flow in a
  differentially heated inclined box}. {\em Journal of Fluid Mechanics\/} {\bf
  47}~(03), 547.

\bibitem[Hart(1971{\natexlab{{\em b\/}}})]{Hart1971b}
{\sc Hart, J.~E.} 1971{\natexlab{{\em b\/}}} {Transition to a wavy vortex
  r{\'{e}}gime in convective flow between inclined plates}. {\em Journal of
  Fluid Mechanics\/} {\bf 48}~(2), 265--271.

\bibitem[Hof(2004)]{Hof2004}
{\sc Hof, B.} 2004 {Experimental Observation of Nonlinear Traveling Waves in
  Turbulent Pipe Flow}. {\em Science\/} {\bf 305}~(5690), 1594--1598.

\bibitem[Hollands \& Konicek(1973)]{Hollands1973}
{\sc Hollands, K. \& Konicek, L.} 1973 {Experimental study of the stability of
  differentially heated inclined air layers}. {\em International Journal of
  Heat and Mass Transfer\/} {\bf 16}~(7), 1467--1476.

\bibitem[Kawahara {\em et~al.\/}(2012)Kawahara, Uhlmann \& van
  Veen]{Kawahara2012}
{\sc Kawahara, G., Uhlmann, M. \& van Veen, L.} 2012 {The Significance of
  Simple Invariant Solutions in Turbulent Flows}. {\em Annual Review of Fluid
  Mechanics\/} {\bf 44}~(1), 203--225.

\bibitem[Kelly(1994)]{Kelly1994}
{\sc Kelly, R.~E.} 1994 {The onset and development of thermal convection in
  fully developed shear flows}. {\em Advances in applied mechanics\/} {\bf 31},
  35--112.

\bibitem[Kerr(1996)]{Kerr1996}
{\sc Kerr, R.~M.} 1996 {Rayleigh number scaling in numerical convection}. {\em
  Journal of Fluid Mechanics\/} {\bf 310}, 139--179.

\bibitem[Kerswell(2005)]{Kerswell2005}
{\sc Kerswell, R.~R.} 2005 {Recent progress in understanding the transition to
  turbulence in a pipe}. {\em Nonlinearity\/} {\bf 18}~(6), R17--R44.

\bibitem[Kleiser \& Schumann(1980)]{Kleiser1980}
{\sc Kleiser, L. \& Schumann, U.} 1980 {Treatment of incompressibility and
  boundary conditions in 3-D numerical spectral simulations of plane channel
  flows}. In {\em Proceedings of the Third GAMM — Conference on Numerical
  Methods in Fluid Mechanics\/} (ed. E.~Hirschel), pp. 165----173. Viewweg,
  Braunschweig.

\bibitem[Krupa(1997)]{Krupa1997}
{\sc Krupa, M.} 1997 {Robust heteroclinic cycles}. {\em Journal of Nonlinear
  Science\/} {\bf 7}~(2), 129--176.

\bibitem[Krupa \& Melbourne(1995)]{Krupa1995}
{\sc Krupa, M. \& Melbourne, I.} 1995 {Asymptotic stability of heteroclinic
  cycles in systems with symmetry}. {\em Ergodic Theory and Dynamical
  Systems\/} {\bf 15}~(1), 121--147.

\bibitem[Lanford(1982)]{Lanford1982}
{\sc Lanford, O.~E.} 1982 {The Strange Attractor Theory of Turbulence}. {\em
  Annual Review of Fluid Mechanics\/} {\bf 14}, 347--364.

\bibitem[Malkus(1964)]{Malkus1964}
{\sc Malkus, W. V.~R.} 1964 {Boussinesq equations}. {\em Notes on the 1964
  Summer Study Program in Geophysical Fluid Dynamics at the Woods Hole
  Oceanographic Institution Vol. 1\/} pp. 1--12.

\bibitem[Melnikov {\em et~al.\/}(2014)Melnikov, Kreilos \&
  Eckhardt]{Melnikov2014}
{\sc Melnikov, K., Kreilos, T. \& Eckhardt, B.} 2014 {Long-wavelength
  instability of coherent structures in plane Couette flow}. {\em Physical
  Review E\/} {\bf 89}~(4), 043008.

\bibitem[Mercader {\em et~al.\/}(2002)Mercader, Prat \& Knobloch]{Mercader2002}
{\sc Mercader, I., Prat, J. \& Knobloch, E.} 2002 {Robust heteroclinic cycles 
in two-dimensional Rayleigh-B{\'{e}}nard convection without Boussinesq symmetry}. 
{\em International Journal of Bifurcation and Chaos\/} {\bf
  12}~(11), 2501--2522.

\bibitem[Morris {\em et~al.\/}(1993)Morris, Bodenschatz, Cannell \&
  Ahlers]{Morris1993}
{\sc Morris, S.~W., Bodenschatz, E., Cannell, D.~S. \& Ahlers, G.} 1993 {Spiral
  defect chaos in large aspect ratio Rayleigh-B{\'{e}}nard convection}. {\em
  Physical Review Letters\/} {\bf 71}~(13), 2026--2029.

\bibitem[Nagata(1990)]{Nagata1990}
{\sc Nagata, M.} 1990 {Three-dimensional finite-amplitude solutions in plane
  Couette flow: bifurcation from infinity}. {\em Journal of Fluid Mechanics\/}
  {\bf 217}~(-1), 519--527.

\bibitem[Nore {\em et~al.\/}(2003)Nore, Tuckerman, Daube \& Xin]{Nore2003}
{\sc Nore, C., Tuckerman, L.~S., Daube, O. \& Xin, S.} 2003 {The 1:2 mode
  interaction in exactly counter-rotating von K{\'{a}}rm{\'{a}}n swirling
  flow}. {\em Journal of Fluid Mechanics\/} {\bf 477}~(477), 51--88.

\bibitem[Nusselt(1908)]{Nusselt1908}
{\sc Nusselt, W.} 1908 {Die W{\"{a}}rmeleitf{\"{a}}higkeit von
  W{\"{a}}rmeisolierstoffen}. PhD thesis.

\bibitem[Peyret(2002)]{Peyret2002}
{\sc Peyret, R.} 2002 {\em {Spectral Methods for Incompressible Flows}\/}.
  Springer-Verlag.

\bibitem[Prigent {\em et~al.\/}(2002)Prigent, Gr{\'{e}}goire, Chat{\'{e}},
  Dauchot \& van Saarloos]{Prigent2002}
{\sc Prigent, A., Gr{\'{e}}goire, G., Chat{\'{e}}, H., Dauchot, O. \& van
  Saarloos, W.} 2002 {Large-Scale Finite-Wavelength Modulation within Turbulent
  Shear Flows}. {\em Physical Review Letters\/} {\bf 89}~(1), 014501.

\bibitem[Proctor \& Jones(1988)]{Proctor1988}
{\sc Proctor, M. R.~E. \& Jones, C.~A.} 1988 {The interaction of two spatially
  resonant patterns in thermal convection. Part 1. Exact 1:2 resonance}. {\em
  Journal of Fluid Mechanics\/} {\bf 188}, 301--335.

\bibitem[Reetz \& Schneider(2020)]{Reetz2020d}
{\sc Reetz, F. \& Schneider, T.~M.} 2020 {Invariant state space structure of
  weakly turbulent inclined layer convection}. {\em in preparation\/} .

\bibitem[Reetz {\em et~al.\/}(2020)Reetz, Subramanian \& Schneider]{Reetz2020b}
{\sc Reetz, F., Subramanian, P. \& Schneider, T.~M.} 2020 {Invariant states in
  inclined layer convection. Part 2. Bifurcations and connections between
  branches of invariant states}. {\em under revision at JFM\/} pp. 1--36.

\bibitem[Ruth {\em et~al.\/}(1980)Ruth, Raithby \& Hollands]{Ruth1980a}
{\sc Ruth, D.~W., Raithby, G.~D. \& Hollands, K. G.~T.} 1980 {On the secondary
  instability in inclined air layers}. {\em Journal of Fluid Mechanics\/} {\bf
  96}~(3), 481--492.

\bibitem[Schneider {\em et~al.\/}(2007)Schneider, Eckhardt \&
  Yorke]{Schneider2007b}
{\sc Schneider, T.~M., Eckhardt, B. \& Yorke, J.~A.} 2007 {Turbulence
  transition and the edge of chaos in pipe flow}. {\em Physical Review
  Letters\/} {\bf 99}, 34502.

\bibitem[Schneider {\em et~al.\/}(2008)Schneider, Gibson, Lagha, {De Lillo} \&
  Eckhardt]{Schneider2008b}
{\sc Schneider, T.~M., Gibson, J.~F., Lagha, M., {De Lillo}, F. \& Eckhardt,
  B.} 2008 {Laminar-turbulent boundary in plane Couette flow}. {\em Physical
  Review E\/} {\bf 78}~(3), 037301.

\bibitem[Shadid \& Goldstein(1990)]{Shadid1990}
{\sc Shadid, J.~N. \& Goldstein, R.~J.} 1990 {Visualization of longitudinal
  convection roll instabilities in an inclined enclosure heated from below}.
  {\em Journal of Fluid Mechanics\/} {\bf 215}, 61--84.

\bibitem[Skufca {\em et~al.\/}(2006)Skufca, Yorke \& Eckhardt]{Skufca2006}
{\sc Skufca, J.~D., Yorke, J.~A. \& Eckhardt, B.} 2006 {Edge of chaos in a
  parallel shear flow}. {\em Physical Review Letters\/} {\bf 96}, 174101.

\bibitem[Subramanian {\em et~al.\/}(2016)Subramanian, Brausch, Daniels,
  Bodenschatz, Schneider \& Pesch]{Subramanian2016}
{\sc Subramanian, P., Brausch, O., Daniels, K.~E., Bodenschatz, E., Schneider,
  T.~M. \& Pesch, W.} 2016 {Spatio-temporal patterns in inclined layer
  convection}. {\em Journal of Fluid Mechanics\/} {\bf 794}, 719--745.

\bibitem[Suri {\em et~al.\/}(2017)Suri, Tithof, Grigoriev \& Schatz]{Suri2017}
{\sc Suri, B., Tithof, J., Grigoriev, R.~O. \& Schatz, M.~F.} 2017 {Forecasting
  Fluid Flows Using the Geometry of Turbulence}. {\em Physical Review
  Letters\/} {\bf 118}~(11), 114501.

\bibitem[Viswanath(2007)]{Viswanath2007}
{\sc Viswanath, D.} 2007 {Recurrent motions within plane Couette turbulence}.
  {\em Journal of Fluid Mechanics\/} {\bf 580}, 339--358.

\bibitem[Waleffe(1998)]{Waleffe1998}
{\sc Waleffe, F.} 1998 {Three-dimensional coherent states in plane shear
  flows}. {\em Physical Review Letters\/} {\bf 81}, 4140--4143.

\end{thebibliography}

\appendix

\section{Time-stepping algorithm}
\label{sec:timestepper}

Inserting the two-dimensional Fourier mode expansion 
\begin{align}
\left[\bm{u},\theta\right](x,y,z,t) &= \sum_{k_x=-K_x}^{K_x}  \; \sum_{k_y=-K_y}^{K_y} \; 
\left[\bar{\bm{u}},\bar{\theta}\,\right]_{k_x,k_y}(z,t) \; e^{2 \pi i \left(k_x x/L_x + k_y y/L_y \right)}
\label{eq:app:fourierexpansion}
\end{align}
together with the base-fluctuation decomposition $[\bm{U},\mathcal{T}]=[\bm{U}_0,\mathcal{T}_0]+[\bm{u},\theta]$ into (\ref{eq:obe1}-\ref{eq:obe3}) allows writing the governing equations in the form
\begin{equation}
 \frac{\partial }{\partial t} \bar{\bm{x}}(z,t)=\mathcal{L} \bar{\bm{x}}(z,t) -\mathcal{N}_{\bm{x}}(\bar{\bm{x}}(z,t)) \label{eq:app:general}
\end{equation}
with $\bar{\bm{x}}=[\bar{\bm{u}},\bar{\theta}]$. Quantities with an overbar, $\bar{\bm{x}}$, denote Fourier-transformed fields along the $x$- and the $y$-dimension, with the $z$-dimension remaining in physical space. The time-stepper operates on this mixed representation of velocity and temperature. Note that the fully spectral representation, denoted by $\tilde{\bm{x}}$ in (\ref{eq:numexpansion}), is composed of spectral Fourier-Fourier-Chebyshev coefficients in all three space dimensions. Using the mixed operator
\begin{equation}
\bar{\nabla}=\left(2\pi i\left(\frac{k_x}{L_x}\hat{\bm{e}}_x + \frac{k_y}{L_y}\hat{\bm{e}}_y \right)+\frac{\partial}{\partial z}\hat{\bm{e}}_z \right) \ ,
\end{equation}
the linear terms for velocity and temperature fields are defined as
\begin{align}
 \mathcal{L}\bar{\bm{u}}& =  \tilde{\nu} \left(\bar{\nabla}\cdot \bar{\nabla} \right ) \bar{\bm{u}} -\bar{\nabla}\bar{p} \ ,\\
  \mathcal{L}\bar{\theta}& = \tilde{\kappa} \left(\bar{\nabla}\cdot \bar{\nabla} \right)  \bar{\theta} \ ,
\end{align}
and the nonlinear terms are defined as
\begin{align}
 \mathcal{N}_{\bm{u}}(\bar{\bm{x}})&= f\left(f^{-1}\left(\bar{\bm{U}}\right)\cdot f^{-1}\left(\bar{\nabla} \bar{\bm{U}}\right)\right) +C_{\bm{u}} -\sin(\gamma)\bar{\theta}\hat{\bm{e}}_x - \cos(\gamma)\bar{\theta}\hat{\bm{e}}_z  \ , \label{eq:app:nonlin1} \\
  \mathcal{N}_{\theta}(\bar{\bm{x}})&= f\left(f^{-1}\left(\bar{\bm{U}}\right) \cdot f^{-1}\left(\bar{\nabla}\bar{T}\right)\right) +C_{\theta} \ ,\label{eq:app:nonlin2}
\end{align}
with $f()$ and $f^{-1}()$ being the direct and inverse 2D Fourier transform, respectively. In this form the nonlinear terms in (\ref{eq:app:nonlin1}) and (\ref{eq:app:nonlin2}) are evaluated as pointwise products in physical space. This leads to the reduced computational complexity of this `pseudo-spectral' method when compared to fully spectral methods \citep[][p.133ff]{Canuto2006}. The constants $C_{\bm{u},\theta}$ are zero in the present study but allow to consider additional body forces. Different algorithms are implemented in \emph{Channelflow-ILC} to advance $\bar{\bm{x}}^n$ at time step $n$ over $\Delta t$. For this study we treat the linear term $\mathcal{L} \bar{\bm{x}}$ in (\ref{eq:app:general}) fully implicitly by using an implicit-explicit multi-step algorithm of $3rd$ order \citep[p.130]{Peyret2002}. The field at time step $n+1$ is given by
\begin{equation}
 \frac{a_0}{\Delta t}\bar{\bm{x}}^{n+1} -\mathcal{L}\bar{\bm{x}}^{n+1} = \sum_{j=1}^{k} \frac{-a_j}{\Delta t} \bar{\bm{x}}^{n+1-j} -b_{j-1} \mathcal{N}\left(\bar{\bm{x}}^{n+1-j} \right) \ , \label{eq:app:timestep}
\end{equation}
with $k=3$ and coefficients $[a_0,a_1,a_2,a_3]=[11/6,-3,3/2,-1/3]$ and $[b_0,b_1,b_2]=[3,-3,1]$. Since the right-hand side is known, (\ref{eq:app:timestep}) defines a Helmholtz problem along the $z$-dimension that is solved implicitly using a Chebyshev tau method \citep[][p.173ff]{Canuto2006}.

\section{Rayleigh number scaling of heat transfer}
\label{sec:scaling}
\begin{figure}
		\begin{subfigure}[b]{0.9\textwidth}
        \begin{tikzpicture}
    	\draw (0, 0) node[inner sep=0]{\includegraphics[width=\linewidth,trim={0.1cm 0.1cm 0.1cm 0.1cm},clip]{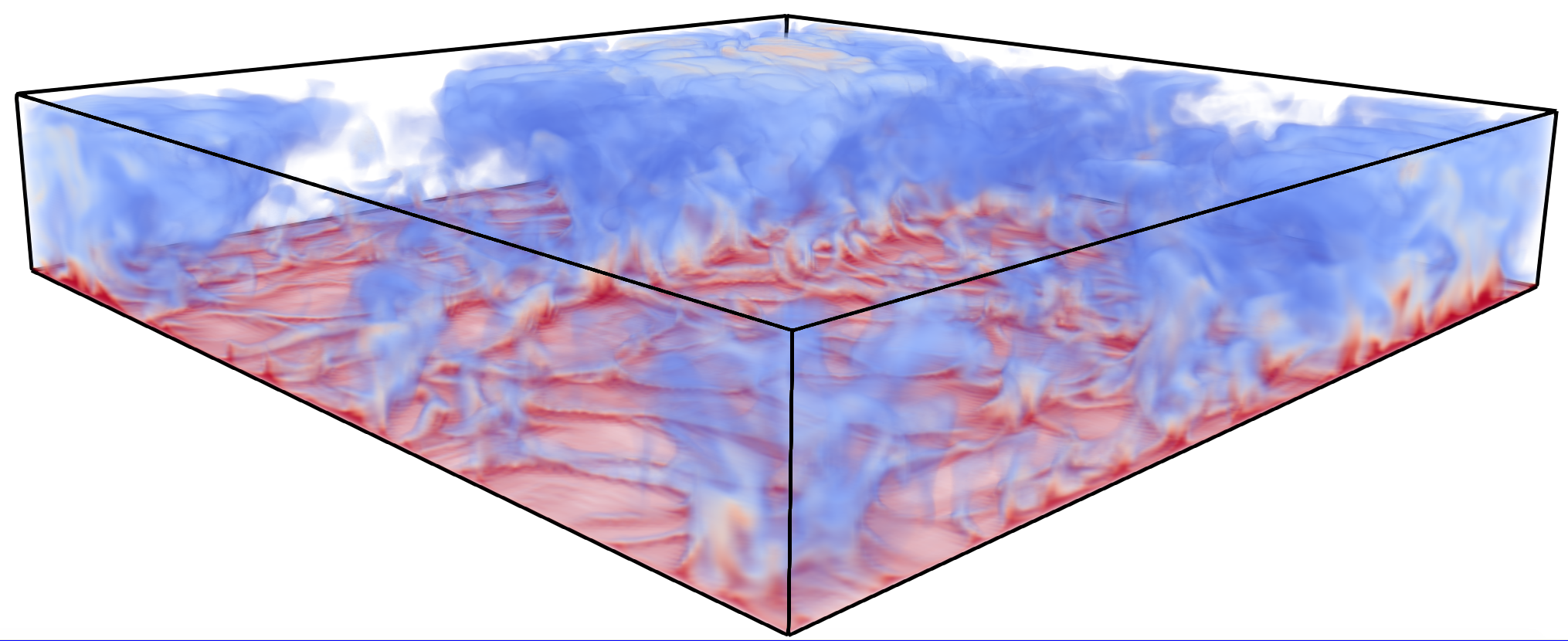}};
    	\draw (-5.2,-0.8) node {\textbf{(a)}};
		\end{tikzpicture}
		\end{subfigure}
		\begin{subfigure}[b]{0.99\textwidth}
        \begin{tikzpicture}
    	\draw (0, 0) node[inner sep=0]{\includegraphics[width=\linewidth,trim={0.1cm 7.1cm 0cm 1cm},clip]{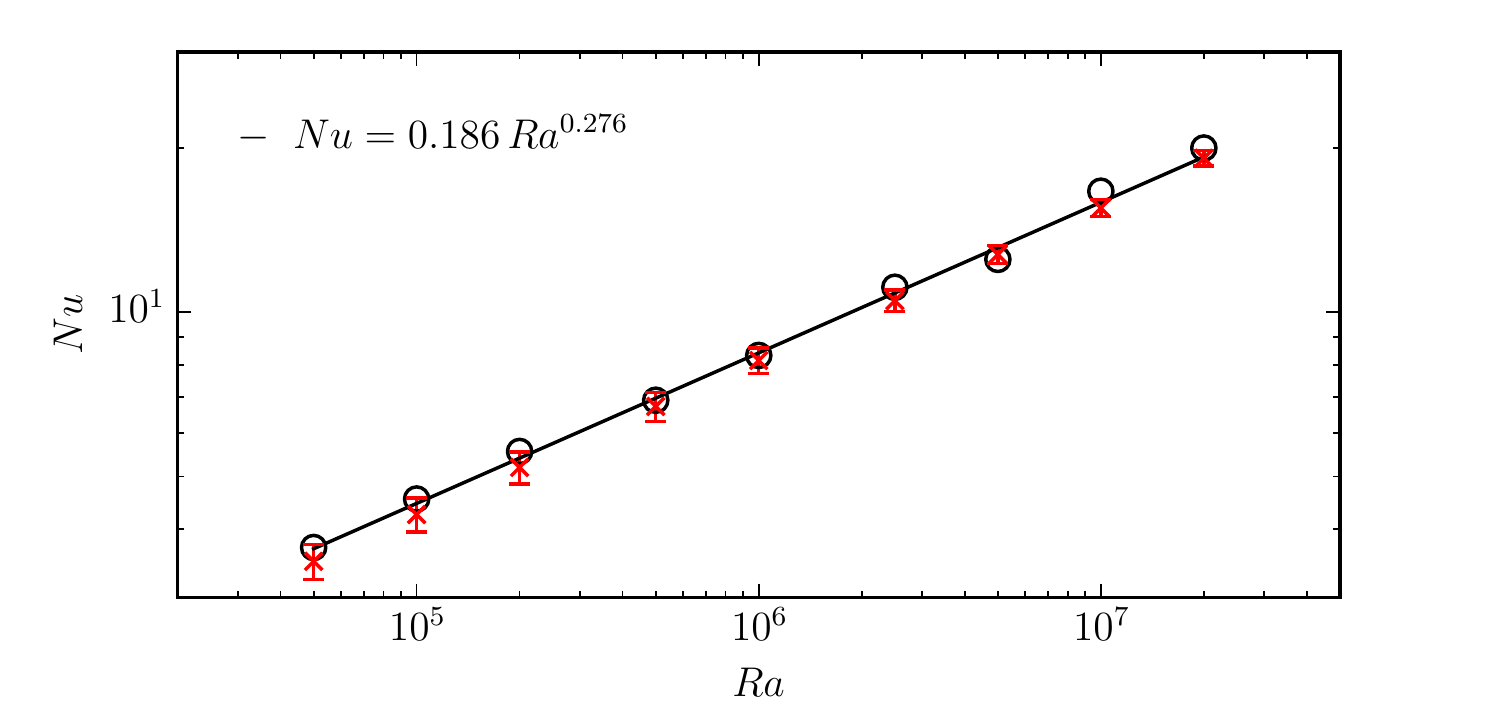}};
    	\draw (-5.9,-2.8) node {\textbf{(b)}};
		\end{tikzpicture}
		\end{subfigure}
       	\caption{\label{fig:turbconv} DNS of turbulent convection in Rayleigh-B\'enard convection at $\mathrm{Pr}=0.7$ using \emph{Channelflow-ILC}. \textbf{(a)} Volumetric rendered temperature field of positive temperature fluctuations around the laminar solution at $\mathrm{Ra}=10^7$ illustrates range of scales of turbulent convection plumes. \textbf{(b)} Scaling of Nusselt number $\mathrm{Nu}$ with Rayleigh number $\mathrm{Ra}$ of the present DNS statistics ($\times $) compare well to the DNS statistics of \citet{Kerr1996} ($\circ$) which fit to $\mathrm{Nu}=0.186\, \mathrm{Ra}^{0.276}$ ($-$). Error bars indicate three standard deviations around the mean of $\mathrm{Nu}(t)$, averaged over $T=2500$ free fall time units. }
\end{figure}
The intention of this article is to numerically study temporal dynamics in ILC at small $\mathrm{Ra}$ to explain the dynamics of weakly turbulent patterns. This requires a numerical implementation of ILC to correctly handle the nonlinearities of the governing equations. In order to demonstrate highly nonlinear behavior, we perform DNS of Rayleigh-B\'enard convection to compare the $\mathrm{Ra}$-scaling of Nusselt number $\mathrm{Nu}$ for turbulent convection with \citet{Kerr1996} where the same type of pseudo-spectral DNS is used as here. The DNS are for $\mathrm{Pr}=0.7$ and $\mathrm{Ra}\in [5\times 10^4,2\times 10^7]$ in a doubly periodic domain of lateral extent $[L_x,L_y]=[6,6]$, discretized by $[N_x,N_y,N_z]=[288,288,96]$ grid points (Figure \ref{fig:turbconv}a). $\mathrm{Nu}$ is calculated at the midplane and averaged over $T=2500$ free fall time units. The present simulated data follows the scaling of \citet{Kerr1996} (Figure \ref{fig:turbconv}b). The DNS at $\mathrm{Ra}=2\times 10^7$ used $11.6 \times 10^3\,CPUh$ requiring a runtime of approximately 1 week using 64 cores in parallel.

\end{document}